\documentclass{aa}  

\usepackage{graphicx}
\usepackage[colorlinks=true, allcolors=blue]{hyperref}
\hypersetup{
     colorlinks   = true,
     citecolor    = blue,
     urlcolor = blue
}
\usepackage{txfonts}
\usepackage{tabularx}

\begin{document}

   \title{Fast radio burst -- persistent radio source systems}
   \subtitle{I. A 1.2~GHz search and catalog definition}

   \author{
          D. Pelliciari
          \inst{1}
          \and
          L. Bruno\inst{1}
          \and
          G. Bernardi\inst{1,2,3}
          \and
          M. Pilia\inst{4}
          \and
          P. Esposito\inst{5,6}
          \and
          L. Beduzzi\inst{1,7}
          \and
          A. Geminardi\inst{4,6,8}
          \and
          O. Smirnov\inst{2,3,1,9,10}
          }

   \institute{INAF-Istituto di Radio Astronomia (IRA), via Piero Gobetti 101, Bologna, Italy\\
              \email{davide.pelliciari@inaf.it}
         \and
             Centre for Radio Astronomy Techniques and Technologies (RATT), Department of Physics and Electronics, Rhodes University, Makhanda 6140, South Africa
         \and
             South African Radio Astronomy Observatory, Cape Town 7700, South Africa
        \and
            INAF-Osservatorio Astronomico di Cagliari, via della Scienza 5, I-09047, Selargius (CA), Italy
        \and
            Scuola Universitaria Superiore IUSS Pavia, Palazzo del Broletto, piazza della Vittoria 15, I-27100 Pavia, Italy
        \and
            INAF–Istituto di Astrofisica Spaziale e Fisica Cosmica di Milano, via Corti 12, I-20133 Milano, Italy
        \and
            Dipartimento di Fisica e Astronomia, Universitá di Bologna, via Gobetti 93/2, 40129 Bologna, Italy
        \and
            Dipartimento di Fisica, Università di Trento, via Sommarive 14, I-38123 Povo (TN), Italy        
        \and
            Astrophysics, Department of Physics, University of Oxford, Keble Road, Oxford, OX1 3RH, UK
        \and
            Breakthrough Listen, Astrophysics, Department of Physics, The University of Oxford, Keble Road, Oxford, OX1 3RH, UK
        }
   \date{Received April XX, 2026}

  \abstract 
{Fast radio bursts (FRBs) are millisecond-duration radio transients observed at extragalactic distances, whose origin remains uncertain. Given their extremely short duration and high implied luminosities, magnetars are commonly identified as sources of FRBs. A small ($\sim 2\%$) fraction of FRBs are known to repeat, but it is still unclear whether repeaters represent a distinct population. Four repeating FRBs present a co-spatial persistent radio source (PRS) that is compact at parsec scales, and is interpreted as a dense and highly magnetized synchrotron-emitting nebula.}
{We investigate the connection between FRBs and PRSs, in particular the PRS occurrence in a sample of localized FRBs.}
{We carried out new $1.26$~GHz observations of 24 FRBs (9 repeaters and 11 one-offs), conducted with the upgraded Giant Meter Wave Radio Telescope (uGMRT) at arc-second angular scale resolution, and combined them with literature data to construct an extended sample of 75 sources with either persistent luminosity measurements or upper limits. Our catalog is complete down to a $10^{29}$~erg~s$^{-1}$~Hz$^{-1}$ luminosity threshold, which corresponds to the value of the first-discovered PRS, associated with FRB 20121102.}
{We detect compact radio emission in six FRB fields, two of which are previously known PRSs and four are new candidates. The four new candidates are associated with three new repeating FRBs and one apparently one-off source. We computed the PRS occurrence using the whole catalog of 75 FRBs and found that PRSs more luminous than $10^{29}$~erg~s$^{-1}$~Hz$^{-1}$ are rare. Unless otherwise stated, all uncertainties and upper limits are reported at the $95\%$ confidence level (CL). If we only consider confirmed PRSs, we find an occurrence of $f_{\rm all} = 0.06^{+0.09}_{-0.03}$, while including PRS candidates as well -- i.e., sources for which parsec-scale constraints are not yet available -- yields $f_{\rm all} = 0.14^{+0.22}_{-0.08}$.

We also compute the PRS occurrence in the two classes of FRBs: repeating and non-repeating. In the conservative case, i.e., when none of the new PRS candidates are considered to be associated with their FRBs, we find no preferred association between the PRSs and FRB class, with an occurrence of $f_{\rm r} = 0.14^{+0.20}_{-0.09}$ for the sample of repeating FRBs and $f_{\rm nr} \leq 0.12$ for the sample of one-off sources.
In the inclusive case, i.e., when all the new candidates are considered confirmed PRSs, we find $f_{\rm r} = 0.27^{+0.21}_{-0.14}$ and $f_{\rm nr} = 0.03^{+0.14}_{-0.02}$, with marginal ($90\%$ CL) evidence that PRSs are preferably associated with repeating FRBs.}
{}


   \maketitle
    \nolinenumbers

\section{Introduction}\label{sec:Intro}
Fast radio bursts \citep[FRBs;][]{cordesChatterjee19,petroff21,Pilia22,Zhang22_rev} are short duration and luminous radio flashes characterized by a large dispersion measure (DM) even at high Galactic latitude \citep[e.g.,][]{cordesChatterjee19}. Approximately $100$ distinct FRB sources have been localized to their host galaxies \citep[e.g.,][]{Gordon23}, and the implied, extremely high brightness temperatures, combined with their short duration, favor compact objects as their progenitors. The most invoked are magnetars, i.e., neutron stars (NSs) powered by the decay of strong ($10^{14}$--$10^{15}$ G) internal magnetic fields \citep{DuncanThompson,ThompsonDuncan,Popov13,Beloborodov19,Liubarsky20,Sobacchi22}.

Nowadays, $\approx 3600$ different sources of FRBs are known; the majority of them are classified as one-off events, which means that, over tens of hours on-source, they have been detected only once. A small ($\sim 2\%$) fraction of the whole FRB population has been seen to repeat \citep{CHIMECat2}. It is not clear whether the latter sources, known as "repeaters", originate from a distinct class of progenitors from one-offs. Bursts from repeaters appear statistically longer in duration and less extended in frequency \citep{pleunis21,Sand25}, while population studies suggest that a large fraction of one-offs may eventually repeat \citep{james22b, james23, chime23}. Differences have been found among the population of repeating FRBs, especially in their level of activity, with burst rates ranging from $10^{-3}\ \rm hr^{-1}$ to $10^2\ \rm hr^{-1}$ \citep{chime23} at Jansky milliseconds fluence threshold. Moreover, deep sensitivity observations at $\approx 1$ mJy ms fluence limits revealed that some FRBs, referred to as ``hyperactive repeaters'', can show periods of activity characterized by burst rates as high as $500$ hr$^{-1}$ \citep[e.g.,][]{Li21,Zhang23}.

The precise localization of FRBs enabled  four repeaters to be robustly linked with persistent radio sources \citep[PRSs;][]{Chatterjee17,Marcote17,Niu21,Bhandari23b,Bruni23,Ibik24,Moroianu26}. The latter resemble synchrotron-emitting nebulae and are too luminous \citep[$L_\nu \geq 10^{29}$ erg s$^{-1}$ Hz$^{-1}$,][]{Law22} and compact \citep[on linear scales of parsecs;][]{Marcote17, Bhandari23b} to originate from star formation processes.

The first discovered PRS is the one associated with FRB 20121102A, hereafter R1, which was the first discovered repeater \citep{Chatterjee17,Marcote17,Chen23}. This is one of the most active FRBs known and has been studied for $\sim 14$ yr up to now \citep[e.g.,][]{Li21,Hewitt22,Jahns23}. R1 is located in a star-forming dwarf galaxy at $z \simeq 0.19273$ \citep{Tendulkar17} and its bursts show a time-varying rotation measure (RM) of $\sim 10^5$ rad m$^{-2}$, which suggests an extremely variable and highly magnetized environment near the source producing the bursts \citep{Michilli18}; this supports the hypothesis that a young magnetar ($10-40$ yrs) is the central engine of R1 \citep{Bassa17,margalitmetzger18}. The associated PRS has a position consistent with the burst's position within $\sim 12$ pc \citep{Snelders26}, and presents a nonthermal emission with a flat spectral index, $\alpha = -0.27 \pm 0.24$, similar to the spectral energy distribution (SED) of the nebula around PSR B0531+21 (Crab pulsar), although with orders of magnitude higher luminosity \citep{Resmi21,Vohl23}. The largely accepted model for a PRS is a strongly magnetized
nebula powered by a young, FRB-emitting magnetar \citep{margalitmetzger18}, but a hypernebula inflated by accretion onto a compact binary
system \citep{Sridhar21,Sridhar24} is also consistent with current observations.

A second, similar PRS was found for FRB 20190520B, often referred to as R1-twin for the similarities it shares with R1. It is another active repeater localized in a dwarf galaxy at $z \approx 0.241$ \citep{Niu21}, with bursts showing RM values of the same orders of magnitude as R1 \citep{Niu21,AnnaThomas23}. In this case  as well, the emission is compact at physical scales $< 9$ pc \citep{Bhandari23b}, and nonthermal, with a spectral index $\alpha = -0.41 \pm 0.04$ \citep{Niu21}. 

More recently, two other PRSs have been discovered, both associated with two repeating FRBs: FRBs 20190417A \citep{Ibik24,Moroianu26} and 20240114A \citep{Bhusare24,Bruni24}. Both the new FRB-PRS systems reside in dwarf galaxies. Moreover, FRB 20190417A exhibits a large and strongly time-variable of RM \citep{McKinven23}, while FRB 20240114A is one of the most active FRBs known \citep[e.g.,][]{Zhang24_hyper} but shows bursts with a significantly smaller RM values \citep{Bruni24}. Interestingly, the spectral luminosities of confirmed PRSs show a positive correlation with the FRB RMs
\citep[e.g.,][]{Bruni23,Ibik24}. This correlation is predicted under the hypothesis that the PRS is a magnetized nebula that is both responsible for the continuum radio emission and the Faraday rotation of the bursts \citep{Yang20, Yang22, Bruni23, Yang26}.

Significant observational efforts have been carried out to search for other PRSs \citep[e.g.,][]{Vohl23,Dong24,Pelliciari24,Bhusare24,Ibik24}. Remarkable cases are the repeating FRBs 20201124A \citep{Bruni23}, 20181030A \citep{Ibik24}, and 20220912A \citep{Pelliciari24,Bhusare24}, for which co-spatial compact radio sources have been reported. We investigate the compactness of the candidate PRS associated with FRB 20181030A in a companion work \citep[Paper II;][]{Pelliciari26_PaperII}. However, in these cases the PRS–FRB association remains tentative because the continuum sources have not been confirmed as compact at parsec scales \citep{Nimmo21,Hewitt24}, which implies they might not be related to the FRB.
 
In this work, we present new observations conducted with the upgraded Giant Meterwave Radio Telescope (uGMRT) at $1.26$~GHz, searching for PRSs in a sample of 24 FRBs that contain both repeating and one-off sources. We combined our results with existing observations to construct the largest catalog of FRB--PRS systems to date and used it to study the PRS occurrence rate. The paper is structured as follows. In Section~\ref{sec: sample_des} we present the FRB sources selected for our observational campaign and outline their properties, and in Section~\ref{sec:calib_reduction} we describe the uGMRT observations and data reduction. In Section~\ref{sec: Discuss} we present the extended FRB--PRS catalog and report the main results of our analysis, including the detection of new candidate PRSs. In Section~\ref{sec: occ} we derive the PRS occurrence rate for the population of sources included in the extended catalog. We summarize our findings and draw conclusions in Section~\ref{sec: conclusions}.

\section{Sample description}
\label{sec: sample_des}

We selected a sample of 24 FRBs, comprising 13 repeaters and 11 one-off sources, which we identify as S1, S2... up to S24. They have different localization accuracies and the position accuracy is better than 1~arcsec for 13 sources (S1, S5, S6, S9, S14, S15, S16, S17, S18, S19, S21, S22, and S24). Among them, 11 sources have a confirmed host galaxy association, with a spectroscopic redshift measured for the latter, but two sources (S21 and S24) do not have a reported redshift. Four other sources (S2, S3, S4, and S8) have less accurate localizations (up to $\approx 1$arcmin), but are still associated with host galaxies. The remaining seven sources have position accuracies better than 1 arcmin with no host association. Six of these were selected because of their measured RMs, which span $|{\rm RM}| \approx 180-1.2 \times 10^{4}$ rad m$^{-2}$ (see Paper III for the individual RM values).
 
The FRB hosts (where known), span a large variety of galaxy types, including star-forming galaxies (SFGs), dwarf galaxies, and systems hosting active galactic nuclei (AGNs). The sources without a secure host galaxy association are S7, S10, S11, S12, S13, S20, S23, and S24. All the relevant properties of the sources of our catalog are reported in Table \ref{table:FRB_sample}, while their localization properties are reported in Table~\ref{tab:loc_prop_sources}. 

We estimated the redshift of the sources without a host galaxy association by using the Macquart relation \citep{Maquart19}, which relates the DM due to the intergalactic medium (IGM) ${\rm DM}_{\rm IGM}$ and the FRB redshift:

\begin{equation}
    {\rm DM}_{\rm IGM}(z) = {\rm DM}_{\rm obs} - {\rm DM}_{\rm ISM} - {\rm DM}_{\rm halo} - {\rm DM}_{\rm host}(z),
\label{eq:macquart}    
\end{equation}where ${\rm DM}_{\rm ISM}$ and ${\rm DM}_{\rm halo}$ are the DM contributions from free electrons in the Milky Way (MW) interstellar medium (ISM) and halo, respectively, and ${\rm DM}_{\rm host}$ is the DM contribution from the FRB host galaxy at redshift $z$ (i.e., not in the host galaxy rest frame of reference):

\begin{equation}
    {\rm DM}_{\rm host}(z) = \frac{{\rm DM}_{\rm host}^{\rm loc}}{1+z}.
\end{equation}We computed ${\rm DM}_{\rm ISM}$ by averaging the two values obtained from the \cite{CordesLazio02} and the YMW16 \citep{YMW16} models, respectively, while ${\rm DM}_{\rm halo}$ was estimated following \cite{Yamasaki}, i.e., modeling the MW hot gas halo with a disk-like and a spherical component constrained by diffuse X-ray observations. We chose ${\rm DM}_{\rm host} = 100$ pc cm$^{-3}$, in agreement with estimates from FRB samples \citep[e.g.,][]{Li20_host,Zhang20b}. We note that Eq.~(\ref{eq:macquart}) should also include the DM contribution from free electrons in intervening halos or galaxy clusters along the line of sight. However, this contribution is currently poorly constrained and any assumed correction would introduce an arbitrary offset in the inferred redshifts, which is already encompassed by the redshift uncertainties derived as described above. We used the Python package \textsc{frb}\footnote{\textsc{frb}: \url{https://github.com/FRBs/FRB/}} to compute the Macquart redshifts as described above: \textsc{frb} provides both the median and mode values of the redshift distribution, and we chose to employ the latter as it better represents the redshift value of asymmetric distributions \citep[see, e.g.,][]{James22_zDM}.

\section{Observations and data reduction}
\label{sec:calib_reduction}

We present new continuum observations of a sample of 24 FRB sources, conducted with the uGMRT at band-5 ($1060$--$1460$~MHz). Each observation has been processed following a standard data reduction using the National Radio Astronomy Observatory (NRAO) Common Astronomy Software Applications ({\tt CASA}; \citealt{McMullin07}) v. 6.7. Primary and secondary calibrators used for each source are listed in Table~\ref{tab:uGMRT_obs}. The flux density scale was set according to the \cite{PerleyButler17} scale.

We carried out flagging, delay, bandpass, and amplitude calibration  for each calibration source. Calibration solutions were then applied to target visibilities that were then Fourier transformed into images using {\tt WSClean} \citep{offringa14,offringa17} v. 3.6, which accounts for wide-field and multifrequency synthesis. Images were deconvolved to generate sky models that were then used to carry out phase and amplitude self-calibration to improve the image fidelity. After self-calibration, visibilities were averaged in time (30~s) and frequency (eight channels).

We generated images with different {\tt briggs} weighting schemes for each target, with {\tt robust} parameters of -0.5, 0, 0.5. Images with   -0.5 robust parameter provide the highest resolution (typically $\sim 2''$) and noise (typically $\sim 30 \; {\rm \mu Jy \; beam^{-1}} $), whereas the choice of robust parameter 0.5 degrades the resolution (typically $\sim 4''$) but improves the noise (typically $\sim 10 \; {\rm \mu Jy \; beam^{-1}} $). In the remainder of the paper, we consider images obtained with {\tt briggs} 0.5 as references for the analysis, unless stated otherwise. The noise levels and angular resolutions of each image are listed in Table~\ref{tab:uGMRT_obs}.

Finally, we computed the $1.26$~GHz spectral luminosity for each source as

\begin{equation}
    L_{1.2} = \frac{4 \pi D_{\rm L}^2 \, S_{1.2}}{(1+z)^{1 - \alpha}},
\label{eq:luminosity}    
\end{equation}where $S_{1.2}$ is either the source flux density (in case of detection) or the upper limit (UL) obtained as two times the r.m.s. noise, and $D_{\rm L}$ is the source luminosity distance obtained from the source redshift using cosmological parameters from the Planck 2018 experiment \citep{Planck20}. We assumed $\alpha = -0.27$, which is the spectral index of the PRS associated with R1 \citep{Chatterjee17}. 

In order to compute luminosity ULs from flux density upper limits, we propagated uncertainties using a Monte Carlo approach. For each source, a luminosity realization was computed using Eq.~\ref{eq:luminosity}, where the redshift value was drawn from a Gaussian distribution centered on the measured value and with standard deviation equal to the measured uncertainty. The flux density values were drawn from a half-normal distribution with zero mean and standard deviation equal to the measured noise (Table~\ref{tab:uGMRT_obs}). The procedure was repeated $10^4$ times in order to obtain a distribution of luminosity values, whose $95^{\rm th}$ percentile was taken to be the luminosity UL (Table~\ref{table:FRB_sample}).

\section{Results and discussion}\label{sec: Discuss}

We first describe the construction of the extended FRB catalog used throughout the analysis, which combines new uGMRT observations at $1.26$~GHz with literature data. We then highlight the main results of the uGMRT observational campaign, while a detailed description of individual FRBs is provided in Appendix~\ref{app: description}. Throughout this work, we only classify as confirmed FRB-PRS systems the four FRBs currently known to be associated with a persistent radio source that is compact on parsec (or sub-parsec) scales and whose association has been firmly established through Very Long Baseline Interferometry (VLBI) observations. These are FRBs 20121102A, 20190417A, 20190520B, and 20240114A \citep{Marcote17,Bhandari23b,Bruni24,Moroianu26}. By contrast, we define a ``PRS candidate'' as a persistent source localized within the $95\%$ CL positional uncertainty of a given FRB, with the prescription that it is unresolved in the highest-resolution uGMRT cleaned images we obtained (see Table~\ref{tab:uGMRT_obs} for details on the resolutions we reached for each source).

\subsection{The extended source catalog}\label{sec: ext_cat}
We complemented our new observations with literature sources in which a search for a persistent radio counterpart was reported. Our final source catalog\footnote{The catalog is also provided at \url{https://github.com/davidepelliciari/PRS-catalogue}.} is presented in Table~\ref{table:FRB_sample} and includes a total of 75 sources, of which 25 are confirmed repeaters and the remaining 50 are (apparently) one-off events. In the same table, we provide references for redshifts and measured flux densities (or ULs on PRS flux density in case of non-detections) for each source.

For a proper comparison across heterogeneous datasets, luminosities and upper limits from the literature measured at a central frequency $\nu$ were scaled to $1.2$~GHz according to

\begin{equation}
    L_{1.2} = L_\nu \left( \frac{1.2\,\mathrm{GHz}}{\nu} \right)^\alpha,
\end{equation}

adopting $\alpha = -0.27$ (i.e.,\ the spectral index of the PRS associated with R1). The only exceptions are R1-twin and FRB~20201124A, for which spectral indices of $-0.41 \pm 0.04$ \citep{Niu21} and $1.0 \pm 0.4$ \citep{Bruni23} have been reported, respectively; we used these values to compute their spectral luminosities at $1.2$~GHz. Beyond the uGMRT sample, we include R1 \citep{Chatterjee17} and R1-twin \citep{Niu21} (the other two confirmed PRSs, i.e.,\ S5 and S22, are already part of the uGMRT sample), and the PRS candidates FRBs 20201124A~\citep{Bruni23} and 20181030A~\citep{Ibik24}.

For sources from the Deep Synoptic Array (DSA) host galaxy catalog \citep{Law23}, we provide deeper luminosity ULs than those reported therein, which were based on single-epoch Very Large Array Sky Survey (VLASS) measurements at 3~GHz \citep{Lacy20}. By stacking the three available VLASS epochs, we reach an average r.m.s.\ noise of $\sim\!80~\mu$Jy beam$^{-1}$, compared with $\sim\!170~\mu$Jy beam$^{-1}$ in \cite{Law23}. The sources that benefit from this improvement are FRBs 20220207C, 20220307B, 20220310F, 20220418A, 20220506D, 20220509G, 20220825A, 20220914A, 20220920A, and 20221012A.

We also incorporated the well-localized one-off FRBs from the 1.4~GHz continuum radio monitoring campaign conducted with MeerKAT by \citet{Mfulwane26}. Their sample comprises 25 one-off FRBs, seven of which overlap with our uGMRT sample (S9, S14, S15, S16, S17, S18, and S19). The remaining sources are listed in Table~\ref{table:FRB_sample} as literature FRBs, with the exception of FRB~20220330, which is excluded due to the lack of a securely identified host galaxy, a measured redshift, and sufficient supporting information. Among the \citet{Mfulwane26} sources not in our uGMRT sample, only FRB~20200906A shows unresolved arc-second-scale radio emission consistent with the FRB position, and therefore qualifies as a PRS candidate under our selection criteria. Reliable confidence levels for the flux-density upper limits of the remaining \citet{Mfulwane26} sources are not available; these entries are marked as ``N/A'' in Table~\ref{table:FRB_sample}, and we adopted the luminosities reported therein as upper limits. Finally, we include FRB~20200428 \citep[i.e.,\ the Galactic FRB-like event from SGR~J1935$+$2154;][]{CHIME20b,Bochenek20a}, treating it as a one-off source\footnote{Although FRB~20200428 had a spectral luminosity marginally consistent with those of extragalactic FRBs \citep{Bochenek20a}, other radio bursts with lower luminosities have been observed from SGR~J1935$+$2154 \citep[e.g.][]{Kirsten21}, making its classification ambiguous. The transition to a pulsar phase has also been observed for this magnetar \citep{Zhu23}. We verified that treating SGR~J1935$+$2154 as a repeater instead does not appreciably change the results presented in Sect.~\ref{sec: occ}.}, following the assumption of \cite{Law22}.

We explicitly exclude from Table~\ref{table:FRB_sample} sources that lack both radio luminosity ULs from a targeted monitoring campaign and a confirmed FRB--host association. This leads us to exclude FRBs 20181119A, 20190110C, 20190604A, 20191114A, 20200223B, 20200619A, and 20200929C from \cite{Ibik24}.

\begin{figure}
    \centering
    \includegraphics[width=1.05\columnwidth]{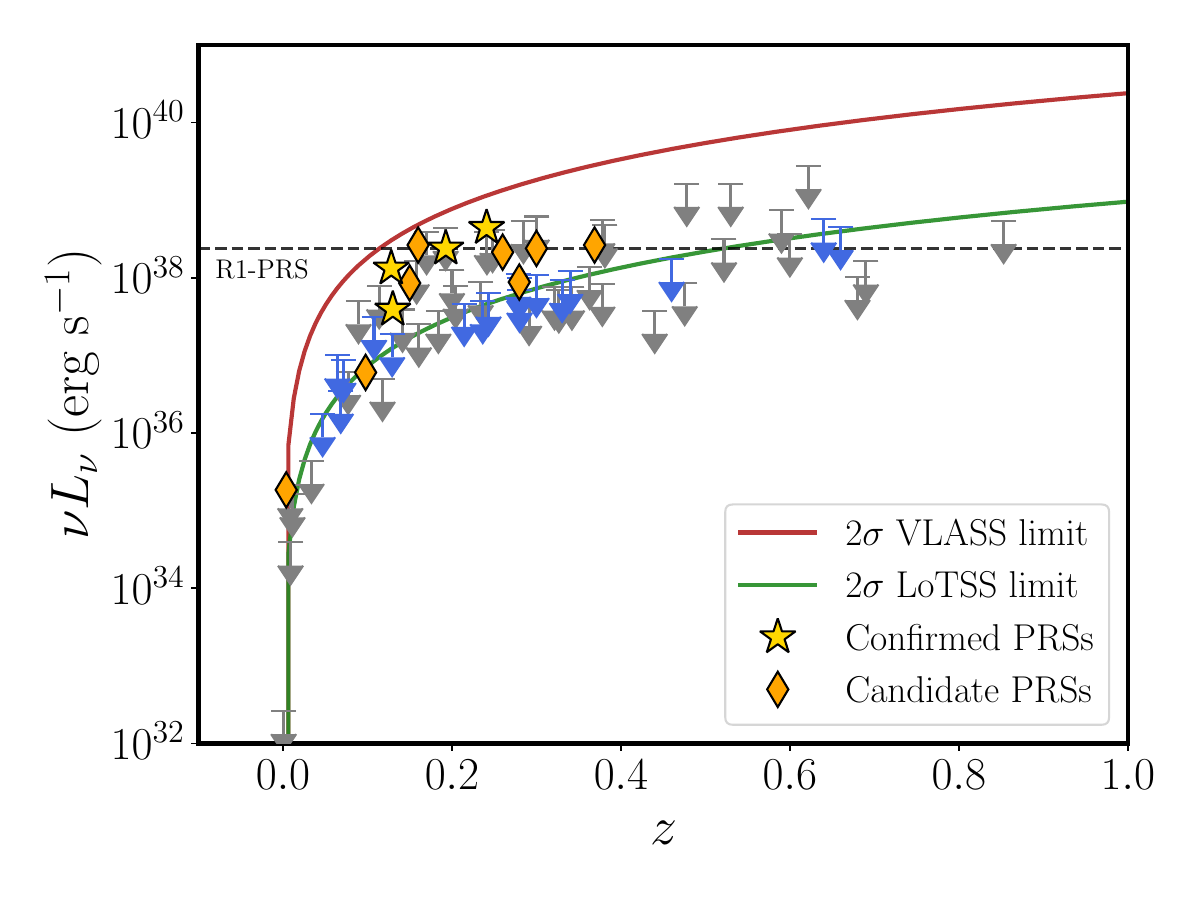}
    \caption{Persistent radio source luminosities and luminosity limits as a function of the FRB redshift. If the host galaxy is not known, the redshift of the FRB is the median value reported in Table~\ref{table:FRB_sample}. Yellow stars indicate confirmed PRSs while orange diamonds indicate PRS candidates. Downward triangles indicate $95\%$ CL. Upper limits on the PRS luminosity from either literature observations (gray) or from our sample (blue). The solid red (green) line represents the $2\sigma$ detection limit of VLASS (LoTSS). The horizontal dashed line marks the observed luminosity of the PRS associated with R1.}
    \label{fig:znuLnu}
\end{figure}

The resulting catalog spans a wide range of redshifts, from the nearby Galactic event FRB~20200428 to FRB~20210912A at $z \approx 1.4$. Figure~\ref{fig:znuLnu} shows the luminosities $L \equiv \nu L_\nu$ and ULs as a function of redshift for the full extended catalog. Confirmed and candidate PRSs occupy a relatively narrow luminosity range around $\nu L_\nu \sim 10^{38}$--$10^{39}$~erg~s$^{-1}$, well above the luminosity limits of most non-detections at $z \geq 0.1$. A total of 50 out of 75 sources have a measured RM, spanning from essentially zero (e.g.,\ R3) to $\sim\!2 \times 10^5$~rad~m$^{-2}$ for R1; this subsample is directly relevant to the test of the $L_\nu$--RM relation, which will be presented in a companion paper (Paper III, Pelliciari et al. in preparation).

\subsection{Results from uGMRT campaign}\label{sec: ugmrt_overview}

We report the detection of persistent radio emission in six FRB fields of our uGMRT sample, namely S5, S11, S12, S13, S20, and S22. Among them, S5 and S22 correspond to FRBs 20190417A and 20240114A, for which a confirmed persistent counterpart was already known \citep{Bruni24,Ibik24,Moroianu26}. The remaining detections represent new PRS candidates associated with repeating FRBs, with the exception of S13, which is an apparently one-off source \citep{PastorMarazuela25}.

\begin{figure*}
\sidecaption
  \includegraphics[width=12cm]{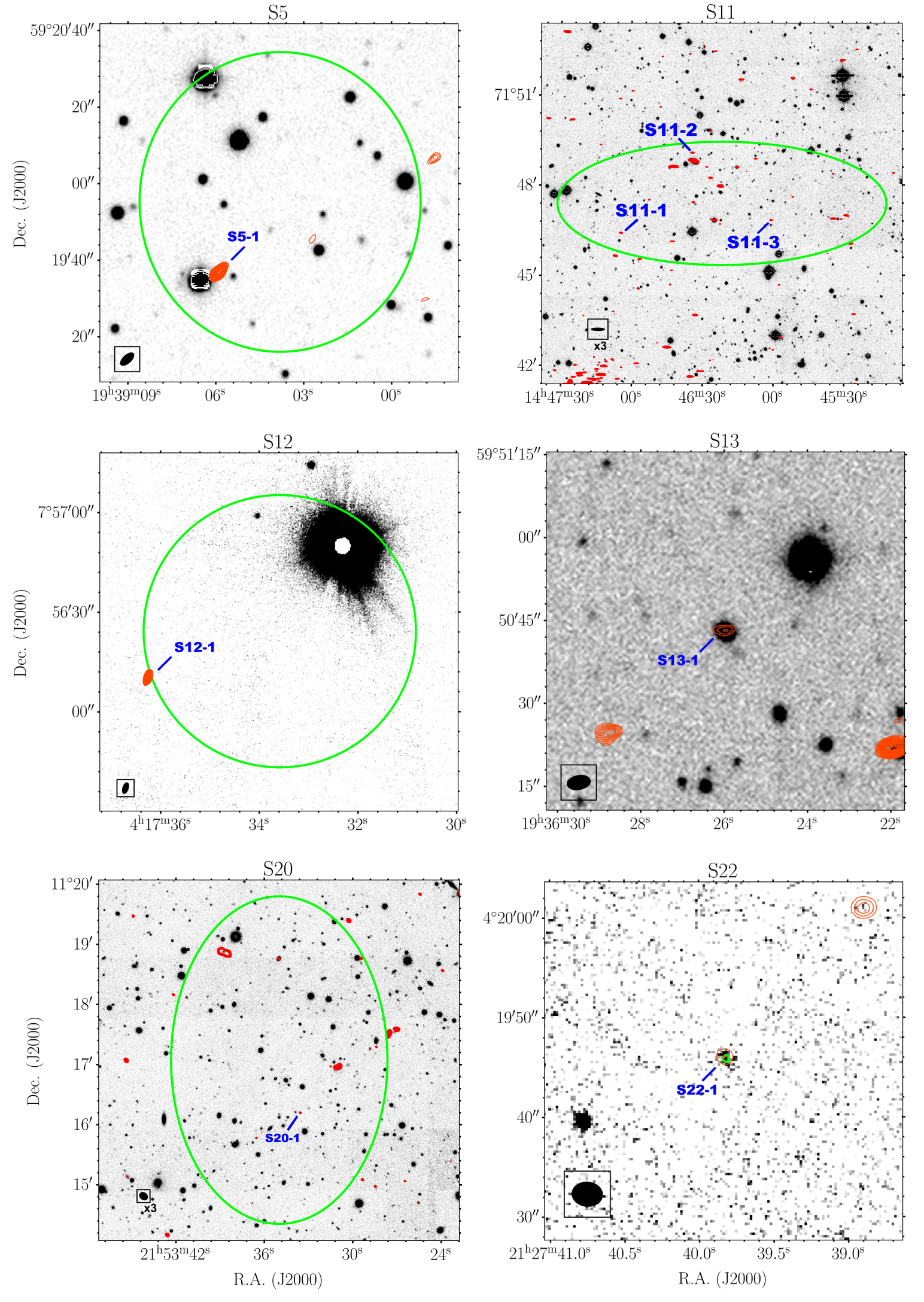}
     \caption{Optical r-filter images of the six FRB sources included in our work for which we report a radio detection. The optical images for S5, S12, and S22 (S11, S13, and S20) are obtained from the Pan-STARRS DR1 \citep{Flewelling20} \citep[DESI;][]{Dey19}. Red contours represent continuum radio emission at $1.26$ GHz as resulting from our uGMRT observations, and are drawn from $4\sigma$, where $\sigma$ is the r.m.s. noise level for each image (see Table \ref{table:FRB_sample}). The green ellipses represent the localization uncertainties of each FRB source at a $95\%$ CL. For S13, the image is centered on the potential S13 host galaxy, and the localization region contour \citep[see][]{PastorMarazuela25}, while still enclosing the galaxy, lies outside the image limits. Blue lines indicate the position of compact persistent sources, which properties are listed in Table \ref{tab: PRS_prop}. The synthesized beam of uGMRT observations is reported in the bottom-left corner of each image as a filled black ellipse. Note that the beams of S11 and S20 have been enlarged  for better visualization.}
     \label{fig:uGMRT_PRS}
\end{figure*}

In order to label a source as compact in our observations, we verify that its integrated and peak flux densities are consistent within $1\sigma$ uncertainties. All detected compact radio sources are shown in Fig.~\ref{fig:uGMRT_PRS}, and their properties are listed in Table~\ref{tab: PRS_prop}. We also derive constraints on the PRS radius using the semimajor axis of the synthesized beam of the uGMRT images obtained with {\sc briggs} weighting and {\sc robust} parameter $-0.5$ (Table~\ref{tab: PRS_prop}; see Table~\ref{tab:uGMRT_obs} for details). S5 and S22 were already constrained at the parsec scale \citep{Bruni24,Moroianu26}. For new PRS candidates, we estimate the chance coincidence probability, $P_{\rm cc}$ , that the detected persistent source is unrelated to the FRB. The values for $P_{\rm cc}$ are listed in Table \ref{tab: PRS_prop}, while we describe the method to obtain and evaluate this probability in Appendix \ref{app: description}. In our sample, S12 and S13 have $P_{\rm cc}<0.1$, whereas the higher values found for S11 and S20 are primarily driven by the comparatively poor localization of their corresponding FRBs (see Table \ref{tab:loc_prop_sources}). While a low $P_{\rm cc}$ strengthens the association between the FRB and the persistent source, a high $P_{\rm cc}$ mainly reflects the limited localization accuracy of the FRB rather than providing evidence against a physical association. Therefore, $P_{\rm cc}$ should be regarded as an indicator of the association robustness and not as a criterion for selecting promising PRS candidates for future follow-up observations.

Recently, \cite{Mfulwane26} reported radio continuum detections spatially coincident with the positions of 14 FRB sources, including four in common with our uGMRT sample (S14, S15, S17, and S18). According to the criteria adopted in this work, only S14 qualifies as a candidate PRS.
Beyond the uGMRT sample, we include R1 \citep{Chatterjee17} and R1-twin \citep{Niu21} (the other two confirmed PRSs,\ S5 and S22, are already part of the uGMRT sample), and the PRS candidates FRBs 20201124A~\citep{Bruni23} and 20181030A~\citep{Ibik24}.

We detect two confirmed PRSs,\ FRBs 20190417A (S5) and 20240114A (S22). We measure a $248 \pm 20~\mu$Jy flux density for S5, $\approx\!3.6\sigma$ higher than \cite{Ibik24}, although the discrepancy is less pronounced ($\approx\!1.3\sigma$) when compared with \cite{Moroianu26}; the differences are further discussed in \cite{Bruno26}. We measure a $53 \pm 11~\mu$Jy flux density for S22, consistent with a previous measurement at comparable frequencies \citep{Zhang25_FRS}.

\subsubsection{Detection of new candidate PRSs associated with repeating FRBs}\label{sec: rep_results}

We find compact radio emission within the $95\%$ localization uncertainties in three other repeating FRBs: S11, S12, and S20. S11 has a large localization uncertainty \citep[$4.8' \times 1.8'$ at $1\sigma$ level;][]{chime23}, and we find three compact sources within its 95$\%$ CL localization region. We label them S11-1, S11-2, and S11-3.

\begin{table*}[htbp]
\centering
\caption{Properties of the compact radio sources we report in this work.}
\label{tab: PRS_prop}
\begin{tabular}{l l c c c c c}
\hline\hline
FRB source & ID & Coordinates & $F_{\rm peak}$ & $L_{1.2}$ & $P_{\rm cc}$ & $R$ \\
   &      &                 ($\alpha_{\rm J2000}$,   $\delta_{\rm J2000}$)    & ($\mu$Jy)          & ($10^{29}\,\mathrm{erg\,s^{-1}\,Hz^{-1}}$) & &  (kpc) \\
\hline
20190417A & & & & & & \\
& S5-1 &
$19^{\rm h}39^{\rm m}05.904^{\rm s} \pm 0.009^{\rm s},\,
+59^\circ19'36.87'' \pm 0.07''$ &
$248 \pm 20$ &
$1.05 \pm 0.08$ & $-$ & $< 5.9$ \\
\hline
20201114A & & & & & & \\
& S11-1 & $14^{\rm h}47^{\rm m}4.475^{\rm s} \pm 0.155^{\rm s}$, $+71^\circ 46'25.118'' \pm 0.152''$ & $100 \pm 15$ & $1.6 \pm 0.7$ & $0.996$ &  $<30$ \\

& S11-2 & $14^{\rm h}46^{\rm m}34.142^{\rm s} \pm 0.182^{\rm s}$, $+71^\circ 49'5.54'' \pm 0.04''$ & $115 \pm 15$ & $1.9^{+0.7}_{-0.8}$ & $0.996$ & $<30$ \\

& S11-3 & $14^{\rm h}46^{\rm m}0.495^{\rm s} \pm 0.158^{\rm s}$, $+71^\circ 46'50.77'' \pm 0.06''$ & $105 \pm 15$ & $1.7^{+0.7}_{-0.8}$ & $0.996$ & $<30$ \\
\hline
20201130A & & & & & & \\
& S12-1 & $4^{\rm h}17^{\rm m}36.256^{\rm s} \pm 0.002^{\rm s}$, $+7^\circ 56'10.387 \pm 0.071''$ & $360 \pm 13$ & $2.1^{+0.9}_{-1.2}$ & $0.066$ & $<7.9$ \\
\hline
20210317A & & & & & \\
& S13-1 & $19^{\rm h}36^{\rm m}25.99^{\rm s} \pm 0.03 s$, $+59^\circ50'43.3'' \pm 0.05''$ & $116 \pm 14$ & $0.68 \pm 0.08$ & $0.041$ & $<9.6$ \\
\hline
20230607A & & & & & \\
& S20-1 & $21^{\rm h}53^{\rm m}33.553^{\rm s} \pm 0.003^{\rm s}$, $+11^\circ 16'11.86 \pm 0.03''$ & $90 \pm 17$ & $1.9 \pm 0.8$ & $0.822$ & $<10$ \\
\hline
20240114A & & & & & & \\
& S22-1 & $21^{\rm h}27^{\rm m}39.823^{\rm s} \pm 0.004^{\rm s}$, $+4^\circ 19'46.134'' \pm 0.052''$ & $53 \pm 11$ & $0.31 \pm 0.04$ & $-$ & $<4.8$ \\
\hline
\end{tabular}
\tablefoot{The first column lists the FRB source name following the TNS convention, while the second provides the corresponding continuum source ID. The third column reports the best-fit coordinates, with associated $1\sigma$ uncertainties. The fourth column gives the peak flux density at $1.2$ GHz, and the fifth the corresponding spectral luminosity, computed using the FRB redshift listed in Table~\ref{table:FRB_sample}. The final column lists the constraint on the linear physical size of each source obtained from our uGMRT observations.}
\end{table*}

The projected angular separations from the FRB localization center are $211''$, $119''$, and $105''$ for S11-1, S11-2, and S11-3, respectively. All these sources are detected with a confidence level between $5\sigma$ and $6\sigma$, and present spectral luminosities consistent with the confirmed PRSs (see Table \ref{tab:uGMRT_obs} and Table \ref{tab: PRS_prop}). We searched in the Panoramic Survey Telescope \& Rapid Response System Dara Release 1 \citep[Pan-STARRS DR1;][]{Flewelling20} and in the Tenth Release of the Dark Energy Spectroscopic Instrument \citep[DESI DR10;][]{Dey19} and found no optical counterparts.

Other radio sources are present within the localization ellipse, but they appear extended, and we therefore exclude them as potential PRS candidates. For S12 there is only one bright compact source just outside the 95$\%$ CL localization uncertainties of the FRB. We detected it at $\sim 20\sigma$ confidence, resulting in a peak flux density of $360 \pm 18$ $\mu$Jy. Like the PRS candidates we report for S11, this source also lacks an optical counterpart and has a spectral luminosity that is consistent with confirmed PRSs.

Finally, there is a compact radio source that lies southeast with respect to the FRB localization ellipse center of S20, within its $68\%$ CL region, with $\approx 90 \pm 17\ \mu$Jy flux density. Remarkably, a $\approx 10^{29}$ erg s$^{-1}$ Hz$^{-1}$ persistent source was predicted, given the high RM of $\sim 1.2 \times 10^4$ rad m$^{-2}$ of the FRB source \citep{Zhou25}. Considering a redshift of $\approx 0.3$ for this source, the flux density of the compact source we find converts to $1.9(8) \times 10^{29}$ erg s$^{-1}$ Hz$^{-1}$. However, a more precise localization of the FRB source is needed in order to associate it with this particular compact radio source.

Although the small number of confirmed systems prevents us from defining a robust FRB--PRS archetype, their shared observational properties (flat spectra, dwarf-galaxy hosts, and generally large RM values) provide useful diagnostics for assessing the candidates. S11 and S20 are particularly interesting given the large RM of their associated FRBs, i.e., $\simeq 1350$ rad m$^{-2}$ \citep{Ng25} and $\simeq -1.2\times10^{4}$ rad m$^{-2}$ \citep{Zhou25} for S11 and S20, respectively. Moreover, their spectral luminosities are broadly consistent with confirmed FRB-PRS systems. However, no information is available regarding the host galaxies of these sources. Follow-up VLBI observations as well as a characterization of their spectral properties will be important to test their spectral properties, which are needed to assess their compactness on parsec scales.

\subsubsection{Candidate PRSs in one-off FRBs}\label{sec: nonrep_results}

\citet{Chibueze21} reported a tentative PRS association with a one-off source, FRB 20190714A (S6 in our nomenclature). The persistent source, detected with MeerKAT at $1.28$~GHz with a flux density of $\sim90~\mu$Jy, is potentially resolved into two components of $86$ and $123$~$\mu$Jy at $1.5$~GHz in higher-resolution observations reported in the same work and conducted with the enhanced Multi Element Remotely Linked Interferometer Network (e-MERLIN) at $1.5$ GHz. Our data, with a $3\sigma$ sensitivity of $39$~$\mu$Jy~beam$^{-1}$, show no continuum emission within the S6 localization region. Considering the flux density measured with MeerKAT, the source would have been detected at $\sim7\sigma$ in our observations, so we conclude that the sources reported by \citet{Chibueze21} are unlikely to be PRSs associated with S6. Further details are provided in Appendix~\ref{app: description}.

Recently, \citet{Mfulwane26} reported persistent radio emission of $32 \pm 7~\mu\mathrm{Jy}$ flux density in correspondence with the one-off S14. The corresponding spectral luminosity is $7(2) \times 10^{28}$~erg~s$^{-1}$~Hz$^{-1}$, which is twice as luminous as the faintest confirmed PRS, FRB 20240114A \citep{Bruni24}. The source is compact at arc-second angular scales, hence we consider this as a candidate PRS in our analysis. The source is marginally detected at $1.8\sigma$ CL in our uGMRT observations, with a flux $F_{1.2} \leq 36~\mu\mathrm{Jy}$.

We find compact radio emission co-spatial with the most probable host galaxy of S13,  PSO J294.1082+59.8453, for which the photometric redshift is known to be $0.15(5)$ \citep{PastorMarazuela25}. We also searched for archival radio data and found no source down to a r.m.s. noise of $0.5$ mJy beam$^{-1}$ in the NRAO VLA Sky Survey \citep[NVSS;][]{Condon98}. Although PSO J294.1082+59.8453 is the galaxy with the highest association probability \citep[computed via a Probabilistic Association of Transients to their Hosts ({\tt PATH}) analysis;][]{Agarwal21}, it still has $P(O|x) = 0.541$ \citep{PastorMarazuela25}, which is not enough to secure the FRB--host link. In the case in which the persistent source is not related to the FRB, and given the low association probability between the FRB and the (putative) host galaxy, we consider $L_\nu \leq 1.4 \times 10^{29}$ erg s$^{-1}$ Hz$^{-1}$, obtained considering a redshift $z = 0.38^{+0.04}_{-0.12}$ computed from the DM of the FRB source, i.e., $466.5$ pc cm$^{-3}$ \citep{PastorMarazuela25}, and considering the flux limit of our uGMRT observations (see Table \ref{table:FRB_sample}).

Finally, the one-off source FRB 20200906A shows co-spatial persistent emission \citep{Mfulwane26}. Although we do not have uGMRT observations of this source, MeerKAT observations indicate that the emission is unresolved at angular scales at 1.4 GHz, and we therefore consider it a PRS candidate as well. This source is particularly interesting because the FRB is well localized within its host galaxy, which is known to be moderately star forming \citep{Bhandari22}.

Similarly to the PRS candidates potentially associated with repeating FRBs, also S13, S14, and the persistent source associated with FRB 20200906A result poorly characterized. With the exception of S13, which has a RM of $\approx 400$ rad m$^{-2}$ in the source reference frame, no RM measurements have been reported for the other PRS candidates potentially associated with one-off FRBs. Deeper observations are needed to investigate their nature, both by better constraining their spectral properties and by improving their localization.

In a forthcoming companion paper (Paper III), we further investigate  the connection between these FRBs, both repeaters and one-offs, and their associated persistent emission by studying the expected $L_\nu$--RM relation for FRB--PRS systems. We exploit the catalog presented in this work.

\section{The PRS occurrence rate}
\label{sec: occ}

We determined the PRS occurrence in two different cases: we considered a scenario where PRS candidates are included (inclusive scenario) and one with only confirmed PRSs (conservative). The inclusive scenario includes nine PRSs associated with repeating FRBs (R1, R1-twin, S5, S11\footnote{Regarding this source, we considered $106 \pm 15$ $\mu$Jy for the flux density of the representative PRS candidate (i.e., the average of the flux densities for the three PRS candidates).}, S12, S20, S22, FRB 20201124A, and FRB 20181030A) and three PRSs associated with one-off FRBs (S13, S14, and FRB 20200906A). The conservative scenario includes four PRSs associated with repeating FRBs (R1, R1-twin, S5, and S22) and none with one-off FRBs. In this case, the spectral luminosities of candidate PRSs are replaced by their corresponding 
$95\%$ CL noise values.

\begin{figure}
    \centering
    \includegraphics[width=1.0\columnwidth]{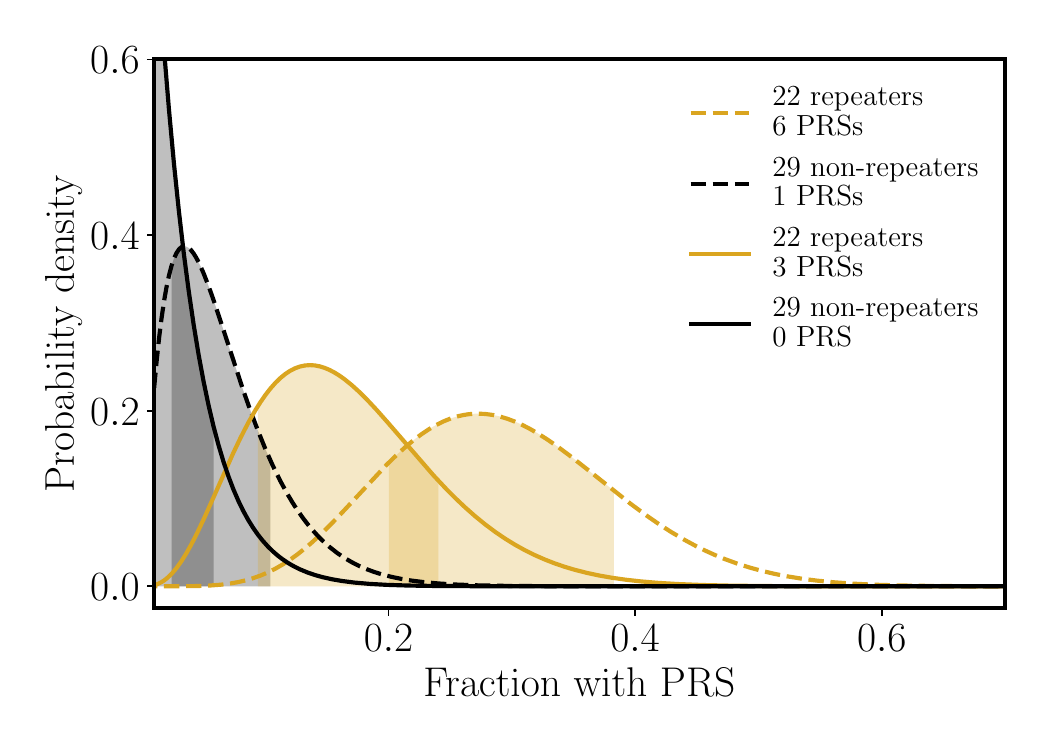}
    \caption{Occurrence of PRSs in the observed population of non-repeaters (nr; black lines) and repeaters (r; orange lines). The solid (dashed) curves represent the conservative (inclusive) scenario considered in this work (see Section \ref{sec: Discuss}). The shaded regions represent the $1\sigma$ range for the occurrence distributions.}
    \label{fig:PRS_occ}
\end{figure}

Following \cite{Law22}, we assumed that the fraction, $f$ , of PRSs associated with either repeating or non-repeating FRBs follows a binomial probability distribution, $p$:

\begin{equation}
    p(f_{\rm x} \mid n_{\rm x}, k_{\rm x}) = \binom{n_{\rm x}}{k_{\rm x}} \, f_{\rm x}^{k_{\rm x}} (1 - f_{\rm x})^{n_{\rm x} - k_{\rm x}},
\end{equation}where $n$ is the number of FRB sources, $k$ is the number of PRSs, and $x$ is either ``nr'' or ``r'' for non-repeaters and repeaters, respectively. From the FRB sample presented in Table \ref{table:FRB_sample}, we have a total of 75 FRB sources, divided into 50 one-off events and 25 repeaters. We restrict the FRB-PRS sample to sources for which the 95\% CL upper limits on the spectral luminosity are lower than $10^{29}$~erg~s$^{-1}$~Hz$^{-1}$. This choice yields $n_{\rm nr} = 29$ and $n_{\rm r} = 22$. 
We also only consider the confirmed PRSs brighter than $10^{29}$~erg~s$^{-1}$~Hz$^{-1}$ and this choice yields  $k_{\rm nr} = 1$ and $k_{\rm r} = 6$ in the inclusive scenario, and $k_{\rm nr} = 0$ and $k_{\rm r} = 3$ in the conservative one. 

We show the resulting probability distributions in Fig.~\ref{fig:PRS_occ}. In the inclusive scenario, we obtain 
$f_{\rm r} = 0.27^{+0.21}_{-0.14}$ and 
$f_{\rm nr} = 0.03^{+0.14}_{-0.02}$ 
($95\%$ C.L.). In the conservative scenario, we obtain 
$f_{\rm r} = 0.14^{+0.20}_{-0.09}$, while for non-repeaters we derive a $95\%$ CL upper limit of 
$f_{\rm nr} \leq 0.12$. Although the occurrence rates of the two populations are not yet statistically distinct, there is tentative evidence for a preferred association between PRSs and repeating FRBs in the inclusive scenario at $\approx 90\%$ CL.

Considering the two populations together, we obtain 
$f_{\rm all} = 0.14^{+0.12}_{-0.08}$ 
($95\%$ CL) in the inclusive scenario, and 
$f_{\rm all} = 0.06^{+0.09}_{-0.03}$ 
($95\%$ CL) in the conservative scenario. In other words, a PRS as luminous as PRS~20121102A is expected to be associated with $\sim 6\%$ of the overall FRB population.

\section{Summary and conclusions}\label{sec: conclusions}

In this work we present new uGMRT observations at $1.26$~GHz of a sample of $24$ FRBs, of which $13$ are confirmed repeaters and the remaining $11$ are one-off sources. Our images reach a noise level in the $10-40$~$\mu$Jy, corresponding to a $1.4 \times 10^{27} - 3.9 \times 10^{29}$ erg s$^{-1}$ Hz$^{-1}$ luminosity range, with arc-second angular resolution.

We identify three new candidate PRSs potentially associated with the repeating FRBs 20201114A (S11), 20201130A (S12), and 20230607A (S20), and one candidate associated with the one-off FRB 20210317A (S13). As these FRB positions are known at $\sim$ arcmin precision, the association with PRS sources remains uncertain. Although the number of confirmed systems is still too small to define a robust FRB--PRS archetype, their shared observational properties (flat radio spectra, dwarf-galaxy hosts, and, in most cases, large RM values) provide useful diagnostics to assess how closely the candidates resemble the confirmed population. Among our candidates, S20 is particularly interesting in this respect, as the associated FRB has a large RM of $\sim 1.2\times10^{4}$ rad m$^{-2}$, comparable to those of several confirmed systems. S11 and S12 are also promising targets given their persistent-source luminosities and, in the case of S12, the relatively low chance-coincidence probability. However, the uncertain localizations of the majority of these FRB sources currently prevents a more detailed investigation. Multifrequency observations of these sources will be particularly important to establish whether their radio spectra are consistent with the approximately flat spectra of the confirmed PRSs, while higher-angular-resolution radio observations (i.e., VLBI) are needed to constrain their compactness at parsec linear scales.

We combined our FRB sample with literature data and obtained a catalog of 75 FRB--PRS systems, comprised of 50 one-off sources and 25 repeaters. We used this catalog to study the PRS occurrence, and find that PRSs luminosities $L_\nu \geq 10^{29}$ erg s$^{-1}$ Hz$^{-1}$ are rare in the overall FRB population, without distinction between repeating and one-off sources. The overall PRS occurrence rate is $f_{\rm all} \sim 0.14^{+0.22}_{-0.08}$ (95\% CL), in the most optimistic scenario.

We find no evidence that repeating FRBs are a preferred host for luminous ($L_\nu \geq 10^{29}$ erg s$^{-1}$ Hz$^{-1}$) PRSs, if we make the conservative assumption that the newly discovered candidates are not associated with FRBs. In this case we find a fractional occurrence $f_{\rm r} = 0.14^{+0.2}_{-0.09}$ ($95\%$ CL) and a 95\% CL upper limit $f_ {\rm nr} \leq 0.12$ for the repeating and non-repeating samples, respectively. In the most inclusive (optimistic) case that all the newly discovered sources are associated with their FRBs, however, we find slight evidence ($\approx 90\%$ CL) that PRSs are preferably associated with repeating FRBs, with a fractional occurrence $f_{\rm r} = 0.27^{+0.21}_{-0.14}$ ($95\%$ CL) and $f_{\rm nr} = 0.03^{+0.14}_{-0.02}$ ($95\%$ CL) for the repeating and non-repeating samples, respectively. Follow-up observations to confirm candidate PRSs and to expand the source sample will increase the statistics of this result.

\section*{Data Availability}

The catalog of sources presented in this work is available in electronic form at \url{https://github.com/davidepelliciari/PRS-catalogue}.

\begin{acknowledgements}
We thank the staff of the GMRT that made these observations possible. GMRT is run by the National Centre for Radio Astrophysics of the Tata Institute of Fundamental Research. The research activities described in this paper were carried out with contribution of the NextGenerationEU funds within the National Recovery and Resilience Plan (PNRR), Mission 4 - Education and Research, Component 2 - From Research to Business (M4C2), Investment Line 3.1 - Strengthening and creation of Research Infrastructures, Project IR0000026 – Next Generation Croce del Nord. This research has made use of the CIRADA cutout service at URL \url{cutouts.cirada.ca}, operated by the Canadian Initiative for Radio Astronomy Data Analysis (CIRADA). CIRADA is funded by a grant from the Canada Foundation for Innovation 2017 Innovation Fund (Project 35999), as well as by the Provinces of Ontario, British Columbia, Alberta, Manitoba and Quebec, in collaboration with the National Research Council of Canada, the US National Radio Astronomy Observatory and Australia’s Commonwealth Scientific and Industrial Research Organisation. The Pan-STARRS1 Surveys (PS1) and the PS1 public science archive have been made possible through contributions by the Institute for Astronomy, the University of Hawaii, the Pan-STARRS Project Office, the Max-Planck Society and its participating institutes, the Max Planck Institute for Astronomy, Heidelberg and the Max Planck Institute for Extraterrestrial Physics, Garching, The Johns Hopkins University, Durham University, the University of Edinburgh, the Queen's University Belfast, the Harvard-Smithsonian Center for Astrophysics, the Las Cumbres Observatory Global Telescope Network Incorporated, the National Central University of Taiwan, the Space Telescope Science Institute, the National Aeronautics and Space Administration under Grant No. NNX08AR22G issued through the Planetary Science Division of the NASA Science Mission Directorate, the National Science Foundation Grant No. AST–1238877, the University of Maryland, Eotvos Lorand University (ELTE), the Los Alamos National Laboratory, and the Gordon and Betty Moore Foundation. OMS's research is supported by the South African Research Chairs Initiative of the Department of Science, Technology and Innovation and the National Research Foundation (grant No. 81737). This work uses the following software packages: \texttt{Astropy} \citep{AstropyCollab13,PriceWhelan18}, \texttt{matplotlib} \citep{Hunter07}, \texttt{NumPy} \citep{vanderWalt11}, \texttt{Scipy} \citep{Virtanen20}. We thank the anonymous referee for the helpful comments and suggestions, which have helped improve the quality of this work.
\end{acknowledgements}

\bibliographystyle{aa}
\bibliography{biblio2}

\begin{thebibliography}{165}
\expandafter\ifx\csname natexlab\endcsname\relax\def\natexlab#1{#1}\fi

\bibitem[{{Abbott} \& {CHIME/FRB Collaboration}(2024)}]{Abbott24}
{Abbott}, T. \& {CHIME/FRB Collaboration}. 2024, The Astronomer's Telegram, 16734, 1

\bibitem[{{Aggarwal} {et~al.}(2021{\natexlab{a}}){Aggarwal}, {Budav{\'a}ri}, {Deller}, {Eftekhari}, {James}, {Prochaska}, \& {Tendulkar}}]{Agarwal21}
{Aggarwal}, K., {Budav{\'a}ri}, T., {Deller}, A.~T., {et~al.} 2021{\natexlab{a}}, \apj, 911, 95

\bibitem[{{Aggarwal} {et~al.}(2021{\natexlab{b}}){Aggarwal}, {Burke-Spolaor}, {Tejos}, {Pignata}, {Xavier Prochaska}, {Ravi}, {Kaczmarek}, \& {Os{\l}owski}}]{Aggarwal21}
{Aggarwal}, K., {Burke-Spolaor}, S., {Tejos}, N., {et~al.} 2021{\natexlab{b}}, \apj, 913, 78

\bibitem[{{An} {et~al.}(2025){An}, {Wang}, {Huang}, {Feng}, {Liu}, {Zhang}, \& {Dai}}]{An25}
{An}, T., {Wang}, A., {Huang}, Y.-C., {et~al.} 2025, \apjl, 991, L20

\bibitem[{{Andrew} \& {CHIME/FRB Collaboration}(2025)}]{Shion25_CHIME}
{Andrew}, S. \& {CHIME/FRB Collaboration}. 2025, The Astronomer's Telegram, 17114, 1

\bibitem[{{Anna-Thomas} {et~al.}(2023){Anna-Thomas}, {Connor}, {Dai}, {Feng}, {Burke-Spolaor}, {Beniamini}, {Yang}, {Zhang}, {Aggarwal}, {Law}, {Li}, {Niu}, {Chatterjee}, {Cruces}, {Duan}, {Filipovic}, {Hobbs}, {Lynch}, {Miao}, {Niu}, {Ocker}, {Tsai}, {Wang}, {Xue}, {Yao}, {Yu}, {Zhang}, {Zhang}, {Zhu}, \& {Zhu}}]{AnnaThomas23}
{Anna-Thomas}, R., {Connor}, L., {Dai}, S., {et~al.} 2023, Science, 380, 599

\bibitem[{{Astropy Collaboration} {et~al.}(2018){Astropy Collaboration}, {Price-Whelan}, {Sip{\H{o}}cz}, {G{\"u}nther}, {Lim}, {Crawford}, {Conseil}, {Shupe}, {Craig}, {Dencheva}, {Ginsburg}, {VanderPlas}, {Bradley}, {P{\'e}rez-Su{\'a}rez}, {de Val-Borro}, {Aldcroft}, {Cruz}, {Robitaille}, {Tollerud}, {Ardelean}, {Babej}, {Bach}, {Bachetti}, {Bakanov}, {Bamford}, {Barentsen}, {Barmby}, {Baumbach}, {Berry}, {Biscani}, {Boquien}, {Bostroem}, {Bouma}, {Brammer}, {Bray}, {Breytenbach}, {Buddelmeijer}, {Burke}, {Calderone}, {Cano Rodr{\'\i}guez}, {Cara}, {Cardoso}, {Cheedella}, {Copin}, {Corrales}, {Crichton}, {D'Avella}, {Deil}, {Depagne}, {Dietrich}, {Donath}, {Droettboom}, {Earl}, {Erben}, {Fabbro}, {Ferreira}, {Finethy}, {Fox}, {Garrison}, {Gibbons}, {Goldstein}, {Gommers}, {Greco}, {Greenfield}, {Groener}, {Grollier}, {Hagen}, {Hirst}, {Homeier}, {Horton}, {Hosseinzadeh}, {Hu}, {Hunkeler}, {Ivezi{\'c}}, {Jain}, {Jenness}, {Kanarek}, {Kendrew}, {Kern}, {Kerzendorf}, {Khvalko}, {King}, {Kirkby}, {Kulkarni},
  {Kumar}, {Lee}, {Lenz}, {Littlefair}, {Ma}, {Macleod}, {Mastropietro}, {McCully}, {Montagnac}, {Morris}, {Mueller}, {Mumford}, {Muna}, {Murphy}, {Nelson}, {Nguyen}, {Ninan}, {N{\"o}the}, {Ogaz}, {Oh}, {Parejko}, {Parley}, {Pascual}, {Patil}, {Patil}, {Plunkett}, {Prochaska}, {Rastogi}, {Reddy Janga}, {Sabater}, {Sakurikar}, {Seifert}, {Sherbert}, {Sherwood-Taylor}, {Shih}, {Sick}, {Silbiger}, {Singanamalla}, {Singer}, {Sladen}, {Sooley}, {Sornarajah}, {Streicher}, {Teuben}, {Thomas}, {Tremblay}, {Turner}, {Terr{\'o}n}, {van Kerkwijk}, {de la Vega}, {Watkins}, {Weaver}, {Whitmore}, {Woillez}, {Zabalza}, \& {Astropy Contributors}}]{PriceWhelan18}
{Astropy Collaboration}, {Price-Whelan}, A.~M., {Sip{\H{o}}cz}, B.~M., {et~al.} 2018, \aj, 156, 123

\bibitem[{{Astropy Collaboration} {et~al.}(2013){Astropy Collaboration}, {Robitaille}, {Tollerud}, {Greenfield}, {Droettboom}, {Bray}, {Aldcroft}, {Davis}, {Ginsburg}, {Price-Whelan}, {Kerzendorf}, {Conley}, {Crighton}, {Barbary}, {Muna}, {Ferguson}, {Grollier}, {Parikh}, {Nair}, {Unther}, {Deil}, {Woillez}, {Conseil}, {Kramer}, {Turner}, {Singer}, {Fox}, {Weaver}, {Zabalza}, {Edwards}, {Azalee Bostroem}, {Burke}, {Casey}, {Crawford}, {Dencheva}, {Ely}, {Jenness}, {Labrie}, {Lim}, {Pierfederici}, {Pontzen}, {Ptak}, {Refsdal}, {Servillat}, \& {Streicher}}]{AstropyCollab13}
{Astropy Collaboration}, {Robitaille}, T.~P., {Tollerud}, E.~J., {et~al.} 2013, \aap, 558, A33

\bibitem[{{Bannister} {et~al.}(2019){Bannister}, {Deller}, {Phillips}, {Macquart}, {Prochaska}, {Tejos}, {Ryder}, {Sadler}, {Shannon}, {Simha}, {Day}, {McQuinn}, {North-Hickey}, {Bhandari}, {Arcus}, {Bennert}, {Burchett}, {Bouwhuis}, {Dodson}, {Ekers}, {Farah}, {Flynn}, {James}, {Kerr}, {Lenc}, {Mahony}, {O'Meara}, {Os{\l}owski}, {Qiu}, {Treu}, {U}, {Bateman}, {Bock}, {Bolton}, {Brown}, {Bunton}, {Chippendale}, {Cooray}, {Cornwell}, {Gupta}, {Hayman}, {Kesteven}, {Koribalski}, {MacLeod}, {McClure-Griffiths}, {Neuhold}, {Norris}, {Pilawa}, {Qiao}, {Reynolds}, {Roxby}, {Shimwell}, {Voronkov}, \& {Wilson}}]{Bannister19}
{Bannister}, K.~W., {Deller}, A.~T., {Phillips}, C., {et~al.} 2019, Science, 365, 565

\bibitem[{{Bassa} {et~al.}(2017){Bassa}, {Tendulkar}, {Adams}, {Maddox}, {Bogdanov}, {Bower}, {Burke-Spolaor}, {Butler}, {Chatterjee}, {Cordes}, {Hessels}, {Kaspi}, {Law}, {Marcote}, {Paragi}, {Ransom}, {Scholz}, {Spitler}, \& {van Langevelde}}]{Bassa17}
{Bassa}, C.~G., {Tendulkar}, S.~P., {Adams}, E.~A.~K., {et~al.} 2017, \apjl, 843, L8

\bibitem[{{Beloborodov}(2020)}]{Beloborodov19}
{Beloborodov}, A.~M. 2020, \apj, 896, 142

\bibitem[{{Bhandari} {et~al.}(2020{\natexlab{a}}){Bhandari}, {Bannister}, {Lenc}, {Cho}, {Ekers}, {Day}, {Deller}, {Flynn}, {James}, {Macquart}, {Mahony}, {Marnoch}, {Moss}, {Phillips}, {Prochaska}, {Qiu}, {Ryder}, {Shannon}, {Tejos}, \& {Wong}}]{Bhandari20a}
{Bhandari}, S., {Bannister}, K.~W., {Lenc}, E., {et~al.} 2020{\natexlab{a}}, \apjl, 901, L20

\bibitem[{{Bhandari} {et~al.}(2022){Bhandari}, {Heintz}, {Aggarwal}, {Marnoch}, {Day}, {Sydnor}, {Burke-Spolaor}, {Law}, {Xavier Prochaska}, {Tejos}, {Bannister}, {Butler}, {Deller}, {Ekers}, {Flynn}, {Fong}, {James}, {Lazio}, {Luo}, {Mahony}, {Ryder}, {Sadler}, {Shannon}, {Han}, {Lee}, \& {Zhang}}]{Bhandari22}
{Bhandari}, S., {Heintz}, K.~E., {Aggarwal}, K., {et~al.} 2022, \aj, 163, 69

\bibitem[{{Bhandari} {et~al.}(2023){Bhandari}, {Marcote}, {Sridhar}, {Eftekhari}, {Hessels}, {Hewitt}, {Kirsten}, {Ould-Boukattine}, {Paragi}, \& {Snelders}}]{Bhandari23b}
{Bhandari}, S., {Marcote}, B., {Sridhar}, N., {et~al.} 2023, \apjl, 958, L19

\bibitem[{{Bhandari} {et~al.}(2020{\natexlab{b}}){Bhandari}, {Sadler}, {Prochaska}, {Simha}, {Ryder}, {Marnoch}, {Bannister}, {Macquart}, {Flynn}, {Shannon}, {Tejos}, {Corro-Guerra}, {Day}, {Deller}, {Ekers}, {Lopez}, {Mahony}, {Nu{\~n}ez}, \& {Phillips}}]{Bhandari20b}
{Bhandari}, S., {Sadler}, E.~M., {Prochaska}, J.~X., {et~al.} 2020{\natexlab{b}}, \apjl, 895, L37

\bibitem[{{Bhardwaj} {et~al.}(2021{\natexlab{a}}){Bhardwaj}, {Gaensler}, {Kaspi}, {Landecker}, {Mckinven}, {Michilli}, {Pleunis}, {Tendulkar}, {Andersen}, {Boyle}, {Cassanelli}, {Chawla}, {Cook}, {Dobbs}, {Fonseca}, {Kaczmarek}, {Leung}, {Masui}, {Mnchmeyer}, {Ng}, {Rafiei-Ravandi}, {Scholz}, {Shin}, {Smith}, {Stairs}, \& {Zwaniga}}]{Bhardwaj21a}
{Bhardwaj}, M., {Gaensler}, B.~M., {Kaspi}, V.~M., {et~al.} 2021{\natexlab{a}}, \apjl, 910, L18

\bibitem[{{Bhardwaj} {et~al.}(2024){Bhardwaj}, {Kirichenko}, \& {Gil de Paz}}]{Bhardwaj24_ATEL}
{Bhardwaj}, M., {Kirichenko}, A., \& {Gil de Paz}, A. 2024, The Astronomer's Telegram, 16613, 1

\bibitem[{{Bhardwaj} {et~al.}(2021{\natexlab{b}}){Bhardwaj}, {Kirichenko}, {Michilli}, {Mayya}, {Kaspi}, {Gaensler}, {Rahman}, {Tendulkar}, {Fonseca}, {Josephy}, {Leung}, {Merryfield}, {Petroff}, {Pleunis}, {Sanghavi}, {Scholz}, {Shin}, {Smith}, \& {Stairs}}]{Bhardwaj21}
{Bhardwaj}, M., {Kirichenko}, A.~Y., {Michilli}, D., {et~al.} 2021{\natexlab{b}}, \apjl, 919, L24

\bibitem[{{Bhusare} {et~al.}(2025){Bhusare}, {Maan}, \& {Kumar}}]{Bhusare24}
{Bhusare}, Y., {Maan}, Y., \& {Kumar}, A. 2025, \apj, 993, 234

\bibitem[{{Bochenek} {et~al.}(2020){Bochenek}, {Ravi}, {Belov}, {Hallinan}, {Kocz}, {Kulkarni}, \& {McKenna}}]{Bochenek20a}
{Bochenek}, C.~D., {Ravi}, V., {Belov}, K.~V., {et~al.} 2020, \nat, 587, 59

\bibitem[{{Bruni} {et~al.}(2025){Bruni}, {Piro}, {Yang}, {Palazzi}, {Nicastro}, {Rossi}, {Savaglio}, {Maiorano}, \& {Zhang}}]{Bruni24}
{Bruni}, G., {Piro}, L., {Yang}, Y.-P., {et~al.} 2025, \aap, 695, L12

\bibitem[{{Bruni} {et~al.}(2024){Bruni}, {Piro}, {Yang}, {Quai}, {Zhang}, {Palazzi}, {Nicastro}, {Feruglio}, {Tripodi}, {O'Connor}, {Gardini}, {Savaglio}, {Rossi}, {Nicuesa Guelbenzu}, \& {Paladino}}]{Bruni23}
{Bruni}, G., {Piro}, L., {Yang}, Y.-P., {et~al.} 2024, \nat, 632, 1014

\bibitem[{{Bruno} {et~al.}(2026){Bruno}, {Pelliciari}, {Bernardi}, {Pilia}, {Beduzzi}, \& {Esposito}}]{Bruno26}
{Bruno}, L., {Pelliciari}, D., {Bernardi}, G., {et~al.} 2026, \aap, 709, L1

\bibitem[{{Butler} {et~al.}(2018){Butler}, {Huynh}, {Delhaize}, {Smol{\v{c}}i{\'c}}, {Kapi{\'n}ska}, {Milakovi{\'c}}, {Novak}, {Baran}, {O'Brien}, {Chiappetti}, {Desai}, {Fotopoulou}, {Horellou}, {Lidman}, \& {Pierre}}]{Butler18}
{Butler}, A., {Huynh}, M., {Delhaize}, J., {et~al.} 2018, \aap, 620, A3

\bibitem[{Caleb {et~al.}(2023)Caleb, Driessen, Gordon, Tejos, Bernales, Qiu, Chibueze, Stappers, Rajwade, Cavallaro, Wang, Kumar, Majid, Wharton, Naudet, Bezuidenhout, Jankowski, Malenta, Morello, Sanidas, Surnis, Barr, Chen, Kramer, Fong, Kilpatrick, Prochaska, Simha, Venter, Heywood, Kundu, \& Schussler}]{Caleb23}
Caleb, M., Driessen, L.~N., Gordon, A.~C., {et~al.} 2023, Monthly Notices of the Royal Astronomical Society, 524, 2064

\bibitem[{{Chatterjee} {et~al.}(2017){Chatterjee}, {Law}, {Wharton}, {Burke-Spolaor}, {Hessels}, {Bower}, {Cordes}, {Tendulkar}, {Bassa}, {Demorest}, {Butler}, {Seymour}, {Scholz}, {Abruzzo}, {Bogdanov}, {Kaspi}, {Keimpema}, {Lazio}, {Marcote}, {McLaughlin}, {Paragi}, {Ransom}, {Rupen}, {Spitler}, \& {van Langevelde}}]{Chatterjee17}
{Chatterjee}, S., {Law}, C.~J., {Wharton}, R.~S., {et~al.} 2017, \nat, 541, 58

\bibitem[{{Chen} {et~al.}(2023){Chen}, {Ravi}, \& {Hallinan}}]{Chen23}
{Chen}, G., {Ravi}, V., \& {Hallinan}, G.~W. 2023, \apj, 958, 185

\bibitem[{{Chen} {et~al.}(2025){Chen}, {Tsai}, {Li}, {Wang}, {Feng}, {Zhang}, {Li}, {Zhang}, {Bao}, {Liao}, {Zhang}, {Zuo}, {Bao}, {Niu}, {Luo}, {Zhu}, {Zou}, {Xue}, \& {Zhang}}]{Chen25}
{Chen}, X.-L., {Tsai}, C.-W., {Li}, D., {et~al.} 2025, \apjl, 980, L24

\bibitem[{{Chibueze} {et~al.}(2022){Chibueze}, {Caleb}, {Spitler}, {Ashkar}, {Sch{\"u}ssler}, {Stappers}, {Venter}, {Heywood}, {Richards}, {Williams}, {Kramer}, {Beswick}, {Bezuidenhout}, {Breton}, {Driessen}, {Jankowski}, {Keane}, {Malenta}, {Mickaliger}, {Morello}, {Qiu}, {Rajwade}, {Sanidas}, {Surnis}, {Scragg}, {Walker}, {Wrigley}, {Aharonian}, {Ait Benkhali}, {Ang{\"u}ner}, {Backes}, {Baghmanyan}, {Barbosa Martins}, {Batzofin}, {Becherini}, {Berge}, {B{\"o}ttcher}, {Boisson}, {Bolmont}, {de Bony de Lavergne}, {Breuhaus}, {Brose}, {Brun}, {Bulik}, {Cangemi}, {Caroff}, {Casanova}, {Catalano}, {Cerruti}, {Chand}, {Chen}, {Chibueze}, {Cotter}, {Cristofari}, {Damascene Mbarubucyeye}, {Devin}, {Djannati-Ata{\"\i}}, {Dmytriiev}, {Egberts}, {Ernenwein}, {Fiasson}, {Fichet de Clairfontaine}, {Fontaine}, {Funk}, {Gabici}, {Ghafourizadeh}, {Giavitto}, {Glawion}, {Grondin}, {H{\"o}rbe}, {Hoischen}, {Holch}, {Huang}, {Jamrozy}, {Jankowsky}, {Joshi}, {Jung-Richardt}, {Kasai}, {Katarzy{\'n}ski}, {Katz}, {Kh{\'e}lifi},
  {Klu{\'z}niak}, {Komin}, {Kosack}, {Kostunin}, {Lemi{\`e}re}, {Lenain}, {Leuschner}, {Lohse}, {Luashvili}, {Lypova}, {Mackey}, {Malyshev}, {Marandon}, {Marchegiani}, {Marcowith}, {Mart{\'\i}-Devesa}, {Marx}, {Mitchell}, {Moderski}, {Mohrmann}, {Moulin}, {Muller}, {Nakashima}, {de Naurois}, {Nayerhoda}, {Niemiec}, {Priyana Noel}, {O'Brien}, {Ohm}, {Olivera-Nieto}, {de Ona Wilhelmi}, {Ostrowski}, {Panny}, {Parsons}, {Pita}, {Poireau}, {Prokhorov}, {Prokoph}, {P{\"u}hlhofer}, {Quirrenbach}, {Reichherzer}, {Reimer}, {Reimer}, {Rowell}, {Rudak}, {Ruiz-Velasco}, {Sahakian}, {Sailer}, {Salzmann}, {Sanchez}, {Santangelo}, {Sasaki}, {Schutte}, {Schwanke}, {Shapopi}, {Specovius}, {Spencer}, {Steenkamp}, {Steinmassl}, {Takahashi}, {Tanaka}, {Thorpe-Morgan}, {Tsuji}, {van Eldik}, {Veh}, {Vink}, {Wagner}, {Wierzcholska}, {Wong}, {Yusafzai}, {Zacharias}, {Zargaryan}, {Zdziarski}, {Zech}, {Zhu}, {Zouari}, \& {{\.Z}ywucka}}]{Chibueze21}
{Chibueze}, J.~O., {Caleb}, M., {Spitler}, L., {et~al.} 2022, \mnras, 515, 1365

\bibitem[{{CHIME/FRB Collaboration} {et~al.}(2026){CHIME/FRB Collaboration}, {Abbott}, {Andersen}, {Andrew}, {Bandura}, {Bhardwaj}, {Bhusare}, {Brar}, {Cassanelli}, {Chatterjee}, {Cliche}, {Cook}, {Curtin}, {Dobbs}, {Dong}, {Eadie}, {Eftekhari}, {Fonseca}, {Gaensler}, {Good}, {Halpern}, {Hessels}, {Ibik}, {Jain}, {Joseph}, {Kader}, {Kaspi}, {Khan}, {Kharel}, {Kumar}, {Landecker}, {Lang}, {Lanman}, {L'Argent}, {Lazda}, {Leung}, {Li}, {Lintott}, {Main}, {Masui}, {Mate}, {McGregor}, {McKinven}, {Mena-Parra}, {Meyers}, {Michilli}, {Ng}, {Ng}, {Nimmo}, {Noble}, {Pandhi}, {Patil}, {Pearlman}, {Pen}, {Pleunis}, {Prochaska}, {Rafiei-Ravandi}, {Ransom}, {Renard}, {Sammons}, {Sand}, {Scholz}, {Shah}, {Shin}, {Siegel}, {Sirota}, {Smith}, {Stairs}, {Stenning}, {Tendulkar}, {Vanderlinde}, {Walmsley}, {Wang}, \& {Wulf}}]{CHIMECat2}
{CHIME/FRB Collaboration}, {Abbott}, T., {Andersen}, B.~C., {et~al.} 2026, \apjs, 283, 34

\bibitem[{{CHIME/FRB Collaboration} {et~al.}(2025){CHIME/FRB Collaboration}, {Amiri}, {Amouyal}, {Andersen}, {Andrew}, {Bandura}, {Bhardwaj}, {Boyle}, {Brar}, {Cassity}, {Chatterjee}, {Curtin}, {Dobbs}, {Dong}, {Dong}, {Eadie}, {Eftekhari}, {Fong}, {Fonseca}, {Gaensler}, {Halpern}, {Hessels}, {Hopkins}, {Ibik}, {Joseph}, {Kaczmarek}, {Kahinga}, {Kaspi}, {Khairy}, {Kilpatrick}, {Lanman}, {Lazda}, {Leung}, {Main}, {Mas-Ribas}, {Masui}, {McKinven}, {Mena-Parra}, {Meyers}, {Michilli}, {Milutinovic}, {Nimmo}, {Noble}, {Pandhi}, {Patil}, {Pearlman}, {Petroff}, {Pleunis}, {Prochaska}, {Rafiei-Ravandi}, {Rahman}, {Renard}, {Sammons}, {Sand}, {Scholz}, {Shah}, {Shin}, {Siegel}, {Simha}, {Smith}, {Stairs}, {Vanderlinde}, {Wang}, {Wulf}, \& {Zegmott}}]{CHIME25}
{CHIME/FRB Collaboration}, {Amiri}, M., {Amouyal}, D., {et~al.} 2025, \apjs, 280, 6

\bibitem[{{CHIME/FRB Collaboration} {et~al.}(2019){CHIME/FRB Collaboration}, {Andersen}, {Bandura}, {Bhardwaj}, {Boubel}, {Boyce}, {Boyle}, {Brar}, {Cassanelli}, {Chawla}, {Cubranic}, {Deng}, {Dobbs}, {Fandino}, {Fonseca}, {Gaensler}, {Gilbert}, {Giri}, {Good}, {Halpern}, {Hill}, {Hinshaw}, {H{\"o}fer}, {Josephy}, {Kaspi}, {Kothes}, {Landecker}, {Lang}, {Li}, {Lin}, {Masui}, {Mena-Parra}, {Merryfield}, {Mckinven}, {Michilli}, {Milutinovic}, {Naidu}, {Newburgh}, {Ng}, {Patel}, {Pen}, {Pinsonneault-Marotte}, {Pleunis}, {Rafiei-Ravandi}, {Rahman}, {Ransom}, {Renard}, {Scholz}, {Siegel}, {Singh}, {Smith}, {Stairs}, {Tendulkar}, {Tretyakov}, {Vanderlinde}, {Yadav}, \& {Zwaniga}}]{CHIME18}
{CHIME/FRB Collaboration}, {Andersen}, B.~C., {Bandura}, K., {et~al.} 2019, \apjl, 885, L24

\bibitem[{{CHIME/FRB Collaboration} {et~al.}(2023){CHIME/FRB Collaboration}, {Andersen}, {Bandura}, {Bhardwaj}, {Boyle}, {Brar}, {Cassanelli}, {Chatterjee}, {Chawla}, {Cook}, {Curtin}, {Dobbs}, {Dong}, {Faber}, {Fandino}, {Fonseca}, {Gaensler}, {Giri}, {Herrera-Martin}, {Hill}, {Ibik}, {Josephy}, {Kaczmarek}, {Kader}, {Kaspi}, {Landecker}, {Lanman}, {Lazda}, {Leung}, {Lin}, {Masui}, {McKinven}, {Mena-Parra}, {Meyers}, {Michilli}, {Ng}, {Pandhi}, {Pearlman}, {Pen}, {Petroff}, {Pleunis}, {Rafiei-Ravandi}, {Rahman}, {Ransom}, {Renard}, {Sand}, {Sanghavi}, {Scholz}, {Shah}, {Shin}, {Siegel}, {Smith}, {Stairs}, {Su}, {Tendulkar}, {Vanderlinde}, {Wang}, {Wulf}, \& {Zwaniga}}]{chime23}
{CHIME/FRB Collaboration}, {Andersen}, B.~C., {Bandura}, K., {et~al.} 2023, \apj, 947, 83

\bibitem[{{CHIME/FRB Collaboration} {et~al.}(2020){CHIME/FRB Collaboration}, {Andersen}, {Bandura}, {Bhardwaj}, {Bij}, {Boyce}, {Boyle}, {Brar}, {Cassanelli}, {Chawla}, {Chen}, {Cliche}, {Cook}, {Cubranic}, {Curtin}, {Denman}, {Dobbs}, {Dong}, {Fandino}, {Fonseca}, {Gaensler}, {Giri}, {Good}, {Halpern}, {Hill}, {Hinshaw}, {H{\"o}fer}, {Josephy}, {Kania}, {Kaspi}, {Landecker}, {Leung}, {Li}, {Lin}, {Masui}, {Mckinven}, {Mena-Parra}, {Merryfield}, {Meyers}, {Michilli}, {Milutinovic}, {Mirhosseini}, {M{\"u}nchmeyer}, {Naidu}, {Newburgh}, {Ng}, {Patel}, {Pen}, {Pinsonneault-Marotte}, {Pleunis}, {Quine}, {Rafiei-Ravandi}, {Rahman}, {Ransom}, {Renard}, {Sanghavi}, {Scholz}, {Shaw}, {Shin}, {Siegel}, {Singh}, {Smegal}, {Smith}, {Stairs}, {Tan}, {Tendulkar}, {Tretyakov}, {Vanderlinde}, {Wang}, {Wulf}, \& {Zwaniga}}]{CHIME20b}
{CHIME/FRB Collaboration}, {Andersen}, B.~C., {Bandura}, K.~M., {et~al.} 2020, \nat, 587, 54

\bibitem[{{Condon} {et~al.}(1998){Condon}, {Cotton}, {Greisen}, {Yin}, {Perley}, {Taylor}, \& {Broderick}}]{Condon98}
{Condon}, J.~J., {Cotton}, W.~D., {Greisen}, E.~W., {et~al.} 1998, \aj, 115, 1693

\bibitem[{{Connor} {et~al.}(2025){Connor}, {Ravi}, {Sharma}, {Ocker}, {Faber}, {Hallinan}, {Harnach}, {Hellbourg}, {Hobbs}, {Hodge}, {Hodges}, {Kosogorov}, {Lamb}, {Law}, {Rasmussen}, {Sherman}, {Somalwar}, {Weinreb}, {Woody}, \& {Konietzka}}]{Connor24}
{Connor}, L., {Ravi}, V., {Sharma}, K., {et~al.} 2025, Nature Astronomy, 9, 1226

\bibitem[{{Connor} {et~al.}(2020){Connor}, {van Leeuwen}, {Oostrum}, {Petroff}, {Maan}, {Adams}, {Attema}, {Bast}, {Boersma}, {D{\'e}nes}, {Gardenier}, {Hargreaves}, {Kooistra}, {Pastor-Marazuela}, {Schulz}, {Sclocco}, {Smits}, {Straal}, {van der Schuur}, {Vohl}, {Adebahr}, {de Blok}, {van Cappellen}, {Coolen}, {Damstra}, {van Diepen}, {Frank}, {Hess}, {Hut}, {Kutkin}, {Loose}, {Lucero}, {Mika}, {Moss}, {Mulder}, {Oosterloo}, {Ruiter}, {Vedantham}, {Vermaas}, {Wijnholds}, \& {Ziemke}}]{Connor20}
{Connor}, L., {van Leeuwen}, J., {Oostrum}, L.~C., {et~al.} 2020, \mnras, 499, 4716

\bibitem[{{Cordes} \& {Chatterjee}(2019)}]{cordesChatterjee19}
{Cordes}, J.~M. \& {Chatterjee}, S. 2019, \araa, 57, 417

\bibitem[{{Cordes} \& {Lazio}(2002)}]{CordesLazio02}
{Cordes}, J.~M. \& {Lazio}, T.~J.~W. 2002, arXiv e-prints, astro, arXiv:astro-ph/0207156

\bibitem[{{Curtin} \& {CHIME/FRB Collaboration}(2024)}]{Curtin24_ATEL}
{Curtin}, A.~P. \& {CHIME/FRB Collaboration}. 2024, The Astronomer's Telegram, 16780, 1

\bibitem[{{D'Amato} {et~al.}(2022){D'Amato}, {Prandoni}, {Gilli}, {Vignali}, {Massardi}, {Liuzzo}, {Jagannathan}, {Brienza}, {Paladino}, {Mignoli}, {Marchesi}, {Peca}, {Chiaberge}, {Mazzolari}, \& {Norman}}]{DAmato22}
{D'Amato}, Q., {Prandoni}, I., {Gilli}, R., {et~al.} 2022, \aap, 668, A133

\bibitem[{{Day} {et~al.}(2020){Day}, {Deller}, {Shannon}, {Qiu(邱昊)}, {Bannister}, {Bhandari}, {Ekers}, {Flynn}, {James}, {Macquart}, {Mahony}, {Phillips}, \& {Xavier Prochaska}}]{Day19}
{Day}, C.~K., {Deller}, A.~T., {Shannon}, R.~M., {et~al.} 2020, \mnras, 497, 3335

\bibitem[{{Dey} {et~al.}(2019){Dey}, {Schlegel}, {Lang}, {Blum}, {Burleigh}, {Fan}, {Findlay}, {Finkbeiner}, {Herrera}, {Juneau}, {Landriau}, {Levi}, {McGreer}, {Meisner}, {Myers}, {Moustakas}, {Nugent}, {Patej}, {Schlafly}, {Walker}, {Valdes}, {Weaver}, {Y{\`e}che}, {Zou}, {Zhou}, {Abareshi}, {Abbott}, {Abolfathi}, {Aguilera}, {Alam}, {Allen}, {Alvarez}, {Annis}, {Ansarinejad}, {Aubert}, {Beechert}, {Bell}, {BenZvi}, {Beutler}, {Bielby}, {Bolton}, {Brice{\~n}o}, {Buckley-Geer}, {Butler}, {Calamida}, {Carlberg}, {Carter}, {Casas}, {Castander}, {Choi}, {Comparat}, {Cukanovaite}, {Delubac}, {DeVries}, {Dey}, {Dhungana}, {Dickinson}, {Ding}, {Donaldson}, {Duan}, {Duckworth}, {Eftekharzadeh}, {Eisenstein}, {Etourneau}, {Fagrelius}, {Farihi}, {Fitzpatrick}, {Font-Ribera}, {Fulmer}, {G{\"a}nsicke}, {Gaztanaga}, {George}, {Gerdes}, {Gontcho}, {Gorgoni}, {Green}, {Guy}, {Harmer}, {Hernandez}, {Honscheid}, {Huang}, {James}, {Jannuzi}, {Jiang}, {Joyce}, {Karcher}, {Karkar}, {Kehoe}, {Kneib}, {Kueter-Young}, {Lan},
  {Lauer}, {Le Guillou}, {Le Van Suu}, {Lee}, {Lesser}, {Perreault Levasseur}, {Li}, {Mann}, {Marshall}, {Mart{\'\i}nez-V{\'a}zquez}, {Martini}, {du Mas des Bourboux}, {McManus}, {Meier}, {M{\'e}nard}, {Metcalfe}, {Mu{\~n}oz-Guti{\'e}rrez}, {Najita}, {Napier}, {Narayan}, {Newman}, {Nie}, {Nord}, {Norman}, {Olsen}, {Paat}, {Palanque-Delabrouille}, {Peng}, {Poppett}, {Poremba}, {Prakash}, {Rabinowitz}, {Raichoor}, {Rezaie}, {Robertson}, {Roe}, {Ross}, {Ross}, {Rudnick}, {Safonova}, {Saha}, {S{\'a}nchez}, {Savary}, {Schweiker}, {Scott}, {Seo}, {Shan}, {Silva}, {Slepian}, {Soto}, {Sprayberry}, {Staten}, {Stillman}, {Stupak}, {Summers}, {Sien Tie}, {Tirado}, {Vargas-Maga{\~n}a}, {Vivas}, {Wechsler}, {Williams}, {Yang}, {Yang}, {Yapici}, {Zaritsky}, {Zenteno}, {Zhang}, {Zhang}, {Zhou}, \& {Zhou}}]{Dey19}
{Dey}, A., {Schlegel}, D.~J., {Lang}, D., {et~al.} 2019, \aj, 157, 168

\bibitem[{{Dong} {et~al.}(2024){Dong}, {Eftekhari}, {Fong}, {Bhandari}, {Berger}, {Ould-Boukattine}, {Hessels}, {Sridhar}, {Reines}, {Margalit}, {Darling}, {Gordon}, {Greene}, {Kilpatrick}, {Marcote}, {Metzger}, {Nimmo}, {Nugent}, {Paragi}, \& {Williams}}]{Dong24}
{Dong}, Y., {Eftekhari}, T., {Fong}, W., {et~al.} 2024, \apj, 973, 133

\bibitem[{{Duncan} \& {Thompson}(1992)}]{DuncanThompson}
{Duncan}, R.~C. \& {Thompson}, C. 1992, \apjl, 392, L9

\bibitem[{{Flewelling} {et~al.}(2020){Flewelling}, {Magnier}, {Chambers}, {Heasley}, {Holmberg}, {Huber}, {Sweeney}, {Waters}, {Calamida}, {Casertano}, {Chen}, {Farrow}, {Hasinger}, {Henderson}, {Long}, {Metcalfe}, {Narayan}, {Nieto-Santisteban}, {Norberg}, {Rest}, {Saglia}, {Szalay}, {Thakar}, {Tonry}, {Valenti}, {Werner}, {White}, {Denneau}, {Draper}, {Hodapp}, {Jedicke}, {Kaiser}, {Kudritzki}, {Price}, {Wainscoat}, {Chastel}, {McLean}, {Postman}, \& {Shiao}}]{Flewelling20}
{Flewelling}, H.~A., {Magnier}, E.~A., {Chambers}, K.~C., {et~al.} 2020, \apjs, 251, 7

\bibitem[{{Galluzzi} {et~al.}(2025){Galluzzi}, {Behiri}, {Giulietti}, \& {Lapi}}]{Galluzzi2025}
{Galluzzi}, V., {Behiri}, M., {Giulietti}, M., \& {Lapi}, A. 2025, Galaxies, 13, 34

\bibitem[{{Gordon} {et~al.}(2023){Gordon}, {Fong}, {Kilpatrick}, {Eftekhari}, {Leja}, {Prochaska}, {Nugent}, {Bhandari}, {Blanchard}, {Caleb}, {Day}, {Deller}, {Dong}, {Glowacki}, {Gourdji}, {Mannings}, {Mahoney}, {Marnoch}, {Miller}, {Paterson}, {Rastinejad}, {Ryder}, {Sadler}, {Scott}, {Sears}, {Shannon}, {Simha}, {Stappers}, \& {Tejos}}]{Gordon23}
{Gordon}, A.~C., {Fong}, W.-f., {Kilpatrick}, C.~D., {et~al.} 2023, \apj, 954, 80

\bibitem[{{Heintz} {et~al.}(2020){Heintz}, {Prochaska}, {Simha}, {Platts}, {Fong}, {Tejos}, {Ryder}, {Aggerwal}, {Bhandari}, {Day}, {Deller}, {Kilpatrick}, {Law}, {Macquart}, {Mannings}, {Marnoch}, {Sadler}, \& {Shannon}}]{Heintz20}
{Heintz}, K.~E., {Prochaska}, J.~X., {Simha}, S., {et~al.} 2020, \apj, 903, 152

\bibitem[{{Hewitt} {et~al.}(2024){Hewitt}, {Bhandari}, {Marcote}, {Hessels}, {Nimmo}, {Kirsten}, {Bach}, {Bezrukovs}, {Bhardwaj}, {Blaauw}, {Bray}, {Buttaccio}, {Corongiu}, {Gawro{\'n}ski}, {Giroletti}, {Keimpema}, {Maccaferri}, {Paragi}, {Trudu}, {Snelders}, {Venturi}, {Wang}, {Williams-Baldwin}, {Wrigley}, {Yang}, \& {Yuan}}]{Hewitt24}
{Hewitt}, D.~M., {Bhandari}, S., {Marcote}, B., {et~al.} 2024, \mnras, 529, 1814

\bibitem[{{Hewitt} {et~al.}(2022){Hewitt}, {Snelders}, {Hessels}, {Nimmo}, {Jahns}, {Spitler}, {Gourdji}, {Hilmarsson}, {Michilli}, {Ould-Boukattine}, {Scholz}, \& {Seymour}}]{Hewitt22}
{Hewitt}, D.~M., {Snelders}, M.~P., {Hessels}, J.~W.~T., {et~al.} 2022, \mnras, 515, 3577

\bibitem[{{Heywood} {et~al.}(2020){Heywood}, {Hale}, {Jarvis}, {Makhathini}, {Peters}, {Sebokolodi}, \& {Smirnov}}]{Heywood20}
{Heywood}, I., {Hale}, C.~L., {Jarvis}, M.~J., {et~al.} 2020, \mnras, 496, 3469

\bibitem[{{Hilmarsson} {et~al.}(2021){Hilmarsson}, {Michilli}, {Spitler}, {Wharton}, {Demorest}, {Desvignes}, {Gourdji}, {Hackstein}, {Hessels}, {Nimmo}, {Seymour}, {Kramer}, \& {Mckinven}}]{Hilmarsson21}
{Hilmarsson}, G.~H., {Michilli}, D., {Spitler}, L.~G., {et~al.} 2021, \apjl, 908, L10

\bibitem[{{Hunter}(2007)}]{Hunter07}
{Hunter}, J.~D. 2007, Computing in Science and Engineering, 9, 90

\bibitem[{{Ibik} {et~al.}(2024{\natexlab{a}}){Ibik}, {Drout}, {Gaensler}, {Scholz}, {Michilli}, {Bhardwaj}, {Kaspi}, {Pleunis}, {Cassanelli}, {Cook}, {Dong}, {Kaczmarek}, {Leung}, {Lu}, {Masui}, {Pearlman}, {Rafiei-Ravandi}, {Sand}, {Shin}, {Smith}, \& {Stairs}}]{Ibik24_b}
{Ibik}, A.~L., {Drout}, M.~R., {Gaensler}, B.~M., {et~al.} 2024{\natexlab{a}}, \apj, 961, 99

\bibitem[{{Ibik} {et~al.}(2024{\natexlab{b}}){Ibik}, {Drout}, {Gaensler}, {Scholz}, {Sridhar}, {Margalit}, {Clarke}, {Law}, {Tendulkar}, {Michilli}, {Eftekhari}, {Bhardwaj}, {Burke-Spolaor}, {Chatterjee}, {Cook}, {Hessels}, {Kirsten}, {Joseph}, {Kaspi}, {Lazda}, {Masui}, {Nimmo}, {Pandhi}, {Pearlman}, {Pleunis}, {Rafiei-Ravandi}, {Shin}, \& {Smith}}]{Ibik24}
{Ibik}, A.~L., {Drout}, M.~R., {Gaensler}, B.~M., {et~al.} 2024{\natexlab{b}}, \apj, 976, 199

\bibitem[{{Isaacson} {et~al.}(2017){Isaacson}, {Siemion}, {Marcy}, {Lebofsky}, {Price}, {MacMahon}, {Croft}, {DeBoer}, {Hickish}, {Werthimer}, {Sheikh}, {Hellbourg}, \& {Enriquez}}]{Isaacson17}
{Isaacson}, H., {Siemion}, A. P.~V., {Marcy}, G.~W., {et~al.} 2017, \pasp, 129, 054501

\bibitem[{{Jahns} {et~al.}(2023){Jahns}, {Spitler}, {Nimmo}, {Hewitt}, {Snelders}, {Seymour}, {Hessels}, {Gourdji}, {Michilli}, \& {Hilmarsson}}]{Jahns23}
{Jahns}, J.~N., {Spitler}, L.~G., {Nimmo}, K., {et~al.} 2023, \mnras, 519, 666

\bibitem[{{James}(2023)}]{james23}
{James}, C.~W. 2023, \pasa, 40, e057

\bibitem[{{James} {et~al.}(2022{\natexlab{a}}){James}, {Prochaska}, {Macquart}, {North-Hickey}, {Bannister}, \& {Dunning}}]{james22b}
{James}, C.~W., {Prochaska}, J.~X., {Macquart}, J.~P., {et~al.} 2022{\natexlab{a}}, \mnras, 510, L18

\bibitem[{{James} {et~al.}(2022{\natexlab{b}}){James}, {Prochaska}, {Macquart}, {North-Hickey}, {Bannister}, \& {Dunning}}]{James22_zDM}
{James}, C.~W., {Prochaska}, J.~X., {Macquart}, J.~P., {et~al.} 2022{\natexlab{b}}, \mnras, 509, 4775

\bibitem[{{Keane} {et~al.}(2016){Keane}, {Johnston}, {Bhandari}, {Barr}, {Bhat}, {Burgay}, {Caleb}, {Flynn}, {Jameson}, {Kramer}, {Petroff}, {Possenti}, {van Straten}, {Bailes}, {Burke-Spolaor}, {Eatough}, {Stappers}, {Totani}, {Honma}, {Furusawa}, {Hattori}, {Morokuma}, {Niino}, {Sugai}, {Terai}, {Tominaga}, {Yamasaki}, {Yasuda}, {Allen}, {Cooke}, {Jencson}, {Kasliwal}, {Kaplan}, {Tingay}, {Williams}, {Wayth}, {Chandra}, {Perrodin}, {Berezina}, {Mickaliger}, \& {Bassa}}]{Keane16}
{Keane}, E.~F., {Johnston}, S., {Bhandari}, S., {et~al.} 2016, \nat, 530, 453

\bibitem[{{Kilpatrick} {et~al.}(2021){Kilpatrick}, {Fong}, {Prochaska}, {Tejos}, {Bhandari}, \& {Day}}]{Kilpatrick21}
{Kilpatrick}, C.~D., {Fong}, W., {Prochaska}, J.~X., {et~al.} 2021, The Astronomer's Telegram, 14516, 1

\bibitem[{{Kirsten} {et~al.}(2022){Kirsten}, {Marcote}, {Nimmo}, {Hessels}, {Bhardwaj}, {Tendulkar}, {Keimpema}, {Yang}, {Snelders}, {Scholz}, {Pearlman}, {Law}, {Peters}, {Giroletti}, {Paragi}, {Bassa}, {Hewitt}, {Bach}, {Bezrukovs}, {Burgay}, {Buttaccio}, {Conway}, {Corongiu}, {Feiler}, {Forss{\'e}n}, {Gawro{\'n}ski}, {Karuppusamy}, {Kharinov}, {Lindqvist}, {Maccaferri}, {Melnikov}, {Ould-Boukattine}, {Possenti}, {Surcis}, {Wang}, {Yuan}, {Aggarwal}, {Anna-Thomas}, {Bower}, {Blaauw}, {Burke-Spolaor}, {Cassanelli}, {Clarke}, {Fonseca}, {Gaensler}, {Gopinath}, {Kaspi}, {Kassim}, {Lazio}, {Leung}, {Li}, {Lin}, {Masui}, {Mckinven}, {Michilli}, {Mikhailov}, {Ng}, {Orbidans}, {Pen}, {Petroff}, {Rahman}, {Ransom}, {Shin}, {Smith}, {Stairs}, \& {Vlemmings}}]{Kirsten22glob}
{Kirsten}, F., {Marcote}, B., {Nimmo}, K., {et~al.} 2022, \nat, 602, 585

\bibitem[{{Kirsten} {et~al.}(2021){Kirsten}, {Snelders}, {Jenkins}, {Nimmo}, {van den Eijnden}, {Hessels}, {Gawro{\'n}ski}, \& {Yang}}]{Kirsten21}
{Kirsten}, F., {Snelders}, M.~P., {Jenkins}, M., {et~al.} 2021, Nature Astronomy, 5, 414

\bibitem[{{Klein} {et~al.}(2018){Klein}, {Lisenfeld}, \& {Verley}}]{Klein18}
{Klein}, U., {Lisenfeld}, U., \& {Verley}, S. 2018, \aap, 611, A55

\bibitem[{{Koss} {et~al.}(2022){Koss}, {Trakhtenbrot}, {Ricci}, {Oh}, {Bauer}, {Stern}, {Caglar}, {den Brok}, {Mushotzky}, {Ricci}, {Mej{\'\i}a-Restrepo}, {Lamperti}, {Treister}, {B{\"a}r}, {Harrison}, {Powell}, {Privon}, {Riffel}, {Rojas}, {Schawinski}, \& {Urry}}]{Koss22}
{Koss}, M.~J., {Trakhtenbrot}, B., {Ricci}, C., {et~al.} 2022, \apjs, 261, 6

\bibitem[{{Kothes} {et~al.}(2018){Kothes}, {Sun}, {Gaensler}, \& {Reich}}]{Kothes18}
{Kothes}, R., {Sun}, X., {Gaensler}, B., \& {Reich}, W. 2018, \apj, 852, 54

\bibitem[{{Kumar} {et~al.}(2023){Kumar}, {Luo}, {Price}, {Shannon}, {Deller}, {Bhandari}, {Feng}, {Flynn}, {Jiang}, {Uttarkar}, {Wang}, \& {Zhang}}]{Kumar23}
{Kumar}, P., {Luo}, R., {Price}, D.~C., {et~al.} 2023, \mnras, 526, 3652

\bibitem[{{Kumar} {et~al.}(2021){Kumar}, {Shannon}, {Flynn}, {Os{\l}owski}, {Bhandari}, {Day}, {Deller}, {Farah}, {Kaczmarek}, {Kerr}, {Phillips}, {Price}, {Qiu}, \& {Thyagarajan}}]{Kumar21}
{Kumar}, P., {Shannon}, R.~M., {Flynn}, C., {et~al.} 2021, \mnras, 500, 2525

\bibitem[{{Lacy} {et~al.}(2020){Lacy}, {Baum}, {Chandler}, {Chatterjee}, {Clarke}, {Deustua}, {English}, {Farnes}, {Gaensler}, {Gugliucci}, {Hallinan}, {Kent}, {Kimball}, {Law}, {Lazio}, {Marvil}, {Mao}, {Medlin}, {Mooley}, {Murphy}, {Myers}, {Osten}, {Richards}, {Rosolowsky}, {Rudnick}, {Schinzel}, {Sivakoff}, {Sjouwerman}, {Taylor}, {White}, {Wrobel}, {Andernach}, {Beasley}, {Berger}, {Bhatnager}, {Birkinshaw}, {Bower}, {Brandt}, {Brown}, {Burke-Spolaor}, {Butler}, {Comerford}, {Demorest}, {Fu}, {Giacintucci}, {Golap}, {G{\"u}th}, {Hales}, {Hiriart}, {Hodge}, {Horesh}, {Ivezi{\'c}}, {Jarvis}, {Kamble}, {Kassim}, {Liu}, {Loinard}, {Lyons}, {Masters}, {Mezcua}, {Moellenbrock}, {Mroczkowski}, {Nyland}, {O'Dea}, {O'Sullivan}, {Peters}, {Radford}, {Rao}, {Robnett}, {Salcido}, {Shen}, {Sobotka}, {Witz}, {Vaccari}, {van Weeren}, {Vargas}, {Williams}, \& {Yoon}}]{Lacy20}
{Lacy}, M., {Baum}, S.~A., {Chandler}, C.~J., {et~al.} 2020, \pasp, 132, 035001

\bibitem[{{Law} {et~al.}(2020){Law}, {Butler}, {Prochaska}, {Zackay}, {Burke-Spolaor}, {Mannings}, {Tejos}, {Josephy}, {Andersen}, {Chawla}, {Heintz}, {Aggarwal}, {Bower}, {Demorest}, {Kilpatrick}, {Lazio}, {Linford}, {Mckinven}, {Tendulkar}, \& {Simha}}]{Law20}
{Law}, C.~J., {Butler}, B.~J., {Prochaska}, J.~X., {et~al.} 2020, \apj, 899, 161

\bibitem[{{Law} {et~al.}(2022){Law}, {Connor}, \& {Aggarwal}}]{Law22}
{Law}, C.~J., {Connor}, L., \& {Aggarwal}, K. 2022, \apj, 927, 55

\bibitem[{{Law} {et~al.}(2024){Law}, {Sharma}, {Ravi}, {Chen}, {Catha}, {Connor}, {Faber}, {Hallinan}, {Harnach}, {Hellbourg}, {Hobbs}, {Hodge}, {Hodges}, {Lamb}, {Rasmussen}, {Sherman}, {Shi}, {Simard}, {Squillace}, {Weinreb}, {Woody}, \& {Yurk}}]{Law23}
{Law}, C.~J., {Sharma}, K., {Ravi}, V., {et~al.} 2024, \apj, 967, 29

\bibitem[{{Lee-Waddell} {et~al.}(2023){Lee-Waddell}, {James}, {Ryder}, {Mahony}, {Bahramian}, {Koribalski}, {Kumar}, {Marnoch}, {North-Hickey}, {Sadler}, {Shannon}, {Tejos}, {Thorne}, {Wang}, \& {Wayth}}]{LeeWaddell23}
{Lee-Waddell}, K., {James}, C.~W., {Ryder}, S.~D., {et~al.} 2023, \pasa, 40, e029

\bibitem[{{Li} {et~al.}(2021){Li}, {Wang}, {Zhu}, {Zhang}, {Zhang}, {Duan}, {Zhang}, {Feng}, {Tang}, {Chatterjee}, {Cordes}, {Cruces}, {Dai}, {Gajjar}, {Hobbs}, {Jin}, {Kramer}, {Lorimer}, {Miao}, {Niu}, {Niu}, {Pan}, {Qian}, {Spitler}, {Werthimer}, {Zhang}, {Wang}, {Xie}, {Yue}, {Zhang}, {Zhi}, \& {Zhu}}]{Li21}
{Li}, D., {Wang}, P., {Zhu}, W.~W., {et~al.} 2021, \nat, 598, 267

\bibitem[{{Li} {et~al.}(2020){Li}, {Gao}, {Wei}, {Yang}, {Zhang}, \& {Zhu}}]{Li20_host}
{Li}, Z., {Gao}, H., {Wei}, J.~J., {et~al.} 2020, \mnras, 496, L28

\bibitem[{{Luo} {et~al.}(2020){Luo}, {Wang}, {Men}, {Zhang}, {Jiang}, {Xu}, {Wang}, {Lee}, {Han}, {Zhang}, {Caballero}, {Chen}, {Chen}, {Gan}, {Guo}, {Hao}, {Huang}, {Jiang}, {Li}, {Li}, {Li}, {Luo}, {Pan}, {Pei}, {Qian}, {Sun}, {Wang}, {Wang}, {Wen}, {Xu}, {Xu}, {Yan}, {Yan}, {Yu}, {Yuan}, {Zhang}, \& {Zhu}}]{Luo22}
{Luo}, R., {Wang}, B.~J., {Men}, Y.~P., {et~al.} 2020, \nat, 586, 693

\bibitem[{{Lyubarsky}(2020)}]{Liubarsky20}
{Lyubarsky}, Y. 2020, \apj, 897, 1

\bibitem[{{Macquart} {et~al.}(2010){Macquart}, {Bailes}, {Bhat}, {Bower}, {Bunton}, {Chatterjee}, {Colegate}, {Cordes}, {D'Addario}, {Deller}, {Dodson}, {Fender}, {Haines}, {Halll}, {Harris}, {Hotan}, {Johnston}, {Jones}, {Keith}, {Koay}, {Lazio}, {Majid}, {Murphy}, {Navarro}, {Phillips}, {Quinn}, {Preston}, {Stansby}, {Stairs}, {Stappers}, {Staveley-Smith}, {Tingay}, {Thompson}, {van Straten}, {Wagstaff}, {Warren}, {Wayth}, {Wen}, \& {CRAFT Collaboration}}]{Macquart10}
{Macquart}, J.-P., {Bailes}, M., {Bhat}, N.~D.~R., {et~al.} 2010, \pasa, 27, 272

\bibitem[{{Macquart} {et~al.}(2020{\natexlab{a}}){Macquart}, {Prochaska}, {McQuinn}, {Bannister}, {Bhandari}, {Day}, {Deller}, {Ekers}, {James}, {Marnoch}, {Os{\l}owski}, {Phillips}, {Ryder}, {Scott}, {Shannon}, \& {Tejos}}]{Maquart19}
{Macquart}, J.~P., {Prochaska}, J.~X., {McQuinn}, M., {et~al.} 2020{\natexlab{a}}, \nat, 581, 391

\bibitem[{{Macquart} {et~al.}(2020{\natexlab{b}}){Macquart}, {Prochaska}, {McQuinn}, {Bannister}, {Bhandari}, {Day}, {Deller}, {Ekers}, {James}, {Marnoch}, {Os{\l}owski}, {Phillips}, {Ryder}, {Scott}, {Shannon}, \& {Tejos}}]{Macquart20}
{Macquart}, J.~P., {Prochaska}, J.~X., {McQuinn}, M., {et~al.} 2020{\natexlab{b}}, \nat, 581, 391

\bibitem[{{Mancuso} {et~al.}(2017){Mancuso}, {Lapi}, {Prandoni}, {Obi}, {Gonzalez-Nuevo}, {Perrotta}, {Bressan}, {Celotti}, \& {Danese}}]{Mancuso17}
{Mancuso}, C., {Lapi}, A., {Prandoni}, I., {et~al.} 2017, \apj, 842, 95

\bibitem[{{Marcote} {et~al.}(2020){Marcote}, {Nimmo}, {Hessels}, {Tendulkar}, {Bassa}, {Paragi}, {Keimpema}, {Bhardwaj}, {Karuppusamy}, {Kaspi}, {Law}, {Michilli}, {Aggarwal}, {Andersen}, {Archibald}, {Bandura}, {Bower}, {Boyle}, {Brar}, {Burke-Spolaor}, {Butler}, {Cassanelli}, {Chawla}, {Demorest}, {Dobbs}, {Fonseca}, {Giri}, {Good}, {Gourdji}, {Josephy}, {Kirichenko}, {Kirsten}, {Landecker}, {Lang}, {Lazio}, {Li}, {Lin}, {Linford}, {Masui}, {Mena-Parra}, {Naidu}, {Ng}, {Patel}, {Pen}, {Pleunis}, {Rafiei-Ravandi}, {Rahman}, {Renard}, {Scholz}, {Siegel}, {Smith}, {Stairs}, {Vanderlinde}, \& {Zwaniga}}]{Marcote20}
{Marcote}, B., {Nimmo}, K., {Hessels}, J.~W.~T., {et~al.} 2020, \nat, 577, 190

\bibitem[{{Marcote} {et~al.}(2017){Marcote}, {Paragi}, {Hessels}, {Keimpema}, {van Langevelde}, {Huang}, {Bassa}, {Bogdanov}, {Bower}, {Burke-Spolaor}, {Butler}, {Campbell}, {Chatterjee}, {Cordes}, {Demorest}, {Garrett}, {Ghosh}, {Kaspi}, {Law}, {Lazio}, {McLaughlin}, {Ransom}, {Salter}, {Scholz}, {Seymour}, {Siemion}, {Spitler}, {Tendulkar}, \& {Wharton}}]{Marcote17}
{Marcote}, B., {Paragi}, Z., {Hessels}, J.~W.~T., {et~al.} 2017, \apjl, 834, L8

\bibitem[{{Margalit} \& {Metzger}(2018)}]{margalitmetzger18}
{Margalit}, B. \& {Metzger}, B.~D. 2018, \apjl, 868, L4

\bibitem[{{Marnoch} {et~al.}(2023){Marnoch}, {Ryder}, {James}, {Gordon}, {Sammons}, {Prochaska}, {Tejos}, {Deller}, {Scott}, {Bhandari}, {Glowacki}, {Mahony}, {McDermid}, {Sadler}, {Shannon}, \& {Qiu}}]{Marnoch23}
{Marnoch}, L., {Ryder}, S.~D., {James}, C.~W., {et~al.} 2023, \mnras, 525, 994

\bibitem[{{Massardi} {et~al.}(2025){Massardi}, {Behiri}, {Galluzzi}, {Giulietti}, {Perrotta}, {Prandoni}, \& {Lapi}}]{Massardi25}
{Massardi}, M., {Behiri}, M., {Galluzzi}, V., {et~al.} 2025, \pasp, 137, 014101

\bibitem[{{Masui} {et~al.}(2015){Masui}, {Lin}, {Sievers}, {Anderson}, {Chang}, {Chen}, {Ganguly}, {Jarvis}, {Kuo}, {Li}, {Liao}, {McLaughlin}, {Pen}, {Peterson}, {Roman}, {Timbie}, {Voytek}, \& {Yadav}}]{Masui15}
{Masui}, K., {Lin}, H.-H., {Sievers}, J., {et~al.} 2015, \nat, 528, 523

\bibitem[{{Matthews} {et~al.}(2021){Matthews}, {Condon}, {Cotton}, \& {Mauch}}]{Matthews21}
{Matthews}, A.~M., {Condon}, J.~J., {Cotton}, W.~D., \& {Mauch}, T. 2021, \apj, 909, 193

\bibitem[{{Mckinven} {et~al.}(2023){Mckinven}, {Gaensler}, {Michilli}, {Masui}, {Kaspi}, {Su}, {Bhardwaj}, {Cassanelli}, {Chawla}, {Dong}, {Fonseca}, {Leung}, {Li}, {Ng}, {Patel}, {Pearlman}, {Petroff}, {Pleunis}, {Rafiei-Ravandi}, {Rahman}, {Sand}, {Shin}, {Stairs}, \& {Tendulkar}}]{McKinven23}
{Mckinven}, R., {Gaensler}, B.~M., {Michilli}, D., {et~al.} 2023, \apj, 951, 82

\bibitem[{{McMullin} {et~al.}(2007){McMullin}, {Waters}, {Schiebel}, {Young}, \& {Golap}}]{McMullin07}
{McMullin}, J.~P., {Waters}, B., {Schiebel}, D., {Young}, W., \& {Golap}, K. 2007, in Astronomical Society of the Pacific Conference Series, Vol. 376, Astronomical Data Analysis Software and Systems XVI, ed. R.~A. {Shaw}, F.~{Hill}, \& D.~J. {Bell}, 127

\bibitem[{{Mfulwane} {et~al.}(2026){Mfulwane}, {Chibueze}, {Letsele}, {Nyambe}, {Venter}, {Bezuidenhout}, {Stappers}, {Spitler}, {Caleb}, {Deller}, {Fortunato}, {Cornejo}, {Sch{\"u}ssler}, {Ashkar}, {Bradascio}, {Kalita}, {Kundu}, {Kramer}, {Keane}, \& {Weltman}}]{Mfulwane26}
{Mfulwane}, L.~L., {Chibueze}, J.~O., {Letsele}, T.~P., {et~al.} 2026, \mnras, 546, stag166

\bibitem[{{Michilli} {et~al.}(2023){Michilli}, {Bhardwaj}, {Brar}, {Gaensler}, {Kaspi}, {Kirichenko}, {Masui}, {Mckinven}, {Ng}, {Patel}, {Sand}, {Scholz}, {Shin}, {Siegel}, {Stairs}, {Cassanelli}, {Cook}, {Dobbs}, {Dong}, {Fonseca}, {Ibik}, {Kaczmarek}, {Leung}, {Pearlman}, {Petroff}, {Pleunis}, {Rafiei-Ravandi}, {Sanghavi}, {Shaw}, \& {Tendulkar}}]{Michilli23}
{Michilli}, D., {Bhardwaj}, M., {Brar}, C., {et~al.} 2023, \apj, 950, 134

\bibitem[{{Michilli} {et~al.}(2018){Michilli}, {Seymour}, {Hessels}, {Spitler}, {Gajjar}, {Archibald}, {Bower}, {Chatterjee}, {Cordes}, {Gourdji}, {Heald}, {Kaspi}, {Law}, {Sobey}, {Adams}, {Bassa}, {Bogdanov}, {Brinkman}, {Demorest}, {Fernandez}, {Hellbourg}, {Lazio}, {Lynch}, {Maddox}, {Marcote}, {McLaughlin}, {Paragi}, {Ransom}, {Scholz}, {Siemion}, {Tendulkar}, {van Rooy}, {Wharton}, \& {Whitlow}}]{Michilli18}
{Michilli}, D., {Seymour}, A., {Hessels}, J.~W.~T., {et~al.} 2018, \nat, 553, 182

\bibitem[{{Moroianu} {et~al.}(2026){Moroianu}, {Bhandari}, {Drout}, {Hessels}, {Hewitt}, {Kirsten}, {Marcote}, {Pleunis}, {Snelders}, {Sridhar}, {Bach}, {Bempong-Manful}, {Bezrukovs}, {Blaauw}, {Bray}, {Buttaccio}, {Chatterjee}, {Corongiu}, {Feiler}, {Gaensler}, {Gawro{\'n}ski}, {Giroletti}, {Ibik}, {Karuppusamy}, {Lazda}, {Leung}, {Lindqvist}, {Masui}, {Michilli}, {Nimmo}, {Ould-Boukattine}, {Pandhi}, {Paragi}, {Pearlman}, {Puchalska}, {Scholz}, {Shin}, {Sluman}, {Trudu}, {Williams-Baldwin}, \& {Yang}}]{Moroianu26}
{Moroianu}, A.~M., {Bhandari}, S., {Drout}, M.~R., {et~al.} 2026, \apjl, 996, L16

\bibitem[{{Ng} {et~al.}(2025){Ng}, {Pandhi}, {Mckinven}, {Curtin}, {Shin}, {Fonseca}, {Gaensler}, {Jow}, {Kaspi}, {Li}, {Main}, {Masui}, {Michilli}, {Nimmo}, {Pleunis}, {Scholz}, {Stairs}, {Bhardwaj}, {Brar}, {Cassanelli}, {Joseph}, {Pearlman}, {Rafiei-Ravandi}, \& {Smith}}]{Ng25}
{Ng}, C., {Pandhi}, A., {Mckinven}, R., {et~al.} 2025, \apj, 982, 154

\bibitem[{{Ng} \& {CHIME/FRB Collaboration}(2025)}]{Ng25_CHIME}
{Ng}, M. \& {CHIME/FRB Collaboration}. 2025, The Astronomer's Telegram, 17081, 1

\bibitem[{{Nimmo} {et~al.}(2022){Nimmo}, {Hewitt}, {Hessels}, {Kirsten}, {Marcote}, {Bach}, {Blaauw}, {Burgay}, {Corongiu}, {Feiler}, {Gawro{\'n}ski}, {Giroletti}, {Karuppusamy}, {Keimpema}, {Kharinov}, {Lindqvist}, {Maccaferri}, {Melnikov}, {Mikhailov}, {Ould-Boukattine}, {Paragi}, {Pilia}, {Possenti}, {Snelders}, {Surcis}, {Trudu}, {Venturi}, {Vlemmings}, {Wang}, {Yang}, \& {Yuan}}]{Nimmo21}
{Nimmo}, K., {Hewitt}, D.~M., {Hessels}, J.~W.~T., {et~al.} 2022, \apjl, 927, L3

\bibitem[{{Niu} {et~al.}(2022){Niu}, {Aggarwal}, {Li}, {Zhang}, {Chatterjee}, {Tsai}, {Yu}, {Law}, {Burke-Spolaor}, {Cordes}, {Zhang}, {Ocker}, {Yao}, {Wang}, {Feng}, {Niino}, {Bochenek}, {Cruces}, {Connor}, {Jiang}, {Dai}, {Luo}, {Li}, {Miao}, {Niu}, {Anna-Thomas}, {Sydnor}, {Stern}, {Wang}, {Yuan}, {Yue}, {Zhou}, {Yan}, {Zhu}, \& {Zhang}}]{Niu21}
{Niu}, C.~H., {Aggarwal}, K., {Li}, D., {et~al.} 2022, \nat, 606, 873

\bibitem[{{Offringa} {et~al.}(2014){Offringa}, {McKinley}, {Hurley-Walker}, {Briggs}, {Wayth}, {Kaplan}, {Bell}, {Feng}, {Neben}, {Hughes}, {Rhee}, {Murphy}, {Bhat}, {Bernardi}, {Bowman}, {Cappallo}, {Corey}, {Deshpande}, {Emrich}, {Ewall-Wice}, {Gaensler}, {Goeke}, {Greenhill}, {Hazelton}, {Hindson}, {Johnston-Hollitt}, {Jacobs}, {Kasper}, {Kratzenberg}, {Lenc}, {Lonsdale}, {Lynch}, {McWhirter}, {Mitchell}, {Morales}, {Morgan}, {Kudryavtseva}, {Oberoi}, {Ord}, {Pindor}, {Procopio}, {Prabu}, {Riding}, {Roshi}, {Shankar}, {Srivani}, {Subrahmanyan}, {Tingay}, {Waterson}, {Webster}, {Whitney}, {Williams}, \& {Williams}}]{offringa14}
{Offringa}, A.~R., {McKinley}, B., {Hurley-Walker}, N., {et~al.} 2014, \mnras, 444, 606

\bibitem[{{Offringa} \& {Smirnov}(2017)}]{offringa17}
{Offringa}, A.~R. \& {Smirnov}, O. 2017, \mnras, 471, 301

\bibitem[{{Orr} {et~al.}(2024){Orr}, {Burkhart}, {Lu}, {Ponnada}, \& {Hummels}}]{Orr2024}
{Orr}, M.~E., {Burkhart}, B., {Lu}, W., {Ponnada}, S.~B., \& {Hummels}, C.~B. 2024, \apjl, 972, L26

\bibitem[{{Os{\l}owski} {et~al.}(2019){Os{\l}owski}, {Shannon}, {Ravi}, {Kaczmarek}, {Zhang}, {Hobbs}, {Bailes}, {Russell}, {van Straten}, {James}, {Jameson}, {Mahony}, {Kumar}, {Andreoni}, {Bhat}, {Burke-Spolaor}, {Dai}, {Dempsey}, {Kerr}, {Manchester}, {Parthasarathy}, {Reardon}, {Sarkissian}, {Spiewak}, {Toomey}, {Wang}, {Zhang}, \& {Zhu}}]{Oslowski19}
{Os{\l}owski}, S., {Shannon}, R.~M., {Ravi}, V., {et~al.} 2019, \mnras, 488, 868

\bibitem[{{Ould-Boukattine} {et~al.}(2024){Ould-Boukattine}, {Dijkema}, {Gawronski}, {Herrmann}, {Hessels}, {Kirsten}, {Snelders}, {Beer}, {Bijlsma}, {Blaauw}, {Boons}, {Boven}, {Buchsteiner}, {Engelskirchen}, {Fischer}, {Hewitt}, {Loge}, {Marcote}, {van der Meer}, {Mulder}, {Munk}, {Nitsche}, {Ovinge}, {Puchalska}, {Sanders}, {Schmitz}, {Sluman}, {Telkamp}, {Wolf}, \& {Yang}}]{OuldBoukattine24}
{Ould-Boukattine}, O.~S., {Dijkema}, T.~J., {Gawronski}, M., {et~al.} 2024, The Astronomer's Telegram, 16565, 1

\bibitem[{{Pandhi} {et~al.}(2026){Pandhi}, {Nimmo}, {Andrew}, {Brar}, {Chatterjee}, {Cook}, {Curtin}, {Gaensler}, {Gawronski}, {Hessels}, {Kaspi}, {Khan}, {Kirsten}, {Lazda}, {Leung}, {Main}, {Masui}, {Mckinven}, {Michilli}, {Ng}, {Ould-Boukattine}, {Pearlman}, {Pleunis}, {Pollak}, {Pradeep E.~T.}, {Puchalska}, {Sammons}, {Scholz}, {Shah}, {Shin}, {Siegel}, \& {Smith}}]{Phandi26}
{Pandhi}, A., {Nimmo}, K., {Andrew}, S., {et~al.} 2026, \apjl, 1000, L53

\bibitem[{{Pastor-Marazuela} {et~al.}(2026){Pastor-Marazuela}, {Gordon}, {Stappers}, {Khrykin}, {Tejos}, {Rajwade}, {Caleb}, {Surnis}, {Driessen}, {Simha}, {Tian}, {Prochaska}, {Barr}, {Buchner}, {Fong}, {Jankowski}, {Kahinga}, {Kilpatrick}, {Kramer}, {Mas-Ribas}, \& {Hennawi}}]{PastorMarazuela25b}
{Pastor-Marazuela}, I., {Gordon}, A.~C., {Stappers}, B., {et~al.} 2026, \mnras, 545, staf2144

\bibitem[{{Pastor-Marazuela} {et~al.}(2025){Pastor-Marazuela}, {van Leeuwen}, {Bilous}, {Connor}, {Maan}, {Oostrum}, {Petroff}, {Vohl}, {Hess}, {Orr{\`u}}, {Sclocco}, \& {Wang}}]{PastorMarazuela25}
{Pastor-Marazuela}, I., {van Leeuwen}, J., {Bilous}, A., {et~al.} 2025, \aap, 693, A279

\bibitem[{{Pelliciari} {et~al.}(2026){Pelliciari}, {Bernardi}, {Margalit}, {Metzger}, {Bruno}, \& {Pilia}}]{Pelliciari26_PaperII}
{Pelliciari}, D., {Bernardi}, G., {Margalit}, B., {et~al.} 2026, \aap, submitted

\bibitem[{{Pelliciari} {et~al.}(2024){Pelliciari}, {Bernardi}, {Pilia}, {Naldi}, {Maccaferri}, {Verrecchia}, {Casentini}, {Perri}, {Kirsten}, {Bianchi}, {Bortolotti}, {Bruno}, {Dallacasa}, {Esposito}, {Geminardi}, {Giarratana}, {Giroletti}, {Lulli}, {Maccaferri}, {Magro}, {Mattana}, {Perini}, {Pupillo}, {Roma}, {Schiaffino}, {Setti}, {Tavani}, {Trudu}, \& {Zanichelli}}]{Pelliciari24}
{Pelliciari}, D., {Bernardi}, G., {Pilia}, M., {et~al.} 2024, \aap, 690, A219

\bibitem[{{Perley} \& {Butler}(2017)}]{PerleyButler17}
{Perley}, R.~A. \& {Butler}, B.~J. 2017, \apjs, 230, 7

\bibitem[{{Petroff} {et~al.}(2017){Petroff}, {Burke-Spolaor}, {Keane}, {McLaughlin}, {Miller}, {Andreoni}, {Bailes}, {Barr}, {Bernard}, {Bhandari}, {Bhat}, {Burgay}, {Caleb}, {Champion}, {Chandra}, {Cooke}, {Dhillon}, {Farnes}, {Hardy}, {Jaroenjittichai}, {Johnston}, {Kasliwal}, {Kramer}, {Littlefair}, {Macquart}, {Mickaliger}, {Possenti}, {Pritchard}, {Ravi}, {Rest}, {Rowlinson}, {Sawangwit}, {Stappers}, {Sullivan}, {Tiburzi}, {van Straten}, {ANTARES Collaboration}, {Albert}, {Andr{\'e}}, {Anghinolfi}, {Anton}, {Ardid}, {Aubert}, {Avgitas}, {Baret}, {Barrios-Mart{\'\i}}, {Basa}, {Bertin}, {Biagi}, {Bormuth}, {Bourret}, {Bouwhuis}, {Bruijn}, {Brunner}, {Busto}, {Capone}, {Caramete}, {Carr}, {Celli}, {Chiarusi}, {Circella}, {Coelho}, {Coleiro}, {Coniglione}, {Costantini}, {Coyle}, {Creusot}, {Deschamps}, {de Bonis}, {Distefano}, {di Palma}, {Donzaud}, {Dornic}, {Drouhin}, {Eberl}, {El Bojaddaini}, {Els{\"a}sser}, {Enzenh{\"o}fer}, {Felis}, {Fusco}, {Galat{\`a}}, {Gay}, {Gei{\ss}els{\"o}der}, {Geyer},
  {Giordano}, {Gleixner}, {Glotin}, {Gr{\'e}goire}, {Gracia-Ruiz}, {Graf}, {Hallmann}, {van Haren}, {Heijboer}, {Hello}, {Hern{\'a}ndez-Rey}, {H{\"o}{\ss}l}, {Hofest{\"a}dt}, {Hugon}, {Illuminati}, {James}, {de Jong}, {Jongen}, {Kadler}, {Kalekin}, {Katz}, {Kie{\ss}ling}, {Kouchner}, {Kreter}, {Kreykenbohm}, {Kulikovskiy}, {Lachaud}, {Lahmann}, {Lef{\`e}vre}, {Leonora}, {Lotze}, {Loucatos}, {Marcelin}, {Margiotta}, {Marinelli}, {Mart{\'\i}nez-Mora}, {Mathieu}, {Mele}, {Melis}, {Michael}, {Migliozzi}, {Moussa}, {Mueller}, {Nezri}, {P{\v{a}}v{\v{a}}la{\c{s}}}, {Pellegrino}, {Perrina}, {Piattelli}, {Popa}, {Pradier}, {Quinn}, {Racca}, {Riccobene}, {Roensch}, {S{\'a}nchez-Losa}, {Salda{\~n}a}, {Salvadori}, {Samtleben}, {Sanguineti}, {Sapienza}, {Schnabel}, {Seitz}, {Sieger}, {Spurio}, {Stolarczyk}, {Taiuti}, {Tayalati}, {Trovato}, {Tselengidou}, {Turpin}, {T{\"o}nnis}, {Vallage}, {Vall{\'e}e}, {van Elewyck}, {Vivolo}, {Vizzoca}, {Wagner}, {Wilms}, {Zornoza}, {Z{\'u}{\~n}iga}, {H.~E.~S.~S. Collaboration},
  {Abdalla}, {Abramowski}, {Aharonian}, {Ait Benkhali}, {Akhperjanian}, {Andersson}, {Ang{\"u}ner}, {Arrieta}, {Aubert}, {Backes}, {Balzer}, {Barnard}, {Becherini}, {Tjus}, {Berge}, {Bernhard}, {Bernl{\"o}hr}, {Blackwell}, {B{\"o}ttcher}, {Boisson}, {Bolmont}, {Bordas}, {Bregeon}, {Brun}, {Brun}, {Bryan}, {Bulik}, {Capasso}, {Casanova}, {Cerruti}, {Chakraborty}, {Chalme-Calvet}, {Chaves}, {Chen}, \& {Chevalier}}]{Petroff17}
{Petroff}, E., {Burke-Spolaor}, S., {Keane}, E.~F., {et~al.} 2017, \mnras, 469, 4465

\bibitem[{{Petroff} {et~al.}(2022){Petroff}, {Hessels}, \& {Lorimer}}]{petroff21}
{Petroff}, E., {Hessels}, J.~W.~T., \& {Lorimer}, D.~R. 2022, \aapr, 30, 2

\bibitem[{{Pilia}(2021)}]{Pilia22}
{Pilia}, M. 2021, Universe, 8, 9

\bibitem[{{Planck Collaboration} {et~al.}(2020){Planck Collaboration}, {Aghanim}, {Akrami}, {Ashdown}, {Aumont}, {Baccigalupi}, {Ballardini}, {Banday}, {Barreiro}, {Bartolo}, {Basak}, {Battye}, {Benabed}, {Bernard}, {Bersanelli}, {Bielewicz}, {Bock}, {Bond}, {Borrill}, {Bouchet}, {Boulanger}, {Bucher}, {Burigana}, {Butler}, {Calabrese}, {Cardoso}, {Carron}, {Challinor}, {Chiang}, {Chluba}, {Colombo}, {Combet}, {Contreras}, {Crill}, {Cuttaia}, {de Bernardis}, {de Zotti}, {Delabrouille}, {Delouis}, {Di Valentino}, {Diego}, {Dor{\'e}}, {Douspis}, {Ducout}, {Dupac}, {Dusini}, {Efstathiou}, {Elsner}, {En{\ss}lin}, {Eriksen}, {Fantaye}, {Farhang}, {Fergusson}, {Fernandez-Cobos}, {Finelli}, {Forastieri}, {Frailis}, {Fraisse}, {Franceschi}, {Frolov}, {Galeotta}, {Galli}, {Ganga}, {G{\'e}nova-Santos}, {Gerbino}, {Ghosh}, {Gonz{\'a}lez-Nuevo}, {G{\'o}rski}, {Gratton}, {Gruppuso}, {Gudmundsson}, {Hamann}, {Handley}, {Hansen}, {Herranz}, {Hildebrandt}, {Hivon}, {Huang}, {Jaffe}, {Jones}, {Karakci}, {Keih{\"a}nen},
  {Keskitalo}, {Kiiveri}, {Kim}, {Kisner}, {Knox}, {Krachmalnicoff}, {Kunz}, {Kurki-Suonio}, {Lagache}, {Lamarre}, {Lasenby}, {Lattanzi}, {Lawrence}, {Le Jeune}, {Lemos}, {Lesgourgues}, {Levrier}, {Lewis}, {Liguori}, {Lilje}, {Lilley}, {Lindholm}, {L{\'o}pez-Caniego}, {Lubin}, {Ma}, {Mac{\'\i}as-P{\'e}rez}, {Maggio}, {Maino}, {Mandolesi}, {Mangilli}, {Marcos-Caballero}, {Maris}, {Martin}, {Martinelli}, {Mart{\'\i}nez-Gonz{\'a}lez}, {Matarrese}, {Mauri}, {McEwen}, {Meinhold}, {Melchiorri}, {Mennella}, {Migliaccio}, {Millea}, {Mitra}, {Miville-Desch{\^e}nes}, {Molinari}, {Montier}, {Morgante}, {Moss}, {Natoli}, {N{\o}rgaard-Nielsen}, {Pagano}, {Paoletti}, {Partridge}, {Patanchon}, {Peiris}, {Perrotta}, {Pettorino}, {Piacentini}, {Polastri}, {Polenta}, {Puget}, {Rachen}, {Reinecke}, {Remazeilles}, {Renzi}, {Rocha}, {Rosset}, {Roudier}, {Rubi{\~n}o-Mart{\'\i}n}, {Ruiz-Granados}, {Salvati}, {Sandri}, {Savelainen}, {Scott}, {Shellard}, {Sirignano}, {Sirri}, {Spencer}, {Sunyaev}, {Suur-Uski}, {Tauber}, {Tavagnacco},
  {Tenti}, {Toffolatti}, {Tomasi}, {Trombetti}, {Valenziano}, {Valiviita}, {Van Tent}, {Vibert}, {Vielva}, {Villa}, {Vittorio}, {Wandelt}, {Wehus}, {White}, {White}, {Zacchei}, \& {Zonca}}]{Planck20}
{Planck Collaboration}, {Aghanim}, N., {Akrami}, Y., {et~al.} 2020, \aap, 641, A6

\bibitem[{{Pleunis} {et~al.}(2021){Pleunis}, {Good}, {Kaspi}, {Mckinven}, {Ransom}, {Scholz}, {Bandura}, {Bhardwaj}, {Boyle}, {Brar}, {Cassanelli}, {Chawla}, {(Adam) Dong}, {Fonseca}, {Gaensler}, {Josephy}, {Kaczmarek}, {Leung}, {Lin}, {Masui}, {Mena-Parra}, {Michilli}, {Ng}, {Patel}, {Rafiei-Ravandi}, {Rahman}, {Sanghavi}, {Shin}, {Smith}, {Stairs}, \& {Tendulkar}}]{pleunis21}
{Pleunis}, Z., {Good}, D.~C., {Kaspi}, V.~M., {et~al.} 2021, \apj, 923, 1

\bibitem[{{Popov} \& {Postnov}(2013)}]{Popov13}
{Popov}, S.~B. \& {Postnov}, K.~A. 2013, arXiv e-prints, arXiv:1307.4924

\bibitem[{{Prandoni} {et~al.}(2018){Prandoni}, {Guglielmino}, {Morganti}, {Vaccari}, {Maini}, {R{\"o}ttgering}, {Jarvis}, \& {Garrett}}]{Prandoni18}
{Prandoni}, I., {Guglielmino}, G., {Morganti}, R., {et~al.} 2018, \mnras, 481, 4548

\bibitem[{{Price} {et~al.}(2019){Price}, {Foster}, {Geyer}, {van Straten}, {Gajjar}, {Hellbourg}, {Karastergiou}, {Keane}, {Siemion}, {Arcavi}, {Bhat}, {Caleb}, {Chang}, {Croft}, {DeBoer}, {de Pater}, {Drew}, {Enriquez}, {Farah}, {Gizani}, {Green}, {Isaacson}, {Hickish}, {Jameson}, {Lebofsky}, {MacMahon}, {M{\"o}ller}, {Onken}, {Petroff}, {Werthimer}, {Wolf}, {Worden}, \& {Zhang}}]{Price19}
{Price}, D.~C., {Foster}, G., {Geyer}, M., {et~al.} 2019, \mnras, 486, 3636

\bibitem[{{Prochaska} {et~al.}(2019){Prochaska}, {Macquart}, {McQuinn}, {Simha}, {Shannon}, {Day}, {Marnoch}, {Ryder}, {Deller}, {Bannister}, {Bhandari}, {Bordoloi}, {Bunton}, {Cho}, {Flynn}, {Mahony}, {Phillips}, {Qiu}, \& {Tejos}}]{Prochaska19}
{Prochaska}, J.~X., {Macquart}, J.-P., {McQuinn}, M., {et~al.} 2019, Science, 366, 231

\bibitem[{{Rajwade} {et~al.}(2024){Rajwade}, {Driessen}, {Barr}, {Pastor-Marazuela}, {Berezina}, {Jankowski}, {Muller}, {Kahinga}, {Stappers}, {Bezuidenhout}, {Caleb}, {Deller}, {Fong}, {Gordon}, {Kramer}, {Malenta}, {Morello}, {Prochaska}, {Sanidas}, {Surnis}, {Tejos}, \& {Wagner}}]{Rajwade24}
{Rajwade}, K.~M., {Driessen}, L.~N., {Barr}, E.~D., {et~al.} 2024, \mnras, 532, 3881

\bibitem[{{Ravi}(2019)}]{Ravi19}
{Ravi}, V. 2019, Nature Astronomy, 3, 928

\bibitem[{{Ravi}(2023)}]{Ravi23_ATEL}
{Ravi}, V. 2023, The Astronomer's Telegram, 16191, 1

\bibitem[{{Ravi} {et~al.}(2023){Ravi}, {Catha}, {Chen}, {Connor}, {Faber}, {Lamb}, {Hallinan}, {Harnach}, {Hellbourg}, {Hobbs}, {Hodge}, {Hodges}, {Law}, {Rasmussen}, {Sharma}, {Sherman}, {Shi}, {Simard}, {Squillace}, {Weinreb}, {Woody}, {Yadlapalli}, {Ahumada}, {Dong}, {Fremling}, {Huang}, {Karambelkar}, \& {Miller}}]{ravi23}
{Ravi}, V., {Catha}, M., {Chen}, G., {et~al.} 2023, \apjl, 949, L3

\bibitem[{{Ravi} {et~al.}(2019){Ravi}, {Catha}, {D'Addario}, {Djorgovski}, {Hallinan}, {Hobbs}, {Kocz}, {Kulkarni}, {Shi}, {Vedantham}, {Weinreb}, \& {Woody}}]{Ravi19_loc}
{Ravi}, V., {Catha}, M., {D'Addario}, L., {et~al.} 2019, \nat, 572, 352

\bibitem[{{Ravi} {et~al.}(2016){Ravi}, {Shannon}, {Bailes}, {Bannister}, {Bhandari}, {Bhat}, {Burke-Spolaor}, {Caleb}, {Flynn}, {Jameson}, {Johnston}, {Keane}, {Kerr}, {Tiburzi}, {Tuntsov}, \& {Vedantham}}]{Ravi16}
{Ravi}, V., {Shannon}, R.~M., {Bailes}, M., {et~al.} 2016, Science, 354, 1249

\bibitem[{{Resmi} {et~al.}(2021){Resmi}, {Vink}, \& {Ishwara-Chandra}}]{Resmi21}
{Resmi}, L., {Vink}, J., \& {Ishwara-Chandra}, C.~H. 2021, \aap, 655, A102

\bibitem[{{Sand} {et~al.}(2025){Sand}, {Curtin}, {Michilli}, {Kaspi}, {Fonseca}, {Nimmo}, {Pleunis}, {Shin}, {Bhardwaj}, {Brar}, {Dobbs}, {Eadie}, {Gaensler}, {Joseph}, {Leung}, {Main}, {Masui}, {Mckinven}, {Pandhi}, {Pearlman}, {Rafiei-Ravandi}, {Sammons}, {Smith}, \& {Stairs}}]{Sand25}
{Sand}, K.~R., {Curtin}, A.~P., {Michilli}, D., {et~al.} 2025, \apj, 979, 160

\bibitem[{{Seymour} {et~al.}(2008){Seymour}, {Dwelly}, {Moss}, {McHardy}, {Zoghbi}, {Rieke}, {Page}, {Hopkins}, \& {Loaring}}]{Seymour08}
{Seymour}, N., {Dwelly}, T., {Moss}, D., {et~al.} 2008, \mnras, 386, 1695

\bibitem[{{Shaji} {et~al.}(2026){Shaji}, {Tian}, {Caleb}, {Rajwade}, {Stappers}, {Pastor-Marazuela}, {Murphy}, {Barr}, {Gulati}, {Jankowski}, {Kramer}, {Lee}, \& {Uttarkar}}]{Shaji26}
{Shaji}, K., {Tian}, J., {Caleb}, M., {et~al.} 2026, \mnras, 545, staf2222

\bibitem[{{Shannon} {et~al.}(2025){Shannon}, {Bannister}, {Bera}, {Bhandari}, {Day}, {Deller}, {Dial}, {Dobie}, {Ekers}, {Fong}, {Glowacki}, {Gordon}, {Gourdji}, {Jaini}, {James}, {Kumar}, {Mahony}, {Marnoch}, {Muller}, {Prochaska}, {Qiu}, {Ryder}, {Sadler}, {Scott}, {Tejos}, {Uttarkar}, \& {Wang}}]{Shannon25}
{Shannon}, R.~M., {Bannister}, K.~W., {Bera}, A., {et~al.} 2025, \pasa, 42, e036

\bibitem[{{Sherman} {et~al.}(2024){Sherman}, {Connor}, {Ravi}, {Law}, {Chen}, {Catha}, {Faber}, {Hallinan}, {Harnach}, {Hellbourg}, {Hobbs}, {Hodge}, {Hodges}, {Lamb}, {Rasmussen}, {Sharma}, {Shi}, {Simard}, {Somalwar}, {Squillace}, {Weinreb}, {Woody}, {Yadlapalli}, \& {The Deep Synoptic Array team}}]{Sherman24}
{Sherman}, M.~B., {Connor}, L., {Ravi}, V., {et~al.} 2024, \apj, 964, 131

\bibitem[{{Shin} \& {CHIME/FRB Collaboration}(2024)}]{Shin24_14A}
{Shin}, K. \& {CHIME/FRB Collaboration}. 2024, The Astronomer's Telegram, 16420, 1

\bibitem[{{Snelders} {et~al.}(2024){Snelders}, {Bhandari}, {Kirsten}, {Hessels}, {Marcote}, {Hewitt}, {Gawronski}, {Puchalska}, {Ould-Boukattine}, {Gopinath}, {Nimmo}, {Karuppusamy}, {Herrmann}, {Yang}, {Blaauw}, {Buttaccio}, {Maccaferri}, {Bach}, {Feiler}, {Bray}, {Williams}, {Wrigley}, {Keimpema}, {Paragi}, {Burgay}, {Corongiu}, {Giroletti}, {Kramer}, {Pilia}, {Spitler}, {Surcis}, {Trudu}, {Yuan}, {Wang}, \& {Bezrukovs}}]{Snelders24}
{Snelders}, M.~P., {Bhandari}, S., {Kirsten}, F., {et~al.} 2024, The Astronomer's Telegram, 16542, 1

\bibitem[{{Snelders} {et~al.}(2025){Snelders}, {Hessels}, {Huang}, {Sridhar}, {Marcote}, {Moroianu}, {Ould-Boukattine}, {Kirsten}, {Bhandari}, {Hewitt}, {Pelliciari}, {Rhodes}, {Anna-Thomas}, {Bach}, {Bempong-Manful}, {Bezrukovs}, {Bray}, {Buttaccio}, {Cognard}, {Corongiu}, {Feiler}, {Gawro{\'n}ski}, {Giroletti}, {Guillemot}, {Karuppusamy}, {Lindqvist}, {Nimmo}, {Possenti}, {Puchalska}, \& {Williams-Baldwin}}]{Snelders26}
{Snelders}, M.~P., {Hessels}, J.~W.~T., {Huang}, J., {et~al.} 2025, arXiv e-prints, arXiv:2510.11352

\bibitem[{{Sobacchi} {et~al.}(2022){Sobacchi}, {Lyubarsky}, {Beloborodov}, \& {Sironi}}]{Sobacchi22}
{Sobacchi}, E., {Lyubarsky}, Y., {Beloborodov}, A.~M., \& {Sironi}, L. 2022, \mnras, 511, 4766

\bibitem[{{Sridhar} {et~al.}(2021){Sridhar}, {Metzger}, {Beniamini}, {Margalit}, {Renzo}, {Sironi}, \& {Kovlakas}}]{Sridhar21}
{Sridhar}, N., {Metzger}, B.~D., {Beniamini}, P., {et~al.} 2021, \apj, 917, 13

\bibitem[{{Sridhar} {et~al.}(2024){Sridhar}, {Metzger}, \& {Fang}}]{Sridhar24}
{Sridhar}, N., {Metzger}, B.~D., \& {Fang}, K. 2024, \apj, 960, 74

\bibitem[{{Tendulkar} {et~al.}(2017){Tendulkar}, {Bassa}, {Cordes}, {Bower}, {Law}, {Chatterjee}, {Adams}, {Bogdanov}, {Burke-Spolaor}, {Butler}, {Demorest}, {Hessels}, {Kaspi}, {Lazio}, {Maddox}, {Marcote}, {McLaughlin}, {Paragi}, {Ransom}, {Scholz}, {Seymour}, {Spitler}, {van Langevelde}, \& {Wharton}}]{Tendulkar17}
{Tendulkar}, S.~P., {Bassa}, C.~G., {Cordes}, J.~M., {et~al.} 2017, \apjl, 834, L7

\bibitem[{{Tendulkar} {et~al.}(2021){Tendulkar}, {Gil de Paz}, {Kirichenko}, {Hessels}, {Bhardwaj}, {{\'A}vila}, {Bassa}, {Chawla}, {Fonseca}, {Kaspi}, {Keimpema}, {Kirsten}, {Lazio}, {Marcote}, {Masui}, {Nimmo}, {Paragi}, {Rahman}, {Pay{\'a}}, {Scholz}, \& {Stairs}}]{Tendulkar21}
{Tendulkar}, S.~P., {Gil de Paz}, A., {Kirichenko}, A.~Y., {et~al.} 2021, \apjl, 908, L12

\bibitem[{{Thompson} \& {Duncan}(1995)}]{ThompsonDuncan}
{Thompson}, C. \& {Duncan}, R.~C. 1995, \mnras, 275, 255

\bibitem[{{Tian} {et~al.}(2025){Tian}, {Pastor-Marazuela}, {Rajwade}, {Stappers}, {Shaji}, {Hanmer}, {Caleb}, {Bezuidenhout}, {Jankowski}, {Breton}, {Barr}, {Kramer}, {Groot}, {Bloemen}, {Vreeswijk}, {Pieterse}, {Woudt}, {Fender}, {Wijnands}, \& {Buckley}}]{Tian25}
{Tian}, J., {Pastor-Marazuela}, I., {Rajwade}, K.~M., {et~al.} 2025, \mnras, 540, 1685

\bibitem[{{Tian} {et~al.}(2024){Tian}, {Rajwade}, {Pastor-Marazuela}, {Stappers}, {Bezuidenhout}, {Caleb}, {Jankowski}, {Barr}, \& {Kramer}}]{Tian24}
{Tian}, J., {Rajwade}, K.~M., {Pastor-Marazuela}, I., {et~al.} 2024, \mnras, 533, 3174

\bibitem[{{Uttarkar} {et~al.}(2024){Uttarkar}, {Kumar}, {Lower}, \& {Shannon}}]{Uttarkar24}
{Uttarkar}, P.~A., {Kumar}, P., {Lower}, M.~E., \& {Shannon}, R.~M. 2024, The Astronomer's Telegram, 16430, 1

\bibitem[{{van der Vlugt} {et~al.}(2021){van der Vlugt}, {Algera}, {Hodge}, {Novak}, {Radcliffe}, {Riechers}, {R{\"o}ttgering}, {Smol{\v{c}}i{\'c}}, \& {Walter}}]{VanDerVlugt21}
{van der Vlugt}, D., {Algera}, H.~S.~B., {Hodge}, J.~A., {et~al.} 2021, \apj, 907, 5

\bibitem[{{Van Der Walt} {et~al.}(2011){Van Der Walt}, {Colbert}, \& {Varoquaux}}]{vanderWalt11}
{Van Der Walt}, S., {Colbert}, S.~C., \& {Varoquaux}, G. 2011, Computing in Science and Engineering, 13, 22

\bibitem[{{Virtanen} {et~al.}(2020){Virtanen}, {Gommers}, {Oliphant}, {Haberland}, {Reddy}, {Cournapeau}, {Burovski}, {Peterson}, {Weckesser}, {Bright}, {van der Walt}, {Brett}, {Wilson}, {Millman}, {Mayorov}, {Nelson}, {Jones}, {Kern}, {Larson}, {Carey}, {Polat}, {Feng}, {Moore}, {VanderPlas}, {Laxalde}, {Perktold}, {Cimrman}, {Henriksen}, {Quintero}, {Harris}, {Archibald}, {Ribeiro}, {Pedregosa}, {van Mulbregt}, \& {SciPy 1. 0 Contributors}}]{Virtanen20}
{Virtanen}, P., {Gommers}, R., {Oliphant}, T.~E., {et~al.} 2020, Nature Methods, 17, 261

\bibitem[{{Vohl} {et~al.}(2023){Vohl}, {Vedantham}, {Hessels}, \& {Bassa}}]{Vohl23}
{Vohl}, D., {Vedantham}, H. .~K., {Hessels}, J.~W.~T., \& {Bassa}, C.~G. 2023, arXiv e-prints, arXiv:2303.11967

\bibitem[{{Woodland} {et~al.}(2024){Woodland}, {Mannings}, {Prochaska}, {Ryder}, {Marnoch}, {Jorgenson}, {Simha}, {Tejos}, {Gordon}, {Fong}, {Kilpatrick}, {Deller}, \& {Glowacki}}]{Woodland24}
{Woodland}, M.~N., {Mannings}, A.~G., {Prochaska}, J.~X., {et~al.} 2024, \apj, 973, 64

\bibitem[{{Worden} {et~al.}(2017){Worden}, {Drew}, {Siemion}, {Werthimer}, {DeBoer}, {Croft}, {MacMahon}, {Lebofsky}, {Isaacson}, {Hickish}, {Price}, {Gajjar}, \& {Wright}}]{Worden17}
{Worden}, S.~P., {Drew}, J., {Siemion}, A., {et~al.} 2017, Acta Astronautica, 139, 98

\bibitem[{{Xu} {et~al.}(2022){Xu}, {Niu}, {Chen}, {Lee}, {Zhu}, {Dong}, {Zhang}, {Jiang}, {Wang}, {Xu}, {Zhang}, {Fu}, {Filippenko}, {Peng}, {Zhou}, {Zhang}, {Wang}, {Feng}, {Li}, {Brink}, {Li}, {Lu}, {Yang}, {Caballero}, {Cai}, {Chen}, {Dai}, {Djorgovski}, {Esamdin}, {Gan}, {Guhathakurta}, {Han}, {Hao}, {Huang}, {Jiang}, {Li}, {Li}, {Li}, {Li}, {Li}, {Liu}, {Luo}, {Men}, {Niu}, {Peng}, {Qian}, {Song}, {Stern}, {Stockton}, {Sun}, {Wang}, {Wang}, {Wang}, {Wang}, {Wu}, {Xiao}, {Xiong}, {Xu}, {Xu}, {Yang}, {Yang}, {Yao}, {Yi}, {Yue}, {Yu}, {Yu}, {Yuan}, {Zhang}, {Zhang}, {Zhang}, {Zhao}, {Zheng}, {Zhu}, \& {Zou}}]{Xu2022}
{Xu}, H., {Niu}, J.~R., {Chen}, P., {et~al.} 2022, \nat, 609, 685

\bibitem[{{Yamasaki} \& {Totani}(2020)}]{Yamasaki}
{Yamasaki}, S. \& {Totani}, T. 2020, \apj, 888, 105

\bibitem[{{Yang}(2026)}]{Yang26}
{Yang}, Y.-P. 2026, arXiv e-prints, arXiv:2603.17615

\bibitem[{{Yang} {et~al.}(2020){Yang}, {Li}, \& {Zhang}}]{Yang20}
{Yang}, Y.-P., {Li}, Q.-C., \& {Zhang}, B. 2020, \apj, 895, 7

\bibitem[{{Yang} {et~al.}(2022){Yang}, {Lu}, {Feng}, {Zhang}, \& {Li}}]{Yang22}
{Yang}, Y.-P., {Lu}, W., {Feng}, Y., {Zhang}, B., \& {Li}, D. 2022, \apjl, 928, L16

\bibitem[{{Yao} {et~al.}(2017){Yao}, {Manchester}, \& {Wang}}]{YMW16}
{Yao}, J.~M., {Manchester}, R.~N., \& {Wang}, N. 2017, \apj, 835, 29

\bibitem[{{Zhang}(2023)}]{Zhang22_rev}
{Zhang}, B. 2023, Reviews of Modern Physics, 95, 035005

\bibitem[{{Zhang} {et~al.}(2020){Zhang}, {Yu}, {He}, \& {Wang}}]{Zhang20b}
{Zhang}, G.~Q., {Yu}, H., {He}, J.~H., \& {Wang}, F.~Y. 2020, \apj, 900, 170

\bibitem[{{Zhang} {et~al.}(2024){Zhang}, {Wu}, {Cao}, {Zhu}, {Zhang}, {Niu}, {Xie}, {Zhou}, {Wang}, {Zhu}, {Zhang}, {Wang}, {Niu}, {Di Li}, {Han}, {Lee}, {Wang}, {Gao}, {Feng}, {Jiang}, {Jing}, {Li}, {Lu}, {Luo}, {Lyu}, {Wang}, {Xu}, {Yang}, {Yu}, {Zhang}, \& {Project}}]{Zhang24_hyper}
{Zhang}, J., {Wu}, Q., {Cao}, S., {et~al.} 2024, The Astronomer's Telegram, 16505, 1

\bibitem[{{Zhang} \& {Yu}(2024{\natexlab{a}})}]{ZhangYu24}
{Zhang}, X. \& {Yu}, W. 2024{\natexlab{a}}, The Astronomer's Telegram, 16695, 1

\bibitem[{{Zhang} \& {Yu}(2024{\natexlab{b}})}]{ZhangWu24_ATEL}
{Zhang}, X. \& {Yu}, W. 2024{\natexlab{b}}, The Astronomer's Telegram, 16695, 1

\bibitem[{{Zhang} {et~al.}(2025){Zhang}, {Yu}, {Yan}, {Xing}, \& {Zhang}}]{Zhang25_FRS}
{Zhang}, X., {Yu}, W., {Yan}, Z., {Xing}, Y., \& {Zhang}, B. 2025, arXiv e-prints, arXiv:2501.14247

\bibitem[{{Zhang} {et~al.}(2023){Zhang}, {Li}, {Zhang}, {Cao}, {Feng}, {Wang}, {Qu}, {Niu}, {Zhu}, {Han}, {Jiang}, {Lee}, {Li}, {Luo}, {Niu}, {Tsai}, {Wang}, {Wang}, {Wu}, {Xu}, {Yang}, {Zhang}, {Zhou}, \& {Zhu}}]{Zhang23}
{Zhang}, Y.-K., {Li}, D., {Zhang}, B., {et~al.} 2023, \apj, 955, 142

\bibitem[{{Zhou} {et~al.}(2025){Zhou}, {Han}, {Zhang}, {Zhu}, {Wang}, {Yang}, {Qu}, {Zhang}, {Yan}, {Jing}, {Cao}, {Xie}, {Yang}, {Tian}, {Li}, {Li}, {Niu}, {Wu}, {Wu}, {Feng}, {Wang}, \& {Wang}}]{Zhou25}
{Zhou}, D., {Han}, J.~L., {Zhang}, B., {et~al.} 2025, \apj, 988, 41

\bibitem[{{Zhu} {et~al.}(2023){Zhu}, {Xu}, {Zhou}, {Lin}, {Wang}, {Wang}, {Zhang}, {Niu}, {Chen}, {Li}, {Meng}, {Lee}, {Zhang}, {Feng}, {Ge}, {G{\"o}{\u{g}}{\"u}{\c{s}}}, {Guan}, {Han}, {Jiang}, {Jiang}, {Kouveliotou}, {Li}, {Miao}, {Miao}, {Men}, {Niu}, {Wang}, {Wang}, {Xu}, {Xu}, {Xue}, {Yang}, {Yu}, {Yuan}, {Yue}, {Zhang}, \& {Zhang}}]{Zhu23}
{Zhu}, W., {Xu}, H., {Zhou}, D., {et~al.} 2023, Science Advances, 9, eadf6198

\end{thebibliography}

\begin{appendix}
\nolinenumbers
\section{Detailed description of observations and source catalog}\label{app: description}

In the following, we report all the specifics of the uGMRT observations conducted. We list in Table \ref{tab:uGMRT_obs} all the 24 FRB sources, including their coordinates, primary and secondary calibrators employed, date of observations and on source time, along with the r.m.s. noise reached after the imaging procedure. The latter is provided for three {\tt briggs} parameter values used when choosing a {\tt robust} weighting scheme for the source imaging. Moreover, the localization properties (e.g. localization area and host galaxy properties) of the uGMRT sources are shown in Table \ref{tab:loc_prop_sources}. For new PRS candidates, we estimate the probability that the detected persistent source is an unrelated to the FRB as:

\begin{equation}
P_{\rm cc} = 1 - e^{-\mu}\ ,
\end{equation}

i.e. we quantify the likelihood of a chance coincidence as the Poisson probability that a background radio source aligns with the FRB localization region. Here $\mu = \Sigma(>F) \times \Omega$ is the expected number of extragalactic radio sources brighter than a given flux density $F$ within a solid angle $\Omega$, and $\Sigma(>F)$ is the sky surface density of radio sources brighter than $F$, and $\Omega = \pi ab$, where are the minor and major semi-axis of the FRB localization area, the latter listed for each FRB source in Table \ref{tab: PRS_prop}.

We consider the cumulative distribution of extragalactic source number counts at $1.4$ GHz reported in \cite{Mancuso17, Galluzzi2025}, which is obtained considering various surveys \citep[e.g.][]{Seymour08, Butler18, Prandoni18, Heywood20, Matthews21, VanDerVlugt21, DAmato22, Massardi25}. The chance alignement probability $P_{\rm cc}$ is reported for each PRS candidate in Table \ref{tab:loc_prop_sources}.

We further provide a detailed description of the FRBs chosen for our sample in the following sections. We refer to Sect. \ref{sec: Discuss} for details of S5, S11 and S12 (i.e. FRBs 20190417A, 20201114A and 20201130A). 

\subsection{FRB 20180301A (S1)}
This is a repeating source discovered during Breakthrough Listen \citep{Isaacson17,Worden17} observations of the Galactic Plane \citep{Price19}, and then localized at sub-arcsecond precision to a star-forming galaxy at $z = 0.3304$ via VLA observations \citep{Bhandari22}. The RM of this source is observed to vary in time, from first observations in 2018 reporting $-3163 \pm 20$ rad m$^{-2}$ \citep{Price19}, passing through $\approx 530$ rad m$^{-2}$ one year later \citep{Luo22} to $-68 \pm 13$ rad m$^{-2}$ \citep{Kumar23} during 2020-2022 observations with Parkes. As stated in \cite{Luo22}, this high absolute RM value could be due to a narrow signal bandwidth of $40$ MHz. The source also presents a possible positive derivative of RM with time \citep{Luo22, Kumar21}. The high observed RM, combined with its variability with time makes S1 an interesting FRB source to be searched for a persistent counterpart. VLA observations at $1.5$ GHz ruled out the presence of a PRS for this source down to a $3\sigma$ flux density limit of $63\ \mu$Jy beam$^{-1}$ \citep{Bhandari22}. With our observations we are able to rule out the presence of a PRS by reaching a $\sim 2$ deeper flux density limit at similar frequencies, i.e. a spectral luminosity $2\sigma$ limit of $L_{1.2} \leq 7.5 \times 10^{28}$ erg s$^{-1}$ Hz$^{-1}$. \\

\subsection{FRB 20180814A (S2) and FRB 20190303A (S3)}
These repeating FRB sources have been discovered by CHIME/FRB and localized at sub-arcminute precision \citep{Michilli23}. The most probable host for S2 is the spiral galaxy J042256.01+733940.7 found in the Panoramic Survey Telescope \& Rapid Response System (PanSTARRS) data archive 1 \citep[PS1;][]{Flewelling20}, at redshift $z = 0.006835$. The bursts of this source show a relatively high RM value of $700 \pm 1$ rad m$^{-2}$ \citep{McKinven23} . Within the localization uncertainties we find an unresolved source, consistent with the position of 20180814A-S5 reported in \cite{Ibik24}, however with a $\approx 2\times$ higher flux density. Although the observations reported in \cite{Ibik24} are conducted with the VLA at a $\approx 3$ times better angular resolution, and that the same source has been reported as extended in their work. Overall, 20180814A-S5 is localized to a galaxy different than the most probable FRB host associated with S2, and so can be excluded as a PRS \citep{Ibik24}. \cite{Ibik24} found a radio source (20180814A-S1 in their work) spatially coincident with PanSTARRS J042256.01+733940.7, not detected in our observations. Still, this radio source appears extended in VLA observations and therefore \cite{Ibik24} ruled out the possibility of it being the continuum counterpart of FRB 20180814A. The authors reported a 5$\sigma$ flux density limit of $17.5\ \mu$Jy beam$^{-1}$, which is consistent with the peak flux density of 20180814A-S1. From our non-detection we place a 2$\sigma$ UL of $24\ \mu$Jy beam$^{-1}$ on the presence of a PRS, giving $L_{1.2} \leq 2.8 \times 10^{27}$ erg s$^{-1}$ Hz$^{-1}$.

Regarding S3, this source is spatially coincident with a pair of merging galaxies \citep[see Fig. 5 of][]{CHIME25}, with the FRB source located in the most south-west galaxy of the pair, i.e. J135159.86+480714.04, at a redshift of $z = 0.0644$. According to \cite{CHIME25}, the merging galaxies are both star-forming. Consistently with \cite{Ibik24}, which observed the field of this source with the VLA at $1.5$ GHz, we also detect extended radio emission from both the galaxies. Given the extended nature of this radio source, \cite{Ibik24} provided a $5\sigma$ flux density UL of $25$ $\mu$Jy beam$^{-1}$ on the presence of a PRS co-located with S3. We provide a more reliable flux density UL by cutting from the S3 image all the baselines sensitive to angular scales $> 10''$. In this way, only few bright ($\sim 6\sigma$) pixels remain, coincident with the bright spot associated with the FRB host galaxy center. From this procedure, we measure a $2\sigma$ flux density limit of $76\ \mu$Jy beam$^{-1}$ which translates to a spectral luminosity of $L_{1.2} \leq 8 \times 10^{27}$ erg s$^{-1}$ Hz$^{-1}$. 

\subsection{FRB 20190523A (S4)}
This is an apparently one-off FRB localized at arcsecond precision and associated with the massive elliptical galaxy PSO J207.0643+72.4708 at $z = 0.66$ \citep{Ravi19}. No search for a continuum counterpart to the FRB has been performed before on this source and we do not detect any persistent emission down to a $2\sigma$ flux limit of $26\ \mu$Jy beam$^{-1}$. This permits to exclude any PRS of spectral luminosity $L_{1.2} \geq 3.6 \times 10^{29}$ erg s$^{-1}$ Hz$^{-1}$. \\

\subsection{FRB 20190714A (S6)}
This source is a seemingly one-off FRB discovered by the Australian Square Kilometre Array Pathfinder (ASKAP) radio telescope and localized at sub-arcsecond precision in a spiral galaxy at $z = 0.2365$ \citep{Heintz20}, with a tentative PRS association \citep{Chibueze21}. The putative PRS is compact at arcsecond angular scales and has a flux density of $\sim 90$ $\mu$Jy at $1.28$ GHz. Higher angular resolution observations conducted with the e-MERLIN revealed two sources compatible with the MeerKAT one, northern and southern with respect to the FRB localization, detected with lower significance ($4.3\sigma$ and $5.3\sigma$ for the northern and southern source, respectively). These sources have reported flux densities of $86$ $\mu$Jy (northern) and $123$ $\mu$Jy (southern) at $1.5$ GHz central frequency. Our observations reached a $2\sigma$ noise of $26$ $\mu$Jy beam$^{-1}$, and we do not find any continuum emission within the S6 localization region. The MeerKAT source would have been detected within our observations at a $\sim 7\sigma$ C.L., henceforth we conclude that the sources found in \cite{Chibueze21} are not ascribable to PRSs associated with S6. This likely means that the MeerKAT source detected in \cite{Chibueze21} is resolved out in our observations and that the e-MERLIN ones were just spurious noise pixels. We also note that the reported peak and integrated flux densities for the MeerKAT observations differ by a factor of $\sim 2$ \citep[see Table 1 and 2 of][]{Chibueze21}, consistently with our hypothesis of the source being extended. \\

\subsection{FRBs 20190804E (S7) and 20191106C (S8)}

These sources were selected from the ``gold sample'' of repeaters presented in \cite{chime23}, i.e. the ones having higher probability to be active sources.

S7 has large (arcmin) uncertainties in its position and its host galaxy is unknown, hence we computed its redshift from its DM, which is $363.68(1)$ pc cm$^{-3}$. We do not report any continuum source within the localization region of S7, obtaining a $2\sigma$ UL on its specific radio luminosity of $8.7 \times 10^{28}$ erg s$^{-1}$ Hz$^{-1}$.

In the case of S8, we found continuum radio emission of $593 \pm 90$ $\mu$Jy integrate and $250 \pm 30$ $\mu$ beam$^{-1}$ peak flux densities, co-located with the most probable host galaxy of the FRB source, which has a spectroscopic redshift of $0.10775(1)$ \citep{Ibik24_b}. Given the inconsistency of the two measured flux densities, we consider this source as extended in our observations, ruling out the PRS hypothesis. \cite{Ibik24_b} reported also the presence of a LoTSS radio source having $3.29 \pm 0.14$ mJy flux density at $144$ MHz. The implied $144-1260$ MHz spectral index for this source results $\alpha = -0.79 \pm 0.07$, compatible either with AGN or star formation origin \citep[e.g.,][]{Klein18}, consistent with our hypothesis. We thus provide a $2\sigma$ UL of $2.5 \times 10^{28}$, considering the r.m.s. noise obtained in the {\sc briggs} $= -0.5$ case, which was enough to resolve out all the extended emission of the radio source.

\subsection{FRB 20191228A (S9)}
This FRB is a seemingly one-off source discovered in the Commensal Real-time ASKAP Fast Transients (CRAFT) Survey \citep{Macquart10} and it is localized at sub-arcsecond precision with VLA 1.5 GHz observations \citep{Bhandari22} in a host galaxy at $z = 0.235$. We do not detect any PRS associated with this FRB down to a $2\sigma$ flux density limit of $30\ \mu$Jy beam$^{-1}$, i.e. $L_{1.2}\leq 5 \times 10^{28}$ erg s$^{-1}$ Hz$^{-1}$ at $1.26$ GHz. \cite{Bhandari22} used the Australia Telescope
Compact Array (ATCA) to search for a PRS at $6.5$ GHz, obtaining a $3\sigma$ flux UL of $22\ \mu$Jy beam$^{-1}$, translating to $L_\nu \leq 3.4 \times 10^{28}$ erg s$^{-1}$ Hz$^{-1}$ \citep{Bhandari22}. This source has also been observed with MeerKAT at 1.4 GHz \citep{Mfulwane26}. The latter observations provide a $1.6$ times deeper flux constrain of $18.9$ $\mu$Jy at $95\%$ CL, corresponding to $L_{1.2} \leq 3.1 \times 10^{28}$ erg s$^{-1}$ Hz$^{-1}$.

 \begin{figure*}
\sidecaption
  \includegraphics[width=12cm]{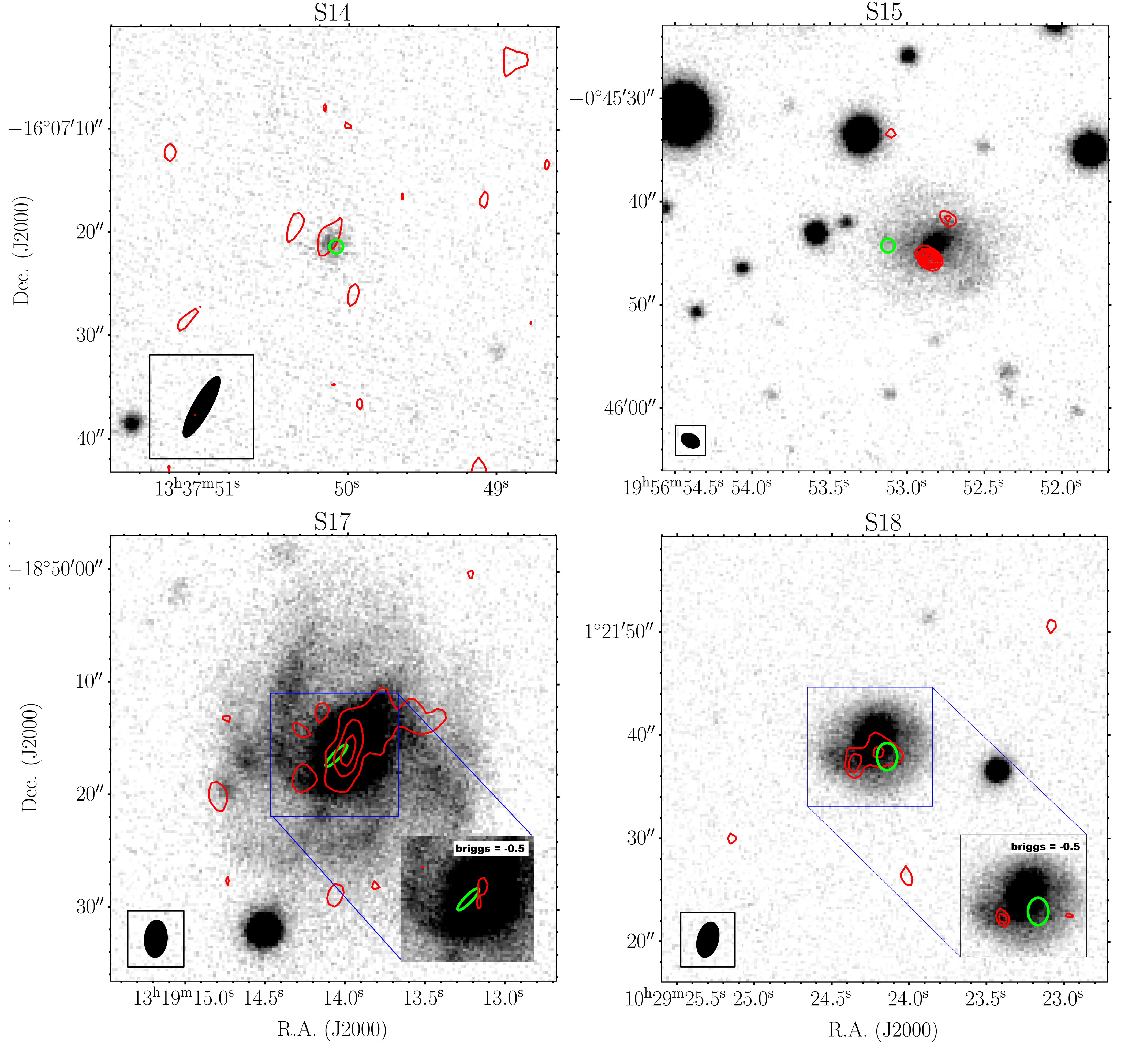}
     \caption{Optical (Pan-STARRS DR1) r-filter image of the S14 (upper left), S15 (upper right), S17 (lower left) and S18 (lower right) localization regions. Red contours represent continuum radio emission at $1.26$ GHz as resulting from our uGMRT observations, and are drawn from $3\sigma$ for all the target fields except for S14, for which instead the only drawn level is at $2\sigma$. Here $\sigma$ is the r.m.s. noise level (see Table \ref{tab:uGMRT_obs}). The green ellipses represent the localization uncertainties of each FRB source at 95$\%$ C.L. A uniform cleaning has been performed for all sources, using a briggs parameter set to $+0.5$, apart from S15 for which instead an image with briggs = $-0.5$ is shown. We provide also the central part of the cleaned image of S17 and S18 as obtained with briggs = $-0.5$ in the inset of each panel. The synthesized beams of uGMRT observations are reported in the bottom left corner of each panel as a filled black ellipse. Their angular extension are listed in Table \ref{tab:uGMRT_obs}.}
     \label{fig: M26_sources}
\end{figure*}

\subsection{FRBs 20200216A (S10) and 20210317A (S13)}
These one-off FRB sources were reported in \citep{PastorMarazuela25}, and were discovered within the ALERT sky survey. These FRBs are the sources with the largest localization uncertainties in our sample, i.e. $21.6'' \times 12.5''$ and $129'' \times 4.8''$ for S10 and S13, respectively \citep{PastorMarazuela25}. These poor localizations were not enough to provide a secure host galaxy association. However, from their observed DM, i.e. $478.7(2)$ pc cm$^{-3}$ (S10) and $ 466.5(1)$ pc cm$^{-3}$ (S13), we estimated their redshift following the method described in Section \ref{sec: sample_des}. Moreover, \cite{PastorMarazuela25} provided the list of galaxies within the FRBs localization uncertainties which also present a compatible (photometric) redshift. In particular, four galaxies were found for S10 and three for S13. Regarding S10, we searched for co-spatial PRSs at the position of these galaxies and we found none down to a limiting $2\sigma$ flux density of $26\ \mu$Jy, which converts to $L_{1.2} \leq 1.4 \times 10^{29}$ erg s$^{-1}$ Hz$^{-1}$. 

Regarding S13, we found compact radio emission in correspondence of PSO J294.1082+59.8453, i.e. its most probable host galaxy having photometric redshift of $z = 0.15 \pm 0.05$. However, the FRB--host association is tentative due to a low {\sc path} probability. The details related to this detection are presented in Section \ref{sec: Discuss}.

\subsection{FRBs 20210320C (S14), 20210807D (S15), 20210117A (S16), 20211127I (S17), 20211212A (S18) and 20220105A (S19)}

These six sources are seemingly one-off FRBs, discovered in the CRAFT survey and localized with sub-arcsecond precision \citep{Gordon23, Woodland24}. Remarkably, S14 presents compact, continuum emission compatible with the arcsecond localization of the FRB \citep{Mfulwane26}. The host galaxy is classified as a star-forming galaxy \citep{Gordon23}. The sensitivity reached in our uGMRT image ($\sigma \approx 18$ $\mu$Jy beam$^{-1}$) did not reached a sufficient sensitivity to detect the persistent source, the latter having a flux density of $32 \pm 7$ $\mu$Jy at $1.4$ GHz. The uGMRT cleaned image of the S14 field is shown in Fig. \ref{fig: M26_sources}, where only a $2\sigma$ contour is present at the nominal position of the FRB. 

Persistent radio emission detection has also been found in \cite{Mfulwane26} for S15, S17 and S18. However, for these sources, the origin of this persistent emission is plausibly attributed to central AGNs, the presence of which is already reported in \cite{Gordon23}. As can be seen in the upper right panel of Fig. \ref{fig: M26_sources}, our uGMRT image reveals that the persistent emission potentially associated with S15 is confined in a region compatible with the host galaxy center. This persistent source is compact and bright, with a flux of $\sim 0.5$ mJy, compatible with the reported $470 \pm 9$ $\mu$Jy at $1.4$ GHz \citep{Mfulwane26}.

In line with \citet{Mfulwane26}, we detected extended radio emission for S17 and S18. For the former, the radio emission is extended, and it is compatible with AGN activity \citep{Gordon23}. We also detect the persistent source, which appears extended as well, resembling the structure of a core with a jet propagating to the north-west direction. Nonetheless, the presence of a non-repeating FRB so close to the central AGN is remarkable, but the implication of it are out of the scope of this work. Regarding S18, we found extended radio emission at the FRB position uncertainties. As can be seen in Fig. \ref{fig: M26_sources}, both the sources (i.e. S17 and S18) are undetected down to $3\sigma$ flux density when we consider the corresponding images obtained with briggs = $-0.5$. We thus conservatively provide in Table \ref{table:FRB_sample}, ULs on the potential PRS fluxes based on the rms noise of images obtained with briggs = $-0.5$, which is slightly higher than the ones obtained with briggs = $+0.5$ (see Table \ref{tab:uGMRT_obs}).

We do not report any continuum source within the positional uncertainties of the remaining FRB sources (i.e. S16 and S19), and we list their implied spectral luminosity limits at $2\sigma$ level in Table \ref{table:FRB_sample}. Also these sources have been observed by MeerKAT, providing flux ULs of $F_{1.4} \leq 20.5$ $\mu$Jy (for S16) and $F_{1.4} \leq 5.2$ $\mu$Jy at $95\%$ CL. These limits are a factor $\sim 1.4$ and $8.5$ deeper than the rms noise of our uGMRT images, for S16 and S19, respectively. Therefore, in the analysis we used these deeper ULs, which translate to $L_{1.2} \leq 2.7 \times 10^{28}$ erg s$^{-1}$ Hz$^{-1}$ (S16) and $L_{1.2} \leq 1.1 \times 10^{28}$ erg s$^{-1}$ Hz$^{-1}$ (S19).

\subsection{FRB 20230607A (S20)}
This repeating source, discovered by the CHIME/FRB Virtual Observatory Event (VOEvent), turned out to be quite active from observations having deep sensitivity, with $565$ bursts detected in $15.6$ h of monitoring conducted with the FAST \citep{Zhou25}. These observations revealed a large average value of $-12249.0 \pm 1.5$ rad m$^{-2}$ for the bursts RM. The expected specific luminosity of the associated PRS is approximately $10^{29}$ erg s$^{-1}$ Hz$^{-1}$ \citep[see also the discussion in][]{Zhou25}. The FRB source is not well-localized, with $1\sigma$ uncertainties of $\sigma_{\rm R.A.} = 44.1''$ and $\sigma_{\rm Dec.} = 66.8''$ \citep{Zhou25}. Furthermore, S20 has not been associated yet with a host galaxy. 

There is a compact radio source, S20-1, of $90 \pm 17$ $\mu$Jy flux density which lies south-east with respect to the FRB localization ellipse center, within the $95\%$ C.L. region (and also within the $68\%$ C.L. ellipse). We consider it to be a potential PRS associated with S20, although this association is only tentative given the large uncertainties on the FRB position. In the case in which S20-1 is not associated with the FRB. Considering the estimated source redshift we obtain an UL on the specific luminosity at $1.26$ GHz of $5.8 \times 10^{28}$ erg s$^{-1}$ Hz$^{-1}$. Finally, there are faint bright pixels in our radio image within the $90\%$ C.L. positional uncertainties but their significance does not reach the $4\sigma$ level. A deeper radio search is encouraged, although a precise localization (e.g. via VLBI observations) is needed to firmly associate any radio source with S20.

\subsection{FRBs 20230814A (S21) and 20240114A (S22)}

S21 is a repeater discovered and localized at arcsecond level by the Deep Synoptic Array \citep[DSA-110;][]{Ravi23_ATEL}. The host galaxy of this FRB is still unknown but from the DM of the bursts, i.e. $\sim 696.4$ pc cm$^{-3}$ \citep{Ravi23_ATEL}, we get an implied redshift $z \approx 0.64$. At this redsfhit, the 2$\sigma$ uGMRT flux density threshold reached of $42\ \mu$Jy converts to $L_{1.2}\leq 4.5 \times 10^{29}$ erg s$^{-1}$ Hz$^{-1}$.

S22 is a repeater discovered by CHIME/FRB \citep{Shin24_14A} and localized to a nearby dwarf galaxy at redshift $z = 0.1306$ \citep{Bhardwaj24_ATEL} at $\sim 0.2$ arcsecond precision \citep{Snelders24}. This FRB showed hyperactivity \citep{Uttarkar24, Zhang24_hyper, OuldBoukattine24} with a burst rate at $1$--$1.5$ GHz that peaked at $\sim 500$ hr$^{-1}$ \citep{Zhang24_hyper}. An unresolved, continuum radio source of $72 \pm 14\ \mu$Jy flux density was initially found in MeerKAT observations at $1.3$ GHz \citep{ZhangYu24}. After this, a compact continuum source was reported with uGMRT at band 4 ($650$ MHz) \citep{Bhusare24}. However, the arcsecond scale resolution achieved in these works did not allow the precise association of the continuum source with the FRB. Very Long Baseline Array (VLBA) observations conducted at $5$ GHz revealed the presence of compact radio emission having peak flux density of $46 \pm 9\ \mu$Jy, co-located with the FRB location \citep{Bruni24}. The angular resolution reached in the latter observations, i.e. $3.7\ {\rm mas} \times 1.7\ {\rm mas}$, was enough to confirm both the PRS nature of this source, as well as the connection with the FRB. Those observations permitted to constrain its size down to $\leq 4$ pc \citep{Bruni24}. We detect the source and measure a peak flux density of $53 \pm 11$ at $1.26$ GHz, consistent with \cite{ZhangWu24_ATEL,Zhang25_FRS}. Our flux measurement provides additional insight into the spectral properties of the source and its temporal variability. \cite{Zhang25_FRS} reported that the source spectrum exhibits significant time variability, with the $800$-$1500$ GHz PRS flux density increasing by a factor of $\approx 2$ in mid-July 2024. Prior to this sudden enhancement, referred to as a “flare” by \cite{Zhang25_FRS}, low-frequency observations carried out in February 2024 (about 170 days earlier) measured a flux density of $64 \pm 14$ $\mu$Jy at 1.4 GHz, compared to $\approx 124$ $\mu$Jy during the flare event. Our S22 observations, conducted on 29 May 2024, indicate that the source was still in a low-flux state at comparable frequencies. Therefore, the flux increase must have occurred on a timescale shorter than $\sim 60$ days, supporting the interpretation of a flare-like event characterized by a rapid and substantial rise in the source luminosity.

\begin{figure}
    \centering
    \includegraphics[width=0.9\columnwidth]{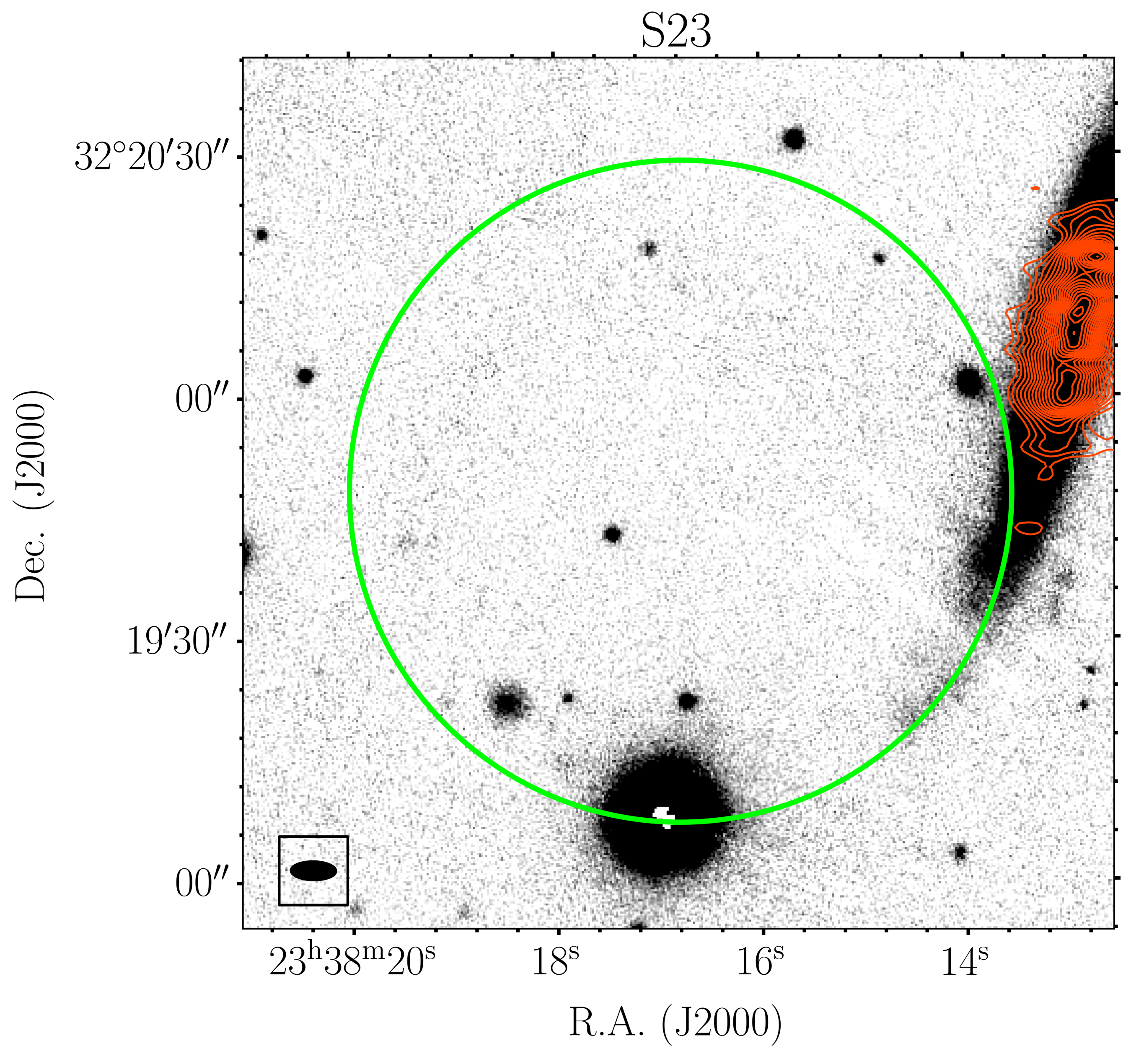}
    \caption{Optical (Pan-STARRS DR1) r-filter image of the S23 localization region. Red contours represent continuum radio emission at $1.26$ GHz as resulting from our uGMRT observations, and are drawn from $3\sigma$, where $\sigma= 16$ $\mu$Jy beam$^{-1}$ is the r.m.s. noise level. The green ellipse represents the FRB localization uncertainties at 95$\%$ C.L \citep{Curtin24_ATEL} The synthesized beam of uGMRT observations is reported in the bottom left corner of the image as a filled black ellipse.}
    \label{fig: S23_loc}
\end{figure}

\subsection{FRBs 20240316A (S23) and 20240619D (S24)}

These are two recently discovered repeating FRBs, presenting moderate bursting activity \citep{Abbott24, Tian25}. S23 is localized with sub-arcminute precision and we do not report any radio emission within its $95\%$ C.L. positional uncertainties. We note that the nominal position of the FRB is $\sim 53''$ south-west with respect to the center of UGC 12714, i.e. an AGN candidate at redshift $z \simeq 0.01613(1)$, which lies in proximity of the $95\%$ C.L. localization of the FRB (see Fig. \ref{fig: S23_loc}). We note that UGC 1714 is an almost completely edge-on galaxy, which could potentially result in a significant contribution to the ${\rm DM}$. The redshift we derived assumed an average host contribution of ${\rm DM}_{\rm host} = 100~{\rm pc~cm^{-3}}$, but this value can be higher for edge-on galaxies \citep[see, e.g.,][]{Orr2024} . Considering the total observed DM of this FRB, ${\rm DM}_{\rm FRB} = 351 \pm 1~{\rm pc~cm^{-3}}$ \citep{Curtin24_ATEL}, and our estimates along the line of sight of ${\rm DM}_{\rm ISM} \approx 48~{\rm pc~cm^{-3}}$ and ${\rm DM}_{\rm halo} \approx 36~{\rm pc~cm^{-3}}$ (see Section \ref{sec: sample_des}), the extragalactic contribution is ${\rm DM}_{\rm EG} = {\rm DM}_{\rm IGM} + {\rm DM}_{\rm host} \simeq 267~{\rm pc~cm^{-3}}$. In the limiting case where the extragalactic DM is entirely due to the host galaxy (i.e. ${\rm DM}_{\rm IGM} = 0$ pc cm$^{-3}$), the implied redshift of the FRB would be $z \geq 0.02$. Given the large positional uncertainties of the FRB and the low overall available information on it, we discard the possibility that the FRB is associated with UGC 12714, and we provide a $2\sigma$ spectral luminosity limit of $5.6 \times 10^{28}$ erg s$^{-1}$ Hz$^{-1}$ for the presence of a PRS, given that no compact sources are present in the source field.

S24 is a repeating FRB initially reported in \cite{Tian25}, which has recently showed hyperactivity and, interestingly, a frequency-dependent burst rate \citep{Tian25}. While the FRB source has been localized at arcsecond precision, its host galaxy is still unknown/undetected \citep{Tian25}. Two unresolved optical sources from the DESI-DR10 catalog \citep{Dey19} are found just outside the edge of the $1\sigma$ FRB localization region \citep[see Fig. 4 of ][]{Tian25}, not providing any secure FRB--host association. From the DM of S24 bursts, i.e. $\sim 470$ pc cm$^{-3}$, we calculate a redshift of $z \approx 0.34$, at which our $2\sigma$ flux density threshold of $36\ \mu$Jy beam$^{-1}$ provide a luminosity UL of $L_{1.2}\leq 9.8 \times 10^{28}$ erg s$^{-1}$ Hz$^{-1}$.

\onecolumn
\begin{center}
\begin{longtable}{clccccc}
\caption{Properties of the FRB sample presented in this work.} \label{table:FRB_sample} \\

\hline\hline
ID & Source name & $z$ & $z$ type & $F_{1.2}$ & $L_{1.2}$ & Refs. \\
& & & & ($\mu$Jy) & ($10^{29}$ erg s$^{-1}$ Hz$^{-1}$) & \\
\hline
\endfirsthead

\multicolumn{6}{c}
{\tablename\ \thetable\ -- Continued} \\
\hline\hline
ID & Source name & $z$ & $z$ type & $F_{1.2}$ & $L_{1.2}$ & Refs. \\
& & & & ($\mu$Jy) & ($10^{29}$ erg s$^{-1}$ Hz$^{-1}$) & \\
\hline
\endhead

\hline \multicolumn{6}{r}{} \\
\endfoot

\hline
\endlastfoot

S1 & r20180301A & $0.3304(1)$ & spec. & $<24$ & $<0.75$ & ($1,2$, this work) \\
S2 & r20180814A & $0.06835(1)$ & spec. & $<24$ & $<2.8 \times 10^{-2}$ & ($3,4$, this work) \\
S3 & r20190303A & $0.06437(1)$ & spec. & $<76$ & $<0.08$ & ($3,4$, this work) \\
S4 & 20190523A & $0.660(2)$ & spec. & $<26$ & $<3.6$ & ($5$, this work) \\
S5 & r20190417A$^{**}$ & $0.12817(2)$ & spec. & $248(20)$ & $1.05(8)$ & ($3,4,6,7$, this work) \\
S6 & 20190714A & $0.2365$ & spec. & $<26$ & $<0.4$ & ($8$, this work) \\
S7 & r20190804E & $0.30^{+0.02}_{-0.10}$ & from DM & $<40$ & $<0.8$ & ($9,10$, this work) \\
S8 & r20191106C & $0.10775(1)$ & spec. & $<84$ & $<0.25$ & ($9,10,11$, this work) \\
S9 & 20191228A & $0.2432(1)$ & spec. & $<30$ & $<0.5$ & ($1$, this work) \\
S10 & 20200216A & $0.46^{+0.02}_{-0.14}$ & from DM & $<26$ & $<1.2$ & ($12$, this work) \\
S11 & r20201114A$^*$ & $0.26^{+0.02}_{-0.08}$ & from DM & $100(15)$ & $1.6(7)$ & ($9,11$, this work) \\
S12 & r20201130A$^*$ & $0.16^{+0.02}_{-0.06}$ & from DM & $360(13)$ & $2.1^{+0.9}_{-1.2}$ & ($9,11$, this work )\\
S13 & 20210317A$^*$ & $0.15(5)$ & phot. & $116(14)$ & $0.68(8)$ & ($12$, this work) \\
S14 & 20210320C$^*$ & $0.2797$ & spec. & $32(7)$ & $0.7(1)$ & ($13,14$) \\
S15 & 20210807D & $0.1292$ & spec. & $<36$ & $<0.15$ & ($15$, this work) \\
S16 & 20210117A & $0.2145$ & spec. & $<30$ & $<0.37$ & ($1,15$, this work) \\
S17 & 20211127I & $0.0469$ & spec. & $<26$ & $<1.4 \times 10^{-2}$ & ($15$, this work) \\
S18 & 20211212A & $0.0715$ & spec. & $<80$ & $<0.1$ & ($15$, this work) \\
S19 & 20220105A & $0.2785$ & spec. & $<44$ & $<0.9$ & ($13,14$, this work) \\
S20 & r20230607A$^*$ & $0.30^{+0.02}_{-0.10}$ & from DM & $90(17)$ & $1.9(8)$ & ($16$, this work) \\
S21 & r20230814A & $0.553$ & spec. & $<42$ & $<3.9$ & ($17,18$, this work) \\
S22 & r20240114A$^{**}$ & $0.1300(2)$ & spec. & $53(11)$ & $0.31(4)$ & ($19,20$, this work) \\
S23 & r20240316A & $0.28^{+0.02}_{-0.10}$ & from DM & $<32$ & $<0.54$ & ($21$, this work) \\
S24 & r20240619D & $0.34^{+0.02}_{-0.10}$ & from DM & $<36$ & $<0.94$ & ($22,23$, this work) \\

\hline
$-$ & 20110523A & $0.6(2)$ & from DM & $<24$ & $<3.9$ & ($24,25$) \\
R1 & r20121102A$^{**}$ & $0.19273(8)$ & spec. & $194(19)$ & $1.9(2)$ & ($26-28$)\\
$-$ & 20150215A & $0.7(2)$ & from DM & $<7.7$ & $<1.6$ & ($29$) \\
$-$ & 20150418A & $0.6(2)$ & from DM & $<49$ & $<7.9$ & ($30$) \\
$-$ & 20150807A & $0.2(1)$ & from DM & $<241$ & $<4.9$ & ($31$) \\
$-$ & r20171019A & $0.44^{+0.04}_{-0.12}$ & from DM & $<10.4$ & $<0.5$ & ($32$) \\
$-$ & 20171020A & $0.0087(5)$ & spec. & $<95$ & $<1.7 \times 10^{-3}$ & ($33$) \\
$-$ & 20180309A & $0.2(1)$ & from DM & $<82$ & $<1.6$ & ($25,34,35$) \\
R3 & r20180916B & $0.0337(2)$ & spec. & $<13$ & $<3.5 \times 10^{-3}$ & ($36,37$) \\
$-$ & 20180924B & $0.3214(2)$ & spec. & $<21$ & $<0.6$ & ($38$) \\
$-$ & r20181030A$^*$ & $0.00385(2)$ & spec. & $417(21)$ & $1.47(7) \times 10^{-3}$ & ($6,39$) \\
$-$ & 20181112A & $0.4755(2)$ & spec. & $<10$ & $<0.68$ & ($8,40,41$)\\
$-$ & 20190102C & $0.2913(2)$ & spec. & $<16$ & $<0.38$ & ($8,41,42$) \\
$-$ & r20190117A & $0.34^{+0.02}_{-0.10}$ & from DM & $<33$ & $<0.85$ & ($3,6$) \\
$-$ & r20190208A & $0.54^{+0.04}_{-0.14}$ & from DM & $<11$ & $<0.8$ & ($3,4,6$) \\
R1-twin & r20190520B$^{**}$ & $0.241(1)$ & spec. & $218(9)$ & $3.4(1)$ & ($43,44$) \\
$-$ & 20190608B & $0.1178$ & spec. & $<11$ & $<4 \times 10^{-2}$ & ($41,45,46$) \\
$-$ & 20190611B & $0.378$ & spec. & $<16$ & $<0.67$ & ($7,42,45$) \\
$-$ & 20190614D & $1.00^{+0.06}_{-0.24}$ & from DM & $<23$ & $<6.4$ & ($47$) \\
$-$ & r20190711A & $0.522$ & spec. & $<30$ & $<2.5$ & ($32,45$) \\
$-$ & 20191001A & $0.2340(1)$ & spec. & $<45$ & $<0.7$ & ($48$) \\
$-$ & 20191108A & $0.5(2)$ & from DM & $<148$ & $<18$ & ($49$) \\
$-$ & r20200120E & $0.00013(6)$ & spec. & $<14$ & $<2.1 \times 10^{-6}$ & ($50-52$) \\
$-$ & 20200428 & MW & $-$ & $1.37(8) \times 10^6$ & $<2 \times 10^{-6}$ & ($53-55$) \\
$-$ & 20200430A & $0.1608$ & spec. & $<17$ & $<0.2$ & ($8,42$) \\
$-$ & 20200906A$^*$ & $0.3688(1)$ & spec. & $52(7)$ & $2.1(3)$ & ($1,42$) \\
$-$ & r20201124A$^*$ & $0.0978(2)$ & spec. & $\approx 2$ & $4.4 \times 10^{-3}$ & ($56-58$) \\
$-$ & 20210410D & $0.1415$ & spec. & N/A$^{\rm a}$ & $<0.31$ & ($42,59$) \\
$-$ & 20210912A & $1.43^{+0.10}_{-0.32}$ & from DM & N/A & $<78$ & ($42,60$) \\
$-$ & 20220207C & $0.04304(1)$ & spec. & $<193$ & $<9 \times 10^{-2}$ & ($61, 62$) \\
$-$ & 20220222C & $0.853$ & spec. & $<18$ & $<4.3$ & ($42,63$) \\
$-$ & 20220307B & $0.2481(1)$ & spec. & $<193$ & $<3.3$ & ($61,62$) \\
$-$ & 20220310F & $0.4779(4)$ & spec. & $<187$ & $<13$ & ($61,62$) \\
$-$ & 20220319D & $0.01123(4)$ & spec. & $<44$ & $<1.3 \times 10^{-3}$ & ($61,62$) \\
$-$ & 20220418A & $0.6220(1)$ & spec. & $<182$ & $<22$ & ($61,62$) \\
$-$ & 20220501C & $0.381$ & spec. & N/A & $<3.8$ & ($14,42$) \\
$-$ & 20220506D & $0.3004(1)$ & spec. & $<195$ & $<4.9$ & ($61,62$) \\
$-$ & 20220509G & $0.0894(1)$ & spec. & $<192$ & $<0.4$ & ($61,62$) \\
$-$ & r20220529A & $0.1839(1)$ & spec. & $<22$ & $<0.3$ & ($64$) \\
$-$ & 20220717A & $0.36295$ & spec. & $<28$ & $<1.1$ &  ($42,65$) \\
$-$ & 20220725A & $0.1926$ & spec. & N/A & $<3.5$ & ($14,42$) \\
$-$ & 20220825A & $0.24139(1)$ & spec. & $<192$ & $<3.1$ & ($61,62$) \\
$-$ & 20220905A & $0.66^{+0.04}_{-0.18}$ & from DM & $<5.2$ & $<0.6$ &  ($42,65$) \\
$-$ & r20220912A & $0.0771(1)$ & spec. & $<33$ & $<4.9 \times 10^{-2}$ & ($66-68$) \\
$-$ & 20220914A & $0.1139$ & spec. & $<189$ & $<0.62$ & ($61$) \\
$-$ & 20220918A & $0.491$ & spec. & N/A & $<6.3$ & ($14,42$) \\
$-$ & 20220920A & $0.1582(2)$ & spec. & $<193$ & $<1.3$ & ($61,62$) \\
$-$ & 20221012A & $0.28467(7)$ & spec. & $<189$ & $<4.3$ & ($61,62$) \\
$-$ & 20221106A & $0.2044$ & spec. & N/A & $<0.63$ & ($14,42$) \\
$-$ & 20230125D & $0.3265$ & spec. & $<18$ & $<0.55$ & ($42,63$) \\
$-$ & 20250316A & $0.009$ & spec. & $<16$ & $<2.1 \times 10^{-4}$ & ($69-71$) \\

\end{longtable}
\tablefoot{Columns are, from left to right, the FRB ID as adopted in the text, its Transient Name Server (TNS) identifier, with confirmed repeaters identified with an `r'' at the beginning of the name, the redshift of the FRB and the redshift type (whether it is spectroscopic, photometric or computed from the FRB DM), the flux density at the FRB localization, and the corresponding spectral luminosity. Flux and luminosity limits are reported at $95\%$ C.L. The table is divided by an horizontal line in sources for which we report new observations in this work and observations from the literature. FRB sources for which a co-spatial persistent emission has been found (either in our observations or in observations reported in the literature) are denoted with a $^*$ superscript, while already confirmed PRSs are denoted with $^{**}$. \\
$^{\rm a}$Literature sources whose flux is denoted with ``N/A'' present co-spatial extended radio emission. For these sources we cannot define an appropriate CL for a flux upper limit on the presence of a compact PRS. The corresponding upper limit on $L_{1.2}$ denotes the luminosity of the peak flux density reported in the reference work.

\textbf{References}. (1) \cite{Bhandari22}, (2) \cite{Kumar23}, (3) \cite{Michilli23}, (4) \cite{McKinven23}, (5) \cite{Ravi19_loc}, (6) \cite{Ibik24}, (7) \cite{Moroianu26}, (8) \cite{Heintz20}, (9) \cite{chime23}, (10) \cite{Ibik24_b}, (11) \cite{Ng25}, (12) \cite{PastorMarazuela25}, (13) \cite{Gordon23}, (14) \cite{Shannon25}, (15) \cite{Woodland24}, (16) \cite{Zhou25}, (17) \cite{Ravi23_ATEL}, (18) \cite{Connor24}, (19) \cite{Snelders24}, (20) \cite{Tian24}, (21) \cite{Curtin24_ATEL}, (22) \cite{Tian25}, (23) \cite{Shaji26}, (24) \cite{Masui15}, (25) \cite{Bruni23}, (26) \cite{Chatterjee17}, (27) \cite{Hilmarsson21}, (28) \cite{Tendulkar21}, (29) \cite{Petroff17}, (30) \cite{Keane16}, (31) \cite{Ravi16}, (32) \cite{Chibueze21}, (33) \cite{LeeWaddell23},  (34) \cite{Aggarwal21}, (35) \cite{Oslowski19}, (36) \cite{Marcote20},  (37) \cite{CHIME18}, (38) \cite{Bannister19}, (39) \cite{Bhardwaj21}, (40) \cite{Prochaska19}, (41) \cite{Bhandari20b}, (42) \cite{Mfulwane26}, (43) \cite{Niu21}, (44) \cite{AnnaThomas23}, (45) \cite{Day19}, (46) \cite{Macquart20}, (47) \cite{Law20}, (48) \cite{Bhandari20a}, (49) \cite{Connor20}, (50) \cite{Bhardwaj21a}, (51) \cite{Kirsten22glob}, (52) \cite{Koss22}, (53) \cite{CHIME20b}, (54) \cite{Bochenek20a}, (55) \cite{Kothes18}, (56) \cite{Kilpatrick21}, (57) \cite{Nimmo21}, (58) \cite{Xu2022}, (59) \cite{Caleb23}, (60) \cite{Marnoch23}, (61) \cite{Law23}, (62) \cite{Sherman24}, (63) \cite{PastorMarazuela25b}, (64) \cite{Phandi26}, (65) \cite{Rajwade24}, (66) \cite{ravi23}, (67) \cite{Zhang23}, (68) \cite{Hewitt24}, (69) \cite{Ng25_CHIME}, (70) \cite{Shion25_CHIME}, (71) \cite{An25}.}
\end{center}

\begin{table*}[h]
\centering
\caption{Localization properties of the sources presented in this work.}
\label{tab:loc_prop_sources}
\begin{tabularx}{\textwidth}{cccXc}
\hline
\hline
ID & Loc. Area & Host galaxy & Notes & Ref. \\
& (arcsec$^2$) & & & \\
\hline
S1  & $1.17$     &  PSO J093.2268+04.6703 & SFG & (1) \\
S2  & $1.1 \times 10^3$     & WISEA J042256.03+733940.7 & Massive and passive red spiral & (2) \\
S3  & $4.2 \times 10^2$      & J135159.86+480714.04 & SFG in an interacting galaxy pair & (2) \\
S4  & $15.7$     &  DSA-10 J134815.6+722811 & Quiescent & (3) \\
S5$^{**}$  & $8 \times 10^{-5}$     & & Low-metallicity, dwarf SFG & (4) \\
S6  & $0.26$     & PSO J121555.0941–130116.004 & Possible spiral & (5) \\
S7  & $2.4 \times 10^3$     & -- & & (6) \\
S8  & $1.4 \times 10^3$     & WISEA J131819.23+425958.9 & High-metallicity, possibly
elliptical, with ongoing star formation& (6,7) \\
S9  & $2.41$     & J225743.40-293539.1 & SFG & (8) \\
S10 & $1.3 \times 10^3$     & -- & & (9) \\
S11$^{*}$ & $3.6 \times 10^4$   & -- & & (6) \\
S12$^{*}$ & $1.5 \times 10^3$     & -- & & (6) \\
S13$^{*}$ & $2.9 \times 10^2$      & -- & & (9) \\
S14$^*$ & $0.66$     & PSO J204.4588-16.1226 & SFG & (10) \\
S15 & $0.5$      & J195652.82-004544.4 & AGN & (1,10) \\
S16 & $5 \times 10^{-2}$   & J223955.07–160905.37 & Dwarf galaxy & (1) \\
S17 & $0.5$      & WALLABY J131913–185018 & SFG (spiral) + AGN & (1,10) \\
S18 & $1.51$     & SDSS J102924.22+012139.2 & SFG (spiral) & (1) \\
S19 & $3.89$     & WISEA J135512.93+222758.6 & AGN & (10) \\
S20$^{*}$ & $9.3 \times 10^3$     & -- & & (11) \\
S21 & $6.28$     & J222354.26+730132.77 & & (11,12) \\
S22$^{**}$ & $0.13$     & SDSS J212739.84+041945.8 & Low-metallicity, dwarf SFG & (13,14) \\
S23 & $4.1 \times 10^3$     & -- & & (15) \\
S24 & $1.54$     & -- & & (16) \\
\hline
\end{tabularx}
\tablefoot{Localization properties for the sources of the catalog we present in this work. In the second column the localization area is reported, computed as $\pi ab$, with $a$ and $b$ being the semi-minor and semi-major axes of the FRB localization ellipse, given at $68\%$ C.L., respectively, as can be found in corresponding Refs. (last column). The third column lists the host galaxy associated with each source; entries are left blank when the host galaxy name is unknown, while a dash (--) is used when no secure association between the FRB and a host galaxy is established. The fourth column reports any notes regarding a given FRB host galaxy. As in Table \ref{tab:uGMRT_obs}, sources for which a co-spatial persistent emission has been found are denoted with a $^*$ superscript, while already confirmed PRSs are denoted with $^{**}$.} 

\tablebib{(1) \cite{Woodland24}, (2) \cite{Michilli23}, (3) \cite{Ravi19_loc}, (4) \cite{Moroianu26}, (5) \cite{Heintz20}, (6) \cite{chime23}, (7) \cite{Ibik24_b}, (8) \cite{Bhandari22}, (9) \cite{PastorMarazuela25}, (10) \cite{Gordon23}, (11) \cite{Zhou25}, (12) \cite{Connor24}, (13) \cite{Snelders24}, (14) \cite{Chen25}, (15) \cite{Curtin24_ATEL}, (16) \cite{Tian25}.}
\end{table*}

\begin{sidewaystable*}  
\centering
\tiny
\caption{Specifics of the observations reported in this work.}
\label{tab:uGMRT_obs}
\begin{tabular}{lccllccccccccc}
\hline
\hline
ID & \multicolumn{2}{c}{Coordinates} & \multicolumn{2}{c}{Calibrators} & Obs. date & On source time &  $\theta_{\rm synth.}$ & RMS ($\mu$Jy beam$^{-1}$) & $\theta_{\rm synth.}$ & RMS ($\mu$Jy beam$^{-1}$) & $\theta_{\rm synth.}$ & RMS ($\mu$Jy beam$^{-1}$) \\
    & $\alpha_{\rm J2000}$ &   $\delta_{\rm J2000}$   &               &                 & (MJD) & (h) & \multicolumn{2}{c}{{\tt briggs} -0.5} &  \multicolumn{2}{c}{{\tt briggs} 0} & \multicolumn{2}{c}{{\tt briggs} +0.5} \\
\hline
S1 & $06^{\rm h}12^{\rm m}54.44^{\rm s}$ & $+04^\circ40'15.8''$ & 3C147 & 0632+103 & $60496$ & $3.5$ & $2.5'' \times 1.5''$ & $22$ & $2.9'' \times 1.9''$ & $15$ & $3.5'' \times 2.6''$ & $12$ \\
S2 & $04^{\rm h}22^{\rm m}44.0^{\rm s}$ & $+73^\circ39'52.0''$ & 3C147 & 0354+729 & $60495$ & $3.9$ & $3.3'' \times 1.3''$ & $19$ & $4.0'' \times 1.6''$ & $14$ & $5.4'' \times 2.0''$ & $12$ \\
S3 & $13^{\rm h}51^{\rm m}59.0^{\rm s}$ & $+48^\circ07'16.0''$ & 3C286 & 1345+497 & $60485$ & $3.5$ & $2.5'' \times 2.0''$ & $42$ & $2.9'' \times 2.4''$ & $30$ & $3.7'' \times 3.2''$ & $27$ \\
S4 & $13^{\rm h}48^{\rm m}15.6^{\rm s}$ & $+72^\circ28'11.0''$ & 3C286 & 1435+760 & $60480$ & $4.3$ & $2.5'' \times 1.5''$ & $18$ & $3.2'' \times 1.8''$ & $14$ & $4.2'' \times 2.3''$ & $13$ \\
S5 & $19^{\rm h}39^{\rm m}03.84^{\rm s}$ & $+59^\circ19'54.99''$ & 3C48 & 1944+548 & $60496$ & $2$ & $2.5'' \times 1.6''$ & $21$ & $3.2'' \times 1.8''$ & $15$ & $4.3'' \times 2.0''$ & $13$ \\
S6 & $12^{\rm h}15^{\rm m}55.13^{\rm s}$ & $-13^\circ01'15.6''$ & 3C147 & 1215-175 & $60497$ & $3.5$ & $2.0'' \times 1.8''$ & $20$ & $2.4'' \times 2.0''$ & $14$ & $3.1'' \times 1.4''$ & $13$ \\
S7 & $17^{\rm h}25^{\rm m}21.60^{\rm s}$ & $+55^\circ04'08.40''$ & 3C286 & 1549+506 & $60866$ & $0.5$ & $6.1'' \times 1.3''$ & $31$ & $6.3'' \times 1.6''$ & $24$ & $6.3'' \times 2.2''$ & $22$ \\
S8 & $13^{\rm h}18^{\rm m}19.20^{\rm s}$ & $+43^\circ00'07.20''$ & 3C286 & 1549+506 & $60866$ & $0.7$ & $1.6'' \times 1.0''$ & $42$ & $2.9'' \times 1.5''$ & $31$ & $4.4'' \times 1.7''$ & $26$ \\
S9 & $22^{\rm h}57^{\rm m}43.3^{\rm s}$ & $-29^\circ35'38.7''$ & 3C48 & 2258-279 & $60495$ & $3.5$ & $3.8'' \times 1.3''$ & $21$ & $3.8'' \times 1.5''$ & $16$ & $4.5'' \times 1.9''$ & $15$ \\
S10 & $22^{\rm h}08^{\rm m}24.7^{\rm s}$ & $+16^\circ35'34.6''$ & 3C48 & 2139+143 & $60731$ & $1.9$ & $2.5'' \times 1.3''$ & $21$ & $2.6'' \times 1.5''$ & $15$ & $3.0'' \times 2.0''$ & $13$ \\
S11 & $14^{\rm h}46^{\rm m}21.60^{\rm s}$ & $+71^\circ47'24.00''$ & 3C286 & 1435+760 & $60866$ & $0.7$ & $6.8'' \times 1.2''$ & $33$ & $7.2'' \times 1.5''$ & $23$ & $8.9'' \times 1.7''$ & $20$ \\
S12 & $4^{\rm h}17^{\rm m}33.60^{\rm s}$ & $+07^\circ56'24.00''$ & 3C48 & 0423-013 & $60848$ & $0.8$ & $2.5'' \times 1.3''$ & $29$ & $2.4'' \times 1.5''$ & $21$ & $3.4'' \times 1.7''$ & $18$ \\
S13 & $19^{\rm h}36^{\rm m}27.4^{\rm s}$ & $+59^\circ51'50.7''$ & 3C147 & 1944+548 & $60730$ & $2.1$ & $2.8'' \times 1.5''$ & $16$ & $3.5'' \times 1.9''$ & $18$ & $4.3'' \times 2.6''$ & $27$ \\
S14 & $13^{\rm h}37^{\rm m}50.10^{\rm s}$ & $-16^\circ07'21.6''$ & 3C286 & 1351-148 & $60866$ & $0.7$ & $3.3'' \times 1.4''$ & $30$ & $4.0'' \times 1.6''$ & $22$ & $6.7'' \times 1.7''$ & $18$ \\
S15 & $19^{\rm h}56^{\rm m}53.14^{\rm s}$ & $-00^\circ45'44.5''$ & 3C286 & 1956-076 & $60478$ & $1.9$ & $1.9'' \times 1.3''$ & $32$ & $2.2'' \times 1.7''$ & $22$ & $7.7'' \times 1.4''$ & $18$ \\
S16 & $22^{\rm h}39^{\rm m}55.04^{\rm s}$ & $-16^\circ09'5.36''$ & 3C48 & 2246-121 & $60496$ & $2$ & $3.0'' \times 1.6''$ & $18$ & $4.0'' \times 1.8''$ & $14$ & $5.0'' \times 2.4''$ & $15$ \\
S17 & $13^{\rm h}19^{\rm m}13.97^{\rm s}$ & $-18^\circ50'16.10''$ & 3C286 & 1311-222 & $60497$ & $3.5$ & $2.6'' \times 1.4''$ & $21$ & $2.7'' \times 1.7''$ & $15$ & $3.3'' \times 2.0''$ & $13$ \\
S18 & $10^{\rm h}29^{\rm m}24.16^{\rm s}$ & $+01^\circ21'37.67''$ & 3C286 & 1041+027 & $60480$ & $3.8$ & $3.1'' \times 1.5''$ & $40$ & $3.0'' \times 1.8''$ & $30$ & $3.5'' \times 2.0''$ & $30$ \\
S19 & $13^{\rm h}55^{\rm m}12.81^{\rm s}$ & $+22^\circ27'58.4''$ & 3C286 & 1327+221 & $60866$ & $0.7$ & $6.7'' \times 1.4''$ & $35$ & $9.4'' \times 1.8''$ & $26$ & $5.3'' \times 1.8''$ & $22$ \\
S20 & $21^{\rm h}53^{\rm m}35.0^{\rm s}$ & $+11^\circ17'04.20''$ & 3C48 & 2130+050 & $60822$ & $4.3$ & $2.1'' \times 1.3''$ & $20$ & $2.5'' \times 1.7''$ & $15$ & $2.9'' \times 2.3''$ & $17$ \\
S21 & $22^{\rm h}23^{\rm m}53.9^{\rm s}$ & $+73^\circ01'33.3''$ & 3C48 & 2250+714 & $60478$ & $2$ & $3.7'' \times 1.5''$ & $34$ & $4.3'' \times 1.7''$ & $25$ & $5.7'' \times 2.0''$ & $21$ \\
S22 & $21^{\rm h}27^{\rm m}39.849^{\rm s}$ & $+04^\circ19'46.00''$ & 3C286 & 2130+0502 & $60459$ & $8.6$ & $2.0'' \times 1.5''$ & $13$ & $2.5'' \times 1.8''$ & $10$ & $3.1'' \times 2.5''$ & $10$ \\
S23 & $23^{\rm h}38^{\rm m}16.8^{\rm s}$ & $+32^\circ19'48.0''$ & 3C147 & 2355+498 & $60546$ & $2.4$ & $3.5'' \times 1.8''$ & $24$ & $4.4'' \times 2.0''$ & $18$ & $5.7'' \times 2.4''$ & $16$ \\
S24 & $19^{\rm h}49^{\rm m}29.2^{\rm s}$ & $-25^\circ12'49.4''$ & 3C286 & 1949-199 & $60729$ & $1.9$ & $2.6'' \times 1.7''$ & $20$ & $3.5'' \times 1.9''$ & $16$ & $4.8'' \times 2.3''$ & $18$ \\
\hline
\end{tabular}
\tablefoot{For each source, we report its ID ($1^{\rm st}$ col.), coordinates ($2^{\rm nd}$ and $3^{\rm rd}$ cols.), the calibrators adopted for the data reduction (primary and secondary calibrators in $4^{\rm th}$ and $5^{\rm th}$ col., respectively), the synthesized beam ($6^{\rm th}$, $8^{\rm th}$ and $10^{\rm th}$ cols.) and RMS noise ($7^{\rm th}$, $9^{\rm th}$ and $11^{\rm th}$ cols.) reached after the imaging prodcedure. We list the latter for three different weighting schemes, depending on the value of the {\tt briggs} parameter used.}
\end{sidewaystable*}

\end{appendix}

\end{document}